\documentclass[12pt,a4paper]{article}
\pdfoutput=1

\usepackage{graphicx,array}
\usepackage{hyperref}
\usepackage[normalem]{ulem} 
\usepackage{cite}
\usepackage{color}
\usepackage{appendix}
\usepackage{epsfig}
\usepackage{latexsym}
\usepackage{amsmath}
\usepackage{amssymb}
\usepackage{bm}
\usepackage{hyperref}
\usepackage{slashed}
\usepackage{bbold}
\usepackage{subfig}
\usepackage{multirow}
\usepackage{adjustbox}
\usepackage{colortbl}
\usepackage{float}

\usepackage{tikz}
\usepackage{calc}
\def\checkmark{\tikz\fill[scale=0.4](0,.35) -- (.25,0) -- (1,.7) -- (.25,.15) -- cycle;} 
\def\checked{\resizebox{\widthof{\checkmark}*\ratio{\widthof{x}}{\widthof{\normalsize x}}}{!}{\checkmark}}

\definecolor{darkblue}{cmyk}{1,0.3,0,0.2}
\definecolor{violet}{cmyk}{0,1,0,0.2}
\hypersetup{colorlinks, bookmarksnumbered, citecolor=darkblue, linkcolor=darkblue, pdfstartview=FitH, urlcolor=darkblue, linktocpage}

\numberwithin{equation}{section}

\newcommand{\be}{\begin{equation}}
\newcommand{\ee}{\end{equation}}
\newcommand{\bea}{\begin{eqnarray}}
\newcommand{\eea}{\end{eqnarray}}

\newcommand{\MeV}{\textrm{ MeV}}
\newcommand{\GeV}{\textrm{ GeV}}
\newcommand{\TeV}{\textrm{ TeV}}
\newcommand{\SU}{\textrm{SU}}

\newcommand{\U}{\textrm{U}}

\newcommand{\SM}{\textrm{SM}}
\newcommand{\gsim}{\lower.7ex\hbox{$\;\stackrel{\textstyle>}{\sim}\;$}}
\newcommand{\lsim}{\lower.7ex\hbox{$\;\stackrel{\textstyle<}{\sim}\;$}}

\newcommand{\LL}{\mathcal{L}}

\newcommand{\OO}{\mathcal{O}}

\newcommand{\NN}{\mathcal{N}}

\newcommand{\Br}{\textrm{Br}}

\newcommand{\ba} {\begin{eqnarray}}
\newcommand{\ea} {\end{eqnarray}}

\renewcommand{\Re}{\mathop{\rm Re}}
\renewcommand{\Im}{\mathop{\rm Im}}

\newcommand{\laL}[1]{\lambda^{#1L}}
\newcommand{\laLT}[1]{\lambda^{#1L \, T}}
\newcommand{\laLdag}[1]{\lambda^{#1L \, \dag}}
\newcommand{\laLst}[1]{\lambda^{#1L \, *}}
\newcommand{\laR}{\lambda^{1R}}

\newcommand{\laRdag}{\lambda^{1R \, \dag}}
\newcommand{\laRst}{\lambda^{1R \, *}}

\begin{document}
 
 \hfill

\begin{flushright}
\hspace{3cm} 
SISSA 20/2020/FISI\\
TUM-HEP-1278/20
\end{flushright}

\vspace{1.0cm}

\begin{center}
{\LARGE\bf Low-energy phenomenology \\[0.3cm] of scalar leptoquarks at one-loop accuracy}
\\ \vspace*{0.5cm}

\bigskip\vspace{1cm}{
{\large Valerio Gherardi$^{a, b}$, David Marzocca$^{b}$, Elena Venturini$^{c}$}
} \\[7mm]

{\em $(a)$ SISSA, Via Bonomea 265, 34136, Trieste, Italy}  \\  
{\em $(b)$ INFN, Sezione di Trieste, SISSA, Via Bonomea 265, 34136, Trieste, Italy}  \\
{\em $(c)$ Technische Universit{\"a}t M{\"u}nchen, Physik-Department, James-Franck-Stra\ss e 1, 85748 Garching, Germany}\\

\vspace*{0.5cm}
   
\end{center}
\vspace*{1.5cm}

\centerline{\large\bf Abstract}
\medskip\noindent 

We perform a complete study of the low-energy phenomenology of $S_1$ and $S_3$ leptoquarks, aimed at addressing the observed deviations in $B$-meson decays and the muon magnetic dipole moment. Leptoquark contributions to observables are computed at one-loop accuracy in an effective field theory approach, using the recently published complete one-loop matching of these leptoquarks to the Standard Model effective field theory. We present several scenarios, discussing in each case the preferred parameter space and the most relevant observables. 

\vspace{0.3cm}

\newpage
\tableofcontents

\newpage
%%%%%%%%%%%%%%%%%%%%%%%%%%%%%%%%%%
\section{Introduction}

The Standard Model (SM) of particle physics provides an excellent description of physical phenomena in a wide range of energies and scales. 
Despite no direct evidence for new physics emerged in direct searches at the LHC, for several years now some low energy measurements continue to show significant deviations from the respective SM predictions, which fuel the hope that some New Physics (NP) might be lurking somewhere at the TeV scale. 
The most significant and robust deviations, that we take into account in this work, are the following:
\begin{itemize}
	\item deviations from the SM predictions in the lepton flavour universality (LFU) ratios of semileptonic $B$-meson decays in $\tau$ vs. light leptons, $R(D^{(*)}) = \Br(B \to D^{(*)} \tau \nu) / \Br(B \to D^{(*)} \ell \nu)$ (where $\ell = \mu, e$)  \cite{Lees:2012xj,Lees:2013uzd,Aaij:2015yra,Huschle:2015rga,Sato:2016svk,Hirose:2016wfn,Hirose:2017dxl,Aaij:2017uff,Aaij:2017deq,Siddi:2018avt,Belle:2019rba},
	\item a deficiency in LFU ratios of rare $B$ decays in muons vs. electrons, $R(K^{(*)}) = \Br(B \to K^{(*)} \mu\mu) / \Br(B \to K^{(*)} ee)$ \cite{Aaij:2014ora,Aaij:2017vbb,Aaij:2019wad,Abdesselam:2019wac},
	\item deviations in differential angular distributions of the $B \to K^* \mu^+ \mu^-$ decay, as well as in several branching ratios of $b \to s \mu \mu$ processes \cite{Aaij:2013qta,Aaij:2015oid,Aaij:2015esa,Aaij:2017vad,Aaboud:2018mst,Aaij:2020nrf},
	\item a longstanding deviation from the SM prediction in the muon anomalous magnetic moment $(g - 2)_\mu$ \cite{Bennett:2006fi,Aoyama:2020ynm}.
\end{itemize}
While also other measurements show deviations from the SM, those above stand out and have been the focus of a large amount of theoretical and experimental effort. In all cases, large theory efforts for improving the SM predictions (often very challenging) have been undertaken, and several experimental endeavours and analyses have been set up for confirming, disproving, or providing cross-checks, for the anomalies. Indeed, new measurements scheduled to appear within the next few years are expected to clarify the nature of all these anomalies. A confirmation for the presence of new physics in any one of these observables would of course be revolutionary in our understanding of physics at the TeV scale.

For the same reasons, an equally large effort has been put into finding possible new physics explanations. In case of the $B$ anomalies, leptoquarks (LQ) at the TeV scale can provide good explanations, even combining neutral and charged-current anomalies. If they couple to both left and right-handed muons, also the muon anomalous magnetic moment could be addressed.
In all scenarios, in order to find a good explanation it is necessary to consider the constraints imposed by a large set of observables generated both at tree-level and radiatively. In some cases, renormalization group evolution (RGE) of the operators generated at the matching scale down to the scale of the observables represent the leading radiative effect \cite{Feruglio:2016gvd,Feruglio:2017rjo,Cornella:2018tfd,Alguero:2019ptt}, however since the logarithm is often just of $\mathcal{O}(1)$, finite contributions can have a relevant impact.

In case of vector leptoquarks, such finite terms are calculable only in ultraviolet-complete models, thus making the analysis necessarily model-dependent (see e.g. \cite{DiLuzio:2017vat,DiLuzio:2018zxy,Fuentes-Martin:2019ign,Fuentes-Martin:2020luw} for analyses of specific gauge models of lepton-quark unification).
On the other hand, scalar leptoquarks can be considered as self-consistent simplified models, and all observables can be computed precisely in terms of the LQ couplings and masses.
A particularly promising set of LQ to address the observed anomalies are the $S_1 = ({\bf \bar 3}, {\bf 1}, 1/3)$ and $S_3 = ({\bf \bar 3}, {\bf 3}, 1/3)$ representations.\footnote{We show the representations under the SM gauge group $\SU(3)_c \times \SU(2)_L \times U(1)_Y$} Several works have been dedicated to study their phenomenology.
The $S_1$ leptoquark has been considered as possible mediator for all anomalies \cite{Davidson:1993qk,Dorsner:2013tla,Bauer:2015knc,Becirevic:2016oho,Cai:2017wry,Angelescu:2018tyl,Azatov:2018kzb,Aydemir:2019ynb,Dorsner:2019itg,Crivellin:2019qnh}, with varying degree of success. $S_3$, instead, has long been recognized to be a very good candidate to address the deviations in the $b \to s \mu \mu$ transition \cite{Gripaios:2009dq,Gripaios:2014tna,Gripaios:2015gra,Bhattacharya:2016mcc,Dorsner:2017ufx,Kumar:2018kmr,deMedeirosVarzielas:2019lgb,Davighi:2020qqa}. 
Finally, the combination of both leptoquarks has been considered as a good combined explanation of charged and neutral-current $B$-anomalies \cite{Crivellin:2017zlb,Buttazzo:2017ixm} and possible ultraviolet (UV) completions have been proposed in terms of a composite Higgs model \cite{Marzocca:2018wcf}, combining flavour anomalies with a solution to the Higgs hierarchy problem, as well as in the framework of asymptotically safe quantum gravity \cite{Kowalska:2020gie}.

More recently, one-loop computations of several observables in this model have been published \cite{Arnan:2019olv,Crivellin:2019dwb,Saad:2020ihm,Crivellin:2020ukd}.
The approach adopted in these works is to compute directly in the model the dominant one-loop contributions to the desired observables. This methodology is however prone to missing possible relevant effects, and is not suitable to be systematically generalizable.

In this work we aim to perform a complete one-loop analysis of the $S_1 + S_3$ model, focussed at addressing the anomalies listed above, while being consistent with all relevant experimental constraints. We adopt an approach based on effective field theories (EFT),  leveraging on our previous work \cite{Gherardi:2020det} where the complete one-loop matching of the $S_1 + S_3$ model to gauge-invariant dimension-six operators of the SMEFT, in the Warsaw basis \cite{Grzadkowski:2010es}, is presented.
The EFT approach is designed to factorize the UV-dependent part of the problem, i.e. the UV matching, from the purely low-energy one.
The latter involves RGE of the EFT coefficients to the energy scale of the observables and the computation of the observables at one-loop, within the EFT, see e.g. Ref.~\cite{Jiang:2018pbd} for a simpler case of a scalar singlet.
As we shall describe, most of these steps are already available in the literature in complete generality.
The complete one-loop UV matching, done manually as in \cite{Gherardi:2020det}, requires a substantial amount of work, however it is possible to proceed systematically without neglecting terms. Furthermore, this step is expected to become automatised in the near future.
This will facilitate extending this work to include more observables, or to apply it to different UV models.
In case of leptoquarks and low-energy observables, the use of EFT approaches is even more justified by the collider bounds from LHC, which put lower bounds on leptoquark masses close to the $\approx 1 \TeV$ scale, see e.g. Refs.~\cite{Marzocca:2018wcf,Saad:2020ihm} for recent reviews of pair production searches of $S_1$ and $S_3$.
Truncating the EFT expansion at dimension-six implies an implicit uncertainty in the evaluation of the NP contributions to observables, due to missing higher-dimension operators, that can be estimated being of $\mathcal{O}(E^2 / M_{\rm LQ}^2)$ or $\mathcal{O}(m_{\rm EW}^2 / M_{\rm LQ}^2)$ compared to the corresponding dimension-six contribution, where $E$ is the typical energy of the process under consideration and $m_{\rm EW}$ an electroweak-scale mass. While former effects are completely negligible, the latter are $\sim 1\%$ for TeV-scale leptoquarks, which do not affect the results in any sizeable way, given present day precision in the observables.\footnote{Dimension-eight terms could be relevant if they generate at tree-level an observable that is instead loop-induced at dimension-six. This, however, does not happen in this UV model for the observables under consideration.}

Our goal is to find interesting scenarios, within the $S_1 + S_3$ setup, capable of addressing one or more of the anomalies listed above, find the preferred region in parameter space, and discuss the most important experimental constraints in each case.
Specifically, we first aim to quantify how well each leptoquark can address which set of anomalies, then we discuss combined explanations with both leptoquarks.
Thanks to the complete one-loop matching, we also discuss limits on leptoquark couplings to the SM Higgs boson, arising from electroweak precision data and Higgs measurements.

In Sec.~\ref{sec:setup} we present the $S_1 + S_3$ model, the methodology employed in the analysis, and present the list of all observables included in the fit, including a discussion of the relevant collider bounds, particularly those from Drell-Yan.
The results for all scenarios considered are collected in Sec.~\ref{sec:scenarios} and a discussion on future prospects can be found in Sec.~\ref{sec:prospects}. We conclude in Sec.~\ref{sec:conclusions}.
In App.~\ref{app:obs} we describe in details the LQ contributions to all the observables considered.

%%%%%%%%%%%%%%%%%%%%%%%%%%%%%%%%%
\section{Setup}
\label{sec:setup}

The Lagrangian for the two leptoquarks is the following
\be\begin{split}
	\LL_{\text{LQ}} &= |D_\mu S_1|^2 + |D_\mu S_3|^2 - M_{1}^2 |S_1|^2 - M_{3} ^2 |S_3|^2 + \\
		& + \left( (\lambda^{1L})_{i\alpha} \bar q^c _i \epsilon \ell _\alpha
			+ (\lambda^{1R})_{i\alpha} \bar u^c _i   e _\alpha  \right) S_1 
			+ (\lambda^{3L})_{i\alpha}\bar q^c _i \epsilon \sigma^I \ell _\alpha S_3^I + \text{h.c.} + \\
		& - \lambda_{H1} |H|^2 |S_1|^2 - \lambda_{H3} |H|^2 |S_3^I|^2 - \left(\lambda_{H13} (H^\dagger \sigma^I H) S_3^{I \dagger} S_1 + \text{h.c.}\right) + \\
		& - \lambda_{\epsilon H 3} i \epsilon^{IJK} (H^\dagger \sigma^I H) S_3^{J \dagger} S_3^{K},
	\label{eq:S1S3Model}
\end{split}\ee
where $\epsilon=i\sigma _2$, $\lambda_{H1}, \lambda_{H3}, \lambda_{\epsilon H 3} \in \mathbb{R}$, $(\lambda^{1L})_{i\alpha}, (\lambda^{1R})_{i\alpha}, (\lambda^{3L})_{i\alpha}, \lambda_{H13} \in \mathbb{C}$. We assume baryon and lepton number conservation\footnote{See Ref.~\cite{Davighi:2020qqa} for an explicit setup forbidding baryon-violating couplings of $S_3$ in a gauge model.} and we neglected quartic self-interactions between leptoquarks.
The convention used for covariant derivatives is
\be 
D_\mu \Phi  = \left(\partial_\mu   + i g^\prime Y_\Phi B_\mu  + ig (t^\Phi _2) ^I W ^I _\mu + i g_s (t^\Phi _3 )^A G^A _\mu \right)\Phi,
\label{eq:g_convention}
\ee
for a generic field $\Phi$ charged under the SM gauge group.
We denote SM quark and lepton fields by $q_i$, $u_i$, $d_i$, $\ell _\alpha$, and $e_\alpha$, while the Higgs doublet is $H$. We adopt latin letters ($i,\,j,\,k,\,\dots$) for quark flavor indices and greek letters ($\alpha,\,\beta,\,\gamma,\,\dots$) for lepton flavor indices.
We work in the down-quark and charged-lepton mass eigenstate basis, where
\be
	q_i = \left( \begin{array}{c} V^*_{ji} u^j_L \\ d^i_L  \end{array}  \right)\,, \qquad
	\ell_\alpha = \left( \begin{array}{c} \nu^\alpha_L \\ e^\alpha_L \end{array} \right)~,
\ee
and $V$ is the CKM matrix. Except for the sign of gauge couplings, here and in the following we use the same notation specified in \cite{Gherardi:2020det}. 

Integrating out at tree-level the two LQ, the following semileptonic operators are generated:
\begin{align}
	[C_{l q}^{(1)}]_{\alpha \beta i j}^{(0)} &= \frac{\lambda^{1L *}_{i \alpha} \lambda^{1L}_{j\beta}}{4 M_1^2} +  \frac{3 \lambda^{3L*}_{i \alpha} \lambda^{3L}_{j\beta}}{4 M_3^2} ~, &
	[C_{l q}^{(3)}]_{\alpha \beta i j}^{(0)} &= - \frac{\lambda^{1L*}_{i \alpha} \lambda^{1L}_{j\beta}}{4 M_1^2} +  \frac{\lambda^{3L *}_{i \alpha} \lambda^{3L}_{j\beta}}{4 M_3^2}  ~, \nonumber\\
	[C_{l equ}^{(1)}]_{\alpha \beta i j}^{(0)} &= \frac{\lambda^{1R}_{j\beta} \lambda_{i\alpha}^{1L*}}{2 M_1^2} ~, &
	[C_{l equ}^{(3)}]_{\alpha \beta i j}^{(0)} &= - \frac{\lambda^{1R}_{j\beta} \lambda_{i\alpha}^{1L*}}{8 M_1^2} ~, \label{eq:EFTS13treematch}\\
	[C_{eu}]_{\alpha \beta i j}^{(0)} &=  \frac{\lambda_{i\alpha}^{1R\,*} \lambda^{1R}_{j\beta}}{2 M_1^2}. \nonumber
\end{align}
The complete one-loop matching between the UV theory and the SMEFT in the Warsaw basis, as well as the definitions for the effective operators, are reported in \cite{Gherardi:2020det}.

%%%%%%%%%%%%%%%%%%%%%%%%%%%%%%%%%
\subsection{Methodology}
\label{sec:method}

Our goal is to study the phenomenology of the $S_1 + S_3$ model described in the previous Section, expressing the low-energy observables as functions of the UV parameters at one-loop level. Given the separation of scales between the LQ masses, assumed to be at the TeV scale, and the typical energy scales of the observables considered, the EFT approach is particularly suited for this goal.
In fact, it allows to separate the complete procedure in a sequence of steps, which can be generalised to be applicable also to other UV scenarios. Going from the ultraviolet to the infrared, the matching procedure allows to pass physical thresholds, i.e. to integrate out heavy fields while defining a new EFT for that energy range, while the renormalization group evolution (RGE) allows to change the scale within an EFT approach.
In our specific case, we have the following steps:
\begin{enumerate}
\item The one-loop matching for the $S_{1,3}$ model into the SMEFT, up to dimension-six operators, resulting by integrating out the two scalar leptoquarks at a scale of the order of their masses $\mu_M \sim M_1,M_3$.  The complete set of matching conditions, obtained with $\overline{\text{MS}}$ renormalization scheme, has been provided in~\cite{Gherardi:2020det}.
\item The RGE of the SMEFT Wilson coefficients from the UV matching scale $\mu_M$ down to the electroweak scale \cite{Jenkins:2013zja,Jenkins:2013wua,Alonso:2013hga};
\item The one-loop matching between the SMEFT and the EFT valid below the electroweak scale, known as Low Energy EFT (LEFT). This results from integrating out the Higgs, the massive electroweak gauge bosons and the top quark and has been done in~\cite{Dekens:2019ept};
\item The RGE of the LEFT Wilson coefficients \cite{Jenkins:2017dyc} from the electroweak scale to the relevant scales of the processes;
\item The expression of the low-energy observables and pseudo-observables in terms of the LEFT Wilson coefficients, taking into account contributions that arise at one-loop level within the LEFT, from the operators generated already at the tree-level.\footnote{In case of observables at the electroweak scale, such as the measurements of $Z$ couplings, the steps 3. and 4. can of course not be considered, since in that case one can work in the SMEFT only.}
\end{enumerate}
By combining everything, we obtain expressions for the observables as a function of the parameters of the scalar leptoquark model at the TeV scale; in such a way, experimental bounds on low-energy data can be used to set constraints on the $S_{1,3}$ couplings. On the other hand, the intermediate steps provide model-independent expressions for observables in terms of EFT Wilson coefficients, which might be exploited in other NP scenarios.

For a generic EFT coefficient we can separate a contribution arising at the tree-level from one arising at one-loop
\be
	C_i = C_i^{(0)} + \frac{1}{(4\pi)^2} C_i^{(1)} ~.
\ee
Working at one-loop accuracy, the RGE, one-loop matching between SMEFT and LEFT, and the one-loop matrix elements to the observables, should only be considered for tree-level generated coefficients, $C_i^{(0)}$ (in our case, those in Eq.~\eqref{eq:EFTS13treematch}).
For the loop-generated coefficients, $C_i^{(1)}$, only the tree-level matching conditions from SMEFT to LEFT, and tree-level matrix elements should be included, the other contributions giving terms which are formally of two-loop order and that could be of the same order as neglected two-loop matching conditions. 

The exception to this is in the RGE due to QCD from the TeV to the GeV scale, for example in four-quark operators contributing to $\Delta F = 2$ observables. In this case the RGE contribution is well known to be important, also due to the large separation of scales, which gives to this effect a parametric enhancement with respect to the neglected two-loop corrections even if four-quark operators are generated at one-loop.

%%%%%%%%%%%%%%%%%%%%%%%%%%%%%%%%%
\subsection{Observables}
\label{sec:obs}

\begin{table}[t]
\begin{center}
\begin{tabular}{ | c | c | c | }
\hline
\textbf{Observable} & \textbf{SM prediction} & \textbf{Experimental bounds} \\
\hline
\hline
\cellcolor[gray]{0.92} $b\to s \ell \ell$ observables & \cellcolor[gray]{0.92} & \cellcolor[gray]{0.92} App.~\ref{app:bsll}\\
\hline
$\Delta\mathcal{C}_{9}^{sb\mu\mu}$ & 0 &  $-0.43\pm 0.09$ \cite{Aebischer:2019mlg} \\
\hline
$\mathcal{C}_{9}^{\text{univ}}$ & 0 &  $-0.48\pm 0.24$ \cite{Aebischer:2019mlg} \\
\hline
\hline
\cellcolor[gray]{0.92} $b\to c \tau(\ell) \nu$ observables & \cellcolor[gray]{0.92}  &  \cellcolor[gray]{0.92} App.~\ref{app:bctaunu},\ref{app:RDmue}\\
\hline
$R_D$ & $0.299\pm 0.003$ \cite{Amhis:2016xyh} & $0.34\pm 0.027\pm 0.013$ \cite{Amhis:2016xyh} \\
\hline
 $R_D^*$ & $0.258\pm 0.005$ \cite{Amhis:2016xyh} & $0.295\pm 0.011\pm 0.008$ \cite{Amhis:2016xyh} \\
 \hline   
 $P_\tau^{D^*}$ & $-0.488\pm 0.018$ \cite{Bordone:2019vic} & $-0.38\pm 0.51\pm 0.2\pm 0.018$ \cite{Hirose:2017dxl} \\
 \hline   
 $F_L$ & $0.470\pm 0.012$ \cite{Bordone:2019vic} & $0.60\pm 0.08\pm 0.038\pm 0.012$ \cite{Abdesselam:2019wbt} \\
 \hline
 $\text{Br}(B_c^+\to \tau ^+ \nu)$ & $2.3 \%$ & $< 10\%$ (95\% CL) \cite{Akeroyd:2017mhr}  \\   
 \hline
	$R_D^{\mu/e}$ & 1 &  $0.978 \pm 0.035$ \cite{Aubert:2008yv,Glattauer:2015teq} \\
 \hline  
\hline
\cellcolor[gray]{0.92} $D$ leptonic decay & \cellcolor[gray]{0.92} & \cellcolor[gray]{0.92} App.~\ref{app:Ds}\\       
\hline
$\text{Br}(D_s\to \tau\nu)$ & $(5.169\pm 0.004)\times 10^{-2}$~\cite{Aoki:2016frl}  &$(5.48 \pm 0.23) \times 10^{-2}$~\cite{Tanabashi:2018oca} \\   
 \hline   
 \hline
 \cellcolor[gray]{0.92} $b\to s \nu\nu$ and $s\to d \nu\nu$ & \cellcolor[gray]{0.92} &  \cellcolor[gray]{0.92} App.~\ref{app:bsnunu}\\
 \hline
 $R_K^\nu$ & 1 \cite{Buras:2014fpa} & $<  4.65$  ~\cite{Grygier:2017tzo} \\
 \hline
  $R_{K^*}^\nu$ & 1 \cite{Buras:2014fpa} & $<  3.22$   ~\cite{Grygier:2017tzo}  \\
 \hline  
$  \text{Br}(K^+ \to \pi^+ \nu\nu)   $ & $8.64 \times 10^{-11}$~\cite{Buras:1998raa} & $ (11.0\pm 4.0) \times 10^{-11}$ \cite{CortinaGil:2020vlo} \\
 \hline
$ \text{Br}(K_L \to \pi^0 \nu\nu) $ & $3.4 \times 10^{-11} $ ~\cite{Buras:1998raa}& $< 3.57 \times 10^{-9}$  \cite{Ahn:2018mvc} \\  
\hline
\hline
\cellcolor[gray]{0.92} $B$ LFV decays & \cellcolor[gray]{0.92} &  \cellcolor[gray]{0.92} App.~\ref{app:BLFVnLFV}\\       
\hline
$\text{Br}(B_d\to \tau^\pm \mu^\mp)$ & 0  &$< 1.4 \times 10^{-5}$ ~\cite{Aaij:2019okb}  \\      
\hline 
$\text{Br}(B_s\to \tau^\pm \mu^\mp)$ &  0 &$< 4.2 \times 10^{-5}$ ~\cite{Aaij:2019okb}  \\      
\hline  
 $\text{Br}(B^+\to K^+ \tau^- \mu^+)$ & 0  &$< 5.4 \times 10^{-5}$ ~\cite{Lees:2012zz}  \\ 
\hline
  \multirow{2}{*}{$\text{Br}(B^+\to K^+ \tau^+ \mu^-)$} &  \multirow{2}{*}{0} &$< 3.3 \times 10^{-5}$ ~\cite{Lees:2012zz}  \\     
 & & $< 4.5 \times 10^{-5}$ ~\cite{Aaij:2020mqb} \\   
\hline        
 \end{tabular}
\caption{ Low-energy \emph{semileptonic} observables with their SM predictions and experimental bounds. Upper limits correspond to 95\%CL. \label{tab:obs}}
\end{center} 
\end{table}

\begin{table}[p]
\begin{center}
\begin{tabular}{ | c | c | c | }
\hline
\textbf{Observable} & \textbf{SM prediction} & \textbf{Experimental bounds} \\
\hline
\hline
\cellcolor[gray]{0.92} $\Delta F=2$ processes& \cellcolor[gray]{0.92} &  \cellcolor[gray]{0.92} App.~\ref{app:DF2} \\
 \hline
 $B^{0}-\overline{B}^{0}$: $|C_{B_d}^1|$ & 0 & $< 9.11\times 10^{-7}$ TeV$^{-2}$ \cite{Bona:2007vi,UTFIT:2016} \\
 \hline 
 $B_s^{0}-\overline{B}_s^{0}$: $|C_{B_s}^1|$ & 0 & $< 2.01\times 10^{-5}$ TeV$^{-2}$ \cite{Bona:2007vi,UTFIT:2016} \\
 \hline  
 $K^{0}-\overline{K}^{0}$: Re[$C_{K}^1$] & 0 & $< 8.04\times 10^{-7}$ TeV$^{-2}$ \cite{Bona:2007vi,UTFIT:2016} \\
 \hline
  $K^{0}-\overline{K}^{0}$: Im[$C_{K}^1$] & 0 & $< 2.95\times 10^{-9}$ TeV$^{-2}$ \cite{Bona:2007vi,UTFIT:2016} \\
 \hline    
 $D^{0}-\overline{D}^{0}$: Re[$C_{D}^1$] & 0 & $< 3.57\times 10^{-7}$ TeV$^{-2}$ \cite{Bona:2007vi,UTFIT:2016}  \\
 \hline
  $D^{0}-\overline{D}^{0}$: Im[$C_{D}^1$] & 0 & $< 2.23\times 10^{-8}$ TeV$^{-2}$ \cite{Bona:2007vi,UTFIT:2016} \\
 \hline    
 $D^{0}-\overline{D}^{0}$: Re[$C_{D}^4$] & 0 & $< 3.22\times 10^{-8}$ TeV$^{-2}$ \cite{Bona:2007vi,UTFIT:2016} \\
 \hline
  $D^{0}-\overline{D}^{0}$: Im[$C_{D}^4$] & 0 & $< 1.17\times 10^{-9}$ TeV$^{-2}$ \cite{Bona:2007vi,UTFIT:2016} \\
 \hline      
 $D^{0}-\overline{D}^{0}$: Re[$C_{D}^5$] & 0 & $< 2.65\times 10^{-7}$ TeV$^{-2}$ \cite{Bona:2007vi,UTFIT:2016} \\
 \hline
  $D^{0}-\overline{D}^{0}$: Im[$C_{D}^5$] & 0 & $< 1.11\times 10^{-8}$ TeV$^{-2}$ \cite{Bona:2007vi,UTFIT:2016} \\
 \hline       
 \hline
\cellcolor[gray]{0.92} LFU in $\tau$ decays & \cellcolor[gray]{0.92} &\cellcolor[gray]{0.92} App.~\ref{app:tauLFU} \\
\hline
$|g_\mu / g_e|^2$ & 1 & $1.0036\pm 0.0028$~\cite{Pich:2013lsa} \\
\hline
$|g_\tau / g_\mu |^2 $ & 1 & $1.0022\pm 0.0030$~\cite{Pich:2013lsa} \\
\hline
$|g_\tau / g_e|^2 $ & 1 & $1.0058\pm 0.0030$~\cite{Pich:2013lsa} \\
\hline
\hline
\cellcolor[gray]{0.92} LFV observables & \cellcolor[gray]{0.92} &  \cellcolor[gray]{0.92}App.~\ref{app:taumuphi}, ~\ref{app:tau3mu}, and~\ref{app:CLFV} \\
\hline
$\text{\Br}(\tau \to \mu \phi) $ & 0 & $ < 1.00\times10^{-7}$   \cite{Miyazaki:2011xe} \\ 
\hline   
$\text{Br}(\tau \to 3\mu)$ & 0 &  $< 2.5\times 10^{-8} $     \cite{Hayasaka:2010np} \\     
\hline
$\text{Br}(\mu\to e\gamma)$ & 0 & $ < 5.00\times10^{-13}$   \cite{TheMEG:2016wtm} \\  
\hline
$\text{Br}(\tau\to \mu\gamma)$ & 0 & $ < 5.24\times10^{-8}$   \cite{Aubert:2009ag}\\  
\hline   
$\text{Br}(\tau\to e\gamma)$ & 0 & $ < 3.93\times10^{-8}$   \cite{Aubert:2009ag} \\  
\hline
\hline
 \cellcolor[gray]{0.92} EDMs & \cellcolor[gray]{0.92} & \cellcolor[gray]{0.92} App.~\ref{app:CLFV}  \\
\hline
$d_e$ & $< 10^{-44}\, e \, cm $~\cite{Pospelov:2013sca,Smith:2017dtz} & $< 1.1 \times 10^{-29}\, e \, cm$~\cite{Andreev:2018ayy}\\
\hline
$d_\mu$ & $< 10^{-42}\, e \, cm $~\cite{Smith:2017dtz} & $< 1.9 \times 10^{-19}\, e \, cm$~\cite{Bennett:2008dy} \\
\hline
$d_\tau$ & $< 10^{-41}\, e \, cm $~\cite{Smith:2017dtz} & $(1.15\pm 1.70) \times 10^{-17}\, e \, cm$~\cite{Inami:2002ah} \\
\hline
\hline 
\cellcolor[gray]{0.92} Anomalous & \cellcolor[gray]{0.92} &  \cellcolor[gray]{0.92}App.~\ref{app:CLFV} \\
 \cellcolor[gray]{0.92} Magnetic Moments & \cellcolor[gray]{0.92} &  \cellcolor[gray]{0.92} \\
\hline
$a_e-a_e^{SM}$ & $\pm 2.3 \times 10^{-13}$~\cite{Keshavarzi:2019abf,Parker:2018vye}& $(-8.9 \pm 3.6)\times 10^{-13}$~\cite{Hanneke:2008tm} \\
\hline
$a_\mu-a_\mu^{SM}$ & $\pm 43 \times 10^{-11}$ \cite{Aoyama:2020ynm} & $ (279 \pm 76)\times 10^{-11}$~\cite{Bennett:2006fi,Aoyama:2020ynm} \\
\hline
$a_\tau-a_\tau^{SM}$ & $\pm 3.9 \times 10^{-8}$~\cite{Keshavarzi:2019abf} & $(-2.1\pm 1.7)\times 10^{-7}$~\cite{Abdallah:2003xd} \\
\hline      
 \end{tabular}
\caption{Meson-mixing and leptonic observables, with their SM predictions and experimental bounds. Upper limits correspond to 95\%CL. \label{tab:obs2}}
\end{center} 
\end{table}

\begin{table}[h!]
\begin{center}
\begin{tabular}{ | c | c | }
\hline
\textbf{Observable} & \textbf{Experimental bounds} \\
\hline
\hline
\cellcolor[gray]{0.92} $Z$ boson couplings & \cellcolor[gray]{0.92} App.~\ref{app:Zcouplings} \\\hline
	$\delta g^Z_{\mu_L}$ 	& $(0.3 \pm 1.1) 10^{-3}$	\cite{ALEPH:2005ab} \\ \hline
	$\delta g^Z_{\mu_R}$	& $(0.2 \pm 1.3) 10^{-3}$	\cite{ALEPH:2005ab} \\ \hline
	$\delta g^Z_{\tau_L}$	& $(-0.11 \pm 0.61) 10^{-3}$	\cite{ALEPH:2005ab} \\ \hline
	$\delta g^Z_{\tau_R}$	& $(0.66 \pm 0.65) 10^{-3}$	\cite{ALEPH:2005ab} \\ \hline
	$\delta g^Z_{b_L}$	& $(2.9 \pm 1.6) 10^{-3}$	\cite{ALEPH:2005ab} \\ \hline
	$\delta g^Z_{c_R}$	& $(-3.3 \pm 5.1) 10^{-3}$	\cite{ALEPH:2005ab} \\ \hline
	$N_\nu$	&	$2.9963 \pm 0.0074$	\cite{Janot:2019oyi} \\
\hline                       
\end{tabular}
\caption{ Limits on the deviations in $Z$ boson couplings to fermions from LEP I. \label{tab:ZcouplLEP}}
\end{center} 
\end{table}

One of our main goals is to provide, with the $S_{1,3}$ model, a combined explanation for the hints of non LFU in the neutral and charged current semileptonic $B$-meson decays, namely to account for the experimental measurements of $R_{K^{(*)}}$ and $R_{D^{(*)}}$, and of the deviation in the muon anomalous magnetic moment $(g-2)_\mu$.
The leptoquark couplings involved in these observable enter also in the other low-energy observables (or pseudo observables), both at tree-level or one-loop level. Therefore, to quantify how the $S_{1,3}$ model can consistently explain the observed anomalies, one should take into account a set of low-energy data as complete as possible. In Tables~\ref{tab:obs}, \ref{tab:obs2}, and \ref{tab:ZcouplLEP}, we show the list of low-energy observables that we analyze, together with their SM predictions  and experimental bounds.

In App.~\ref{app:obs}, these low-energy observables are discussed in length. We will explicitly show, as functions of the parameters of the $S_{1,3}$ model, tree-level contributions together with dominant one-loop effects, while in the numerical analysis the full set of one-loop corrections is considered. Some of the considered observables vanish or are flavor-suppressed at tree-level, for example meson-mixing $\Delta F=2$ processes, $\tau \to 3 \mu$ and $\tau \to \mu \gamma$ LFV interactions or $\tau \to \mu \phi(\eta,\eta^\prime)$ decay; in such cases the inclusion of one-loop contributions is relevant and might bring non negligible changes in a global fit of the low-energy data.

From the observables listed above, and their expression in terms of the parameters of the model, LQ couplings and masses, we build a global likelihood as:
\be
	- 2 \log \LL \equiv \chi^2(\lambda_x, M_x) = \sum_i \frac{ \left(\OO_i(\lambda_x, M_x) - \mu_i\right)^2}{\sigma_i^2}~,
	\label{eq:Global_chiSQ}
\ee
where $\OO_i(\lambda_x, M_x)$ is the expression of the observable as function of the model parameters, $\mu_i$ its experimental central value, and $\sigma_i$ the uncertainty. These are all discussed in App.~\ref{app:obs}.
From the $\chi^2$ built in this way, in each scenario considered we obtain the maximum likelihood point by minimizing the $\chi^2$, which we use to compute the $\Delta \chi^2 \equiv \chi^2 - \chi^2_{\rm min}$. This allows us to obtain the 68, 95, and 99\% CL regions.
In the Standard Model limit we get a $\chi^2_{\SM} = 101.0$, for 50 observables.

For each scenario we get the CL regions in the plane of two real couplings, by profiling the likelihood over all the other couplings. We are often also interested in the values of some observables corresponding to these CL regions. To obtain this, we perform a numerical scan over all the parameter space\footnote{For each numerical scan we collected $O(10^4)$ benchmark points. For our more complex models (\textit{i.e.} with up to ten parameters), this is quite demanding from the computational point of view; in order to efficiently scan the high-dimensional parameter spaces, we employ a Markov Chain Monte Carlo algorithm (Hastings-Metropolis) for the generation of trial points.} and select only the points with a $\Delta \chi^2$ less than the one corresponding to $68$ and $95\%$CL. 
The points obtained in this way also reproduce the CL regions in parameter space obtained by profiling. With this set of parameter-space points we can then plot any observable evaluated on them.

%%%%%%%%%%%%%%%%%%%%%%%%%%%%%%%%%
\subsubsection{Collider constraints}
\label{sec:collider}

Leptoquarks are also actively searched for at high-energy colliders. Their most important signatures can be classified in three categories: \emph{i)} pair production, \emph{ii)} resonant single-production, and \emph{iii)} off-shell t-channel exchange in Drell-Yan processes, $p p \to \ell^+ \ell^-$ or $\ell \nu$. See e.g. Refs.~\cite{Dorsner:2018ynv,Schmaltz:2018nls} for reviews.

The pair production cross section is mostly independent on the LQ couplings to fermions, unless some are very large, and thus provides limits which depend only on the LQ mass and the branching ratios in the relevant search channels. We refer to Refs.~\cite{Marzocca:2018wcf,Angelescu:2018tyl,Schmaltz:2018nls,Saad:2020ihm} for reviews of such searches. Once the branching ratios are taken into account, the most recent ATLAS and CMS searches using an integrated luminosity of $\sim 36 \text{fb}^{-1}$ put a lower bound on the $S_1$ and $S_3$ masses at $\approx 1\TeV$ or less.
At present, limits from single production are not competitive with those from pair production and Drell-Yan \cite{Schmaltz:2018nls}.

Leptoquarks can also be exchanged off-shell in the t-channel in Drell-Yan processes. The final states most relevant to our setup are $\tau \bar\tau$, $\tau \bar\nu$, and $\mu \bar \mu$. The limits on LQ couplings as a function of their mass from neutral-current processes can be taken directly from \cite{Faroughy:2016osc,Angelescu:2018tyl,Schmaltz:2018nls} (see also \cite{Greljo:2017vvb,Afik:2018nlr,Afik:2019htr,Angelescu:2020uug} for other studies of dilepton tails in relation with $B$-anomalies) while the mono-tau channel in relation to the $B$-anomalies has been studied in \cite{Altmannshofer:2017poe,Greljo:2018tzh,Abdullah:2018ets,Brooijmans:2020yij,Fuentes-Martin:2020lea,Marzocca:2020ueu} and at present it doesn't exclude the region of interest.
Using the results from \cite{Angelescu:2018tyl} we get the following 95\% CL upper limits on the couplings relevant to our model for $M_{1,3} = 1\TeV$, taken one at a time:
\be\begin{split}
	& 	\lambda^{1R}_{c\tau} < 1.62~, \qquad
		\lambda^{1R}_{c\mu} < 0.90~, \qquad
	 	\lambda^{1L}_{s\tau} < 1.66~, \qquad 
		\lambda^{1L}_{s\mu} < 0.91~, \\
	& 	\lambda^{3L}_{b\tau} < 1.40~, \qquad 
	 	\lambda^{3L}_{s\tau} < 0.97~, \qquad 
		\lambda^{3L}_{b\mu} < 0.77~, \qquad
		\lambda^{3L}_{s\mu} < 0.56~,
\label{eq:collider_limits}
\end{split}\ee
while the limits on $\lambda^{1L}_{b\tau}$ and $\lambda^{1L}_{b\mu}$ from the neutral-channel are in the non-perturbative regime since they induce only a $|V_{cb}|^2$-suppressed amplitude in $c \bar c \to \tau \bar \tau, ~\mu \bar\mu$. 
Since only the constraint on $\lambda^{1R}_{c\tau} $ turns out to be relevant for our setup, we are justified taking the limits above derived one at a time.

We implement the constraints in Eq.~\eqref{eq:collider_limits} in our global likelihoods by assuming that the signal cross section is dominated by the purely New Physics contribution, thus with a scaling $\propto |\lambda|^4$, and that it follows a Gaussian distribution around zero such that the 95\%CL limits on single-couplings reproduce those in Eq.~\eqref{eq:collider_limits}, i.e. with standard deviation given by $\sigma_x = |\lambda_x^{\rm lim}|^4 / \sqrt{\chi^2_{\rm 95\% CL (1dof)}}$,
where for each coupling $\lambda_x^{\rm lim}$ is the limit in Eq.~\eqref{eq:collider_limits} and $\chi^2_{\rm 95\% CL (1dof)} \approx 3.84$:
\be
	\chi^2_{\rm collider} = \chi^2_{\rm 95\% CL (1dof)} \sum_x \left( |\lambda_x |^4 / |\lambda_x^{\rm lim}|^4 \right)^2~,
\ee
which is added to the global $\chi^2$ from low-energy observables of Eq.~\eqref{eq:Global_chiSQ}.

\newpage
%%%%%%%%%%%%%%%%%%%%%%%%%%%%%%%%%
\section{Scenarios and results}
\label{sec:scenarios}

In this Section we discuss several minimal models within the $S_{1}+S_{3}$
setup, and how well (or bad) each of them is able to address the charged
and/or neutral current anomalies, while remaining compatible with
all the other experimental constraints. We denote the leptoquark
couplings to fermions by:
\begin{equation}
\lambda^{1R}=\begin{pmatrix}\lambda_{ue}^{1R} & \lambda_{u\mu}^{1R} & \lambda_{u\tau}^{1R}\\
\lambda_{ce}^{1R} & \lambda_{c\mu}^{1R} & \lambda_{c\tau}^{1R}\\
\lambda_{te}^{1R} & \lambda_{t\mu}^{1R} & \lambda_{t\tau}^{1R}
\end{pmatrix},\quad\lambda^{1L}=\begin{pmatrix}\lambda_{de}^{1L} & \lambda_{d\mu}^{1L} & \lambda_{d\tau}^{1L}\\
\lambda_{se}^{1L} & \lambda_{s\mu}^{1L} & \lambda_{s\tau}^{1L}\\
\lambda_{be}^{1L} & \lambda_{b\mu}^{1L} & \lambda_{b\tau}^{1L}
\end{pmatrix},\quad\lambda^{3L}=\begin{pmatrix}\lambda_{de}^{3L} & \lambda_{d\mu}^{3L} & \lambda_{d\tau}^{3L}\\
\lambda_{se}^{3L} & \lambda_{s\mu}^{3L} & \lambda_{s\tau}^{3L}\\
\lambda_{be}^{3L} & \lambda_{b\mu}^{3L} & \lambda_{b\tau}^{3L}
\end{pmatrix}~.
\label{eq:Leptoquark couplings}
\end{equation}

The main experimental anomalies driving the fit can be split in three categories:
\begin{itemize}
\item \emph{CC}: deviations in $b \to c \tau \nu$ transitions; 
\item \emph{NC}: deviations in $b \to s \mu \mu$ transitions; 
\item \emph{$(g-2)_\mu$}: deviation in the muon magnetic moment.
\end{itemize}
While our setup allows to keep all the above couplings in a completely general analysis, given the large number of parameters this would preclude a clear understanding of the physics underlying the fit.
Furthermore, it can be interesting to consider only one leptoquark or to focus on one specific experimental anomaly.
For these reasons we take a step-by-step approach by starting with single-leptoquark scenarios and switching on the couplings needed to fit a given set of anomalies.
In all cases, we keep the complete likelihood described in the previous Section, with all the observables.
For instance, if the couplings to muons are set to zero, neutral-current $B$-anomalies and the muon anomalous magnetic moment are automatically \emph{frozen} to the corresponding Standard Model values and do not impact the final fit.

The models are thus defined by the leptoquark content and the set of active couplings, which, for simplicity, we assume to be real.
We have considered the models detailed in Table \ref{tab:Summary of models}, for each of
which we allow the couplings listed in the third column of the table
to be non-vanishing in our global fit.
We first analyze single mediator models and study their potential to address as many anomalies as possible. In each case we point out the main tensions which prevent a combined explanation of all anomalies.
Then, we move on to study models involving both leptoquarks.
In the first we only allow left-handed couplings, $\lambda^{1L}$ and $\lambda^{3L}$, as this possibility has better chances to find motivation in a scenario in which the flavor structure is determined by a flavor symmetry, see e.g. \cite{Buttazzo:2017ixm,Marzocca:2018wcf}.
In the second we switch on also some of the $S_1$ couplings to right-handed fermions, and aim to provide a combined explanation for all three anomalies.
Finally, we study the limits on the leptoquark potential couplings to the Higgs, which is an analysis largely independent on the couplings to fermions and requires to consider different observables than those studied in the main fit, see also \cite{Crivellin:2020ukd}.

\begin{table}[t]
{\setlength\extrarowheight{10pt}%
\begin{tabular}{|c| l |c|c|c|}
\hline 
Model  & Couplings & CC & NC & $(g-2)_\mu$\tabularnewline
\hline 
\hline 
\multirow{2}{*}{$S_{1}^{ ~(CC)}$} & \multirow{2}{*}{$\lambda_{c\tau}^{1R}, \lambda_{b\tau}^{1L}$} & \multirow{2}{*}{\textbf{\textcolor{green}{\LARGE{}$\checked$}}} & \multirow{2}{*}{\textbf{\textcolor{red}{\LARGE{}$\times$}}} & \multirow{2}{*}{\textbf{\textcolor{red}{\LARGE{}$\times$}}} \tabularnewline
\multirow{2}{*}{$S_{1}^{ ~(NC)}$} & \multirow{2}{*}{$\lambda_{b\mu}^{1L}, \lambda_{s\mu}^{1L}$} & \multirow{2}{*}{\textbf{\textcolor{red}{\LARGE{}$\times$}}} & \multirow{2}{*}{\textbf{\textcolor{orange}{\LARGE{}$\otimes$}}} & \multirow{2}{*}{\textbf{\textcolor{red}{\LARGE{}$\times$}}} \tabularnewline
\multirow{2}{*}{$S_{1}^{ ~(a_\mu)}$} & \multirow{2}{*}{$\lambda_{t\mu}^{1R}, \lambda_{b\mu}^{1L}$} & \multirow{2}{*}{\textbf{\textcolor{red}{\LARGE{}$\times$}}} & \multirow{2}{*}{\textbf{\textcolor{red}{\LARGE{}$\times$}}} & \multirow{2}{*}{\textbf{\textcolor{green}{\LARGE{}$\checked$}}}  \tabularnewline
\multirow{2}{*}{$S_{1}^{ ~(CC + a_\mu)}$} & \multirow{2}{*}{$\lambda_{t\tau}^{1R}, \lambda_{c\tau}^{1R} , \lambda_{t\mu}^{1R}, \lambda_{b\tau}^{1L}, \lambda_{b\mu}^{1L}$} & \multirow{2}{*}{\textbf{\textcolor{green}{\LARGE{}$\checked$}}} & \multirow{2}{*}{\textbf{\textcolor{red}{\LARGE{}$\times$}}} & \multirow{2}{*}{\textbf{\textcolor{green}{\LARGE{}$\checked$}}} \tabularnewline
  &  &  &  & \tabularnewline
\hline 
\multirow{2}{*}{$S_{3}^{ ~(CC+NC)}$} & \multirow{2}{*}{$\lambda_{b\tau}^{3L}, \lambda_{s\tau}^{3L},\lambda_{b\mu}^{3L}, \lambda_{s\mu}^{3L}$} & \multirow{2}{*}{\textbf{\textcolor{red}{\LARGE{}$\times$}}} & \multirow{2}{*}{\textbf{\textcolor{green}{\LARGE{}$\checked$}}} & \multirow{2}{*}{\textbf{\textcolor{red}{\LARGE{}$\times$}}}  \tabularnewline
  &  &  &  & \tabularnewline
\hline 
\multirow{2}{*}{$S_1 + S_3^{\rm ~(LH)}$ } & \multirow{2}{*}{$ \lambda_{b\tau}^{1L},  \lambda_{s\tau}^{1L},\lambda_{b\tau}^{3L}, \lambda_{s\tau}^{3L},\lambda_{b\mu}^{3L}, \lambda_{s\mu}^{3L}$} & \multirow{2}{*}{\textbf{\textcolor{green}{\LARGE{}$\checked$}}} & \multirow{2}{*}{\textbf{\textcolor{green}{\LARGE{}$\checked$}}} & \multirow{2}{*}{\textbf{\textcolor{red}{\LARGE{}$\times$}}}  \tabularnewline
  &  &  &  & \tabularnewline
 \hline 
\multirow{2}{*}{$S_1 + S_3 ^{\rm ~(all)}$} & \multirow{2}{*}{$ \lambda_{b\tau}^{1L},  \lambda_{s\tau}^{1L}, \lambda_{b\mu}^{1L}, \lambda_{t\tau}^{1R}, \lambda_{c\tau}^{1R},\lambda_{t\mu}^{1R}, \lambda_{b\tau}^{3L}, \lambda_{s\tau}^{3L},\lambda_{b\mu}^{3L}, \lambda_{s\mu}^{3L}$} &  \multirow{2}{*}{\textbf{\textcolor{green}{\LARGE{}$\checked$}}} & \multirow{2}{*}{\textbf{\textcolor{green}{\LARGE{}$\checked$}}} & \multirow{2}{*}{\textbf{\textcolor{green}{\LARGE{}$\checked$}}}  \tabularnewline
  &  &  &  & \tabularnewline
\hline 
\multirow{2}{*}{$S_1 + S_3^{\rm ~(pot)}$} & \multirow{2}{*}{$ \lambda_{H1}, \lambda_{H3}, \lambda_{H13}, \lambda_{\epsilon H3} $} &  \multirow{2}{*}{--} & \multirow{2}{*}{--} & \multirow{2}{*}{--}  \tabularnewline
  &  &  &  & \tabularnewline
\hline 
\end{tabular}}

\caption{\label{tab:Summary of models}Summary of leptoquark models considered in this work.
The third columns lists the couplings we allow to be different from zero in our global fit.
The last three columns indicate whether the models provide a satisfying fit of each set of anomalies, respectively.}
\end{table}

In any given model there is, of course, no particular reason to
expect the exact flavor structures implied by Table \ref{tab:Summary of models}.
For instance, the couplings we set to zero will be radiatively generated.
In our bottom-up approach we assume them to be small enough at the matching scale that the observables in the fit are not impacted in a sizeable way.
In a more top-down approach one might have expectations on the size of these terms based on the UV picture, such as due to the presence of approximate flavor symmetries or other flavor-protection mechanisms.

In the numerical analysis we fix for concreteness values of leptoquark masses equal to $M_1 = M_3 = 1\TeV$. While this is borderline with the exclusion limits from pair production, discussed previously, the results do not change qualitatively by increasing slightly the masses. Since most of the observables driving the fits scale as $\lambda^2 / M^2$, with a good approximation this scaling can be used to adapt our fits to slightly larger masses.\footnote{The exception to this scaling are $\Delta F = 2$ observables, which scale as $\lambda^4/M^4$, but are relevant only for the fits of Sec.~\ref{sec:S3CCNC} and \ref{sec:S1S3LH}.} We note that the future limits on LQ masses from HL-LHC are expected to not go much above $1.5\TeV$ \cite{Marzocca:2018wcf}.

Concerning our specific benchmarks, the choice of active couplings in each case
is guided by some simple phenomenological observations (more details
on each concrete model can be found in the relevant Subsections below):
\begin{enumerate}
\item Since the observed deviations in $B$-decays involve LQ couplings
to second and third generation, and given the strong constraints on
$s\leftrightarrow d$ quark flavor transitions, couplings to first
generation of down quarks can only play a minor role in the fit of $B$-anomalies and are thus set to zero (note that even in this case, due to the CKM matrix, effects in up-quark observables are present, for instance $D$-meson mixing).
\item Hints to LFU violation in rare $B$-decays, combined with the deviations
observed in $B\to K^{*}\mu^{+}\mu^{-}$, suggest that the LQ couplings
to muons should be larger than those to electrons. We consider, for
simplicity the case in which $b\to s\ell\ell$ anomalies are entirely
explained by muon couplings and set to zero the couplings to electrons. 
\item The $S_{1}$ couplings to $\mu_{R}$ and $c_{R}$ or $t_{R}$ do not
contribute to $b\to s\ell\ell$, nor to $b\to c(\tau/\ell)\nu$, however are relevant for fitting the observed anomaly in the muon anomalous magnetic moment, which gets the main contribution from the couplings to $b_L \mu_L$ and $t_R \mu_R$ (c.f. Eq.~\eqref{eq:amu_num}).
\end{enumerate}
Details for all models are given in the following Subsections.

%------------------------------------------------------------------------------------------------
%------------------------------------------------------------------------------------------------
\subsection{Single-leptoquark $S_1$}
\label{sec:S1}

%---------------------------------------------------------------------------------------
\subsubsection{Addressing CC anomalies}

\begin{figure}[t]
\centering
\includegraphics[width=0.44\hsize]{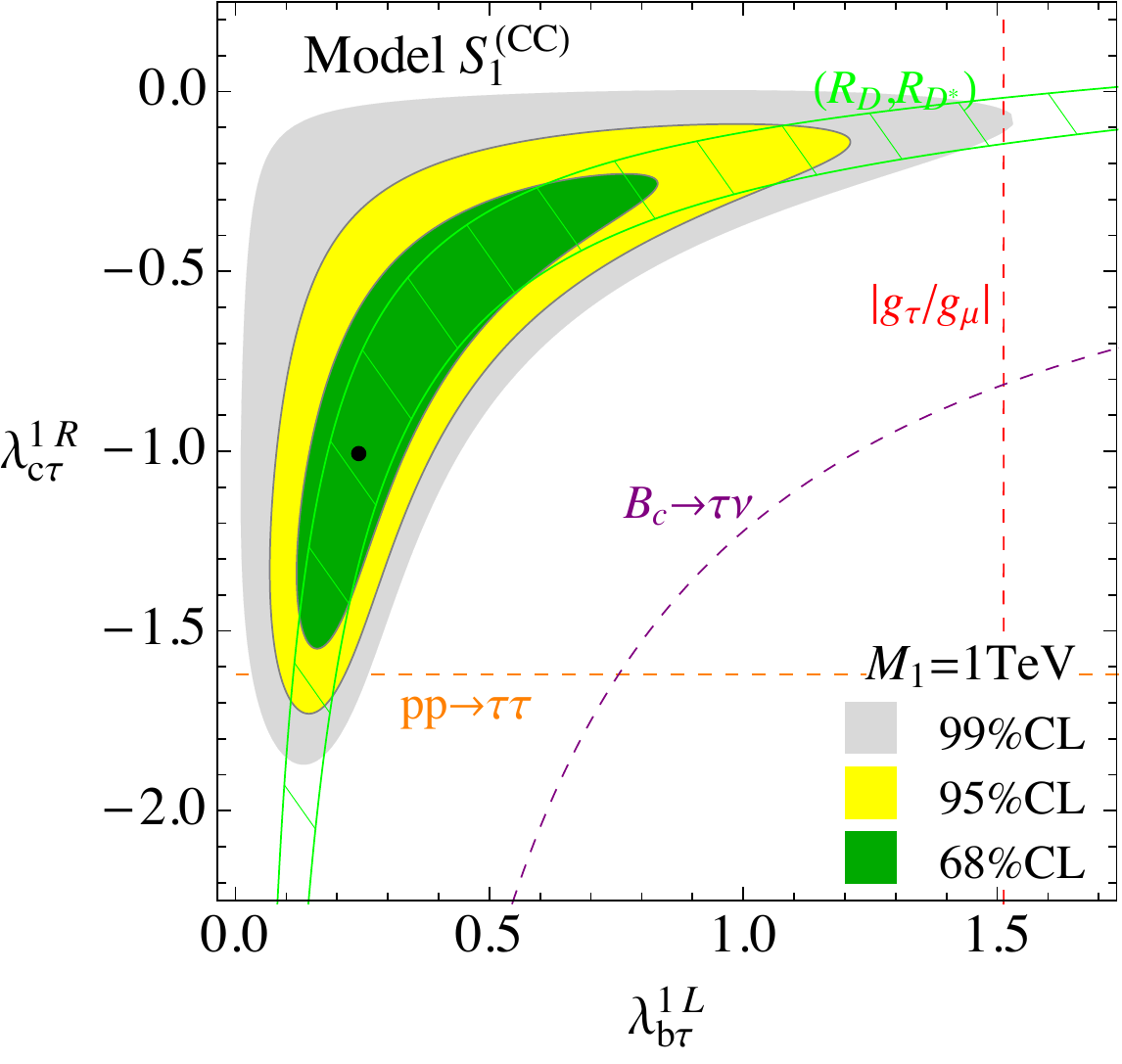} ~
\includegraphics[width=0.51\hsize]{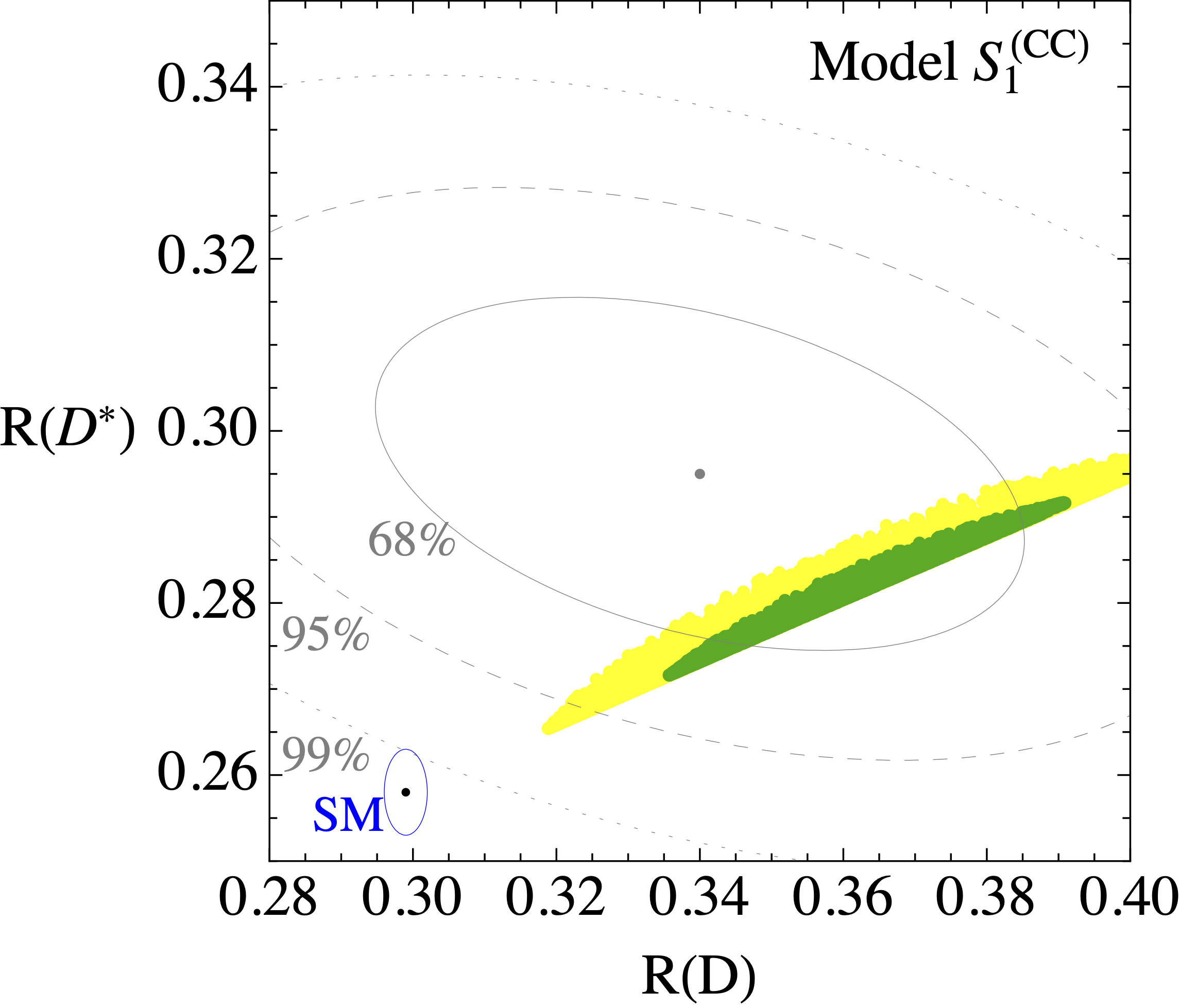}
\caption{\small Result from a fit in the two $S_1$ couplings $\lambda^{1L}_{b\tau}$ and $\lambda^{1R}_{c\tau}$, for a leptoquark of 1 TeV. In the left panel we show the preferred regions in the plane of the two couplings, and the individual $2\sigma$ limits from the most relevant observables (except for $(R(D), R(D^*))$, for which we show the $68\%$CL region). The black dot corresponds to the best-fit point.
In the right panel we show where this preferred region is mapped in the plane of $R(D)-R(D^*)$, together with the experimental combination from HFLAV \cite{Amhis:2016xyh}.}
\label{fig:S1CC}
\end{figure}

This LQ can address the deviations in $R(D)$ and $R(D^*)$ with only two couplings: $\lambda^{1L}_{b\tau}$ and $\lambda^{1R}_{c\tau}$. They generate at tree-level a contribution to the semileptonic scalar and tensor operators $C^{(1,3)}_{lequ}$ at the UV matching scale, Eq.~\eqref{eq:EFTS13treematch}, which then run down to the GeV scale. The best fit region is entirely determined by the following few observables: $R(D)$, $R(D^*)$, $\text{Br}(B_c^+\to \tau ^+ \nu)$ (App.~\ref{app:bctaunu}), $|g_\tau / g_\mu|$ (App.~\ref{app:tauLFU}), and the constraints from $p p \to \tau^+ \tau^-$.

The results from the fit, assuming real couplings and $M_1 = 1 \TeV$, can be seen in Fig.~\ref{fig:S1CC}.
Since all the relevant low-energy observables scale with $\lambda/M$, the fit can be easily adapted to other masses.
The left panel shows confidence level regions for the two couplings.
The dashed lines are 95\% CL constraints from single observables, and help illustrate the role of each observable within the global fit.

In the right panel we show how the 68\% and 95\% CL region from the global fit of the left panel maps in the $R(D^{(*)})$ plane. This is almost degenerate, due to the approximate linear relationship between these two observables in the present model.
We overlay as gray lines the CL ellipses (for 2 degrees of freedom) from the HFLAV global fit of the two observables \cite{Amhis:2016xyh} (specifically, the Spring 2019 update).

The model is successful in fitting the deviation in $R(D^{(*)})$ within the 68\%CL level, with smaller values of $R(D^{*})$ (or larger $R(D)$) preferred by the fit.
Improved measurements of $R(D^{(*)})$ can test this setup due to the precise linear relationship among the two modes predicted by the model, as well as improved Drell-Yan constraints.

The best-fit point, for $M_1 = 1 \TeV$, is found for
\be
	S_1^{~(CC)} |_{\text{best-fit}}: \quad
	\lambda^{1L}_{b\tau} \approx 0.24, \quad
	\lambda^{1R}_{c\tau} \approx -1.00.
\ee

%---------------------------------------------------------------------------------------
\subsubsection{Addressing NC anomalies}

\begin{figure}[t]
\centering
\includegraphics[width=0.44\hsize]{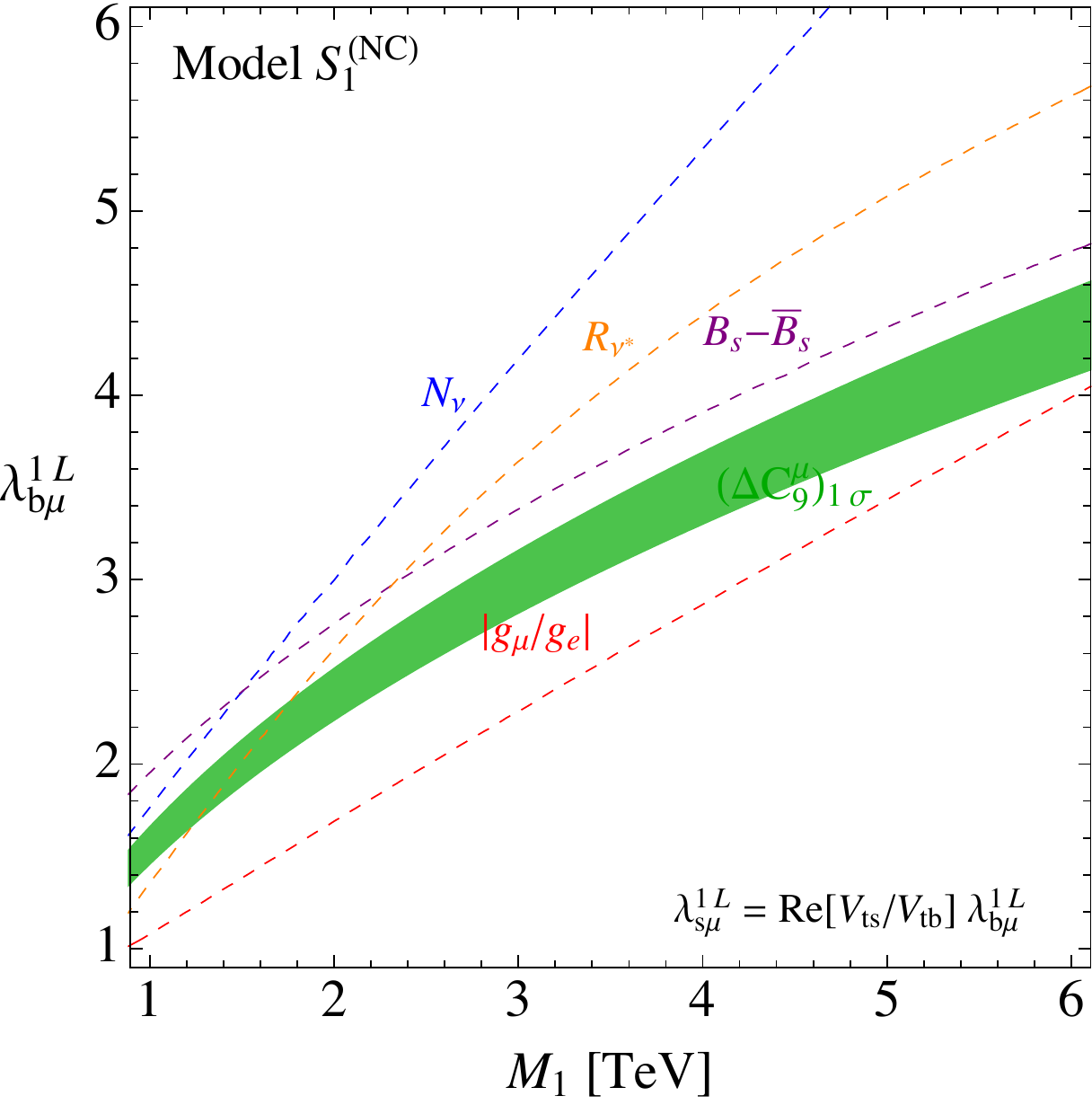}~
\includegraphics[width=0.46\hsize]{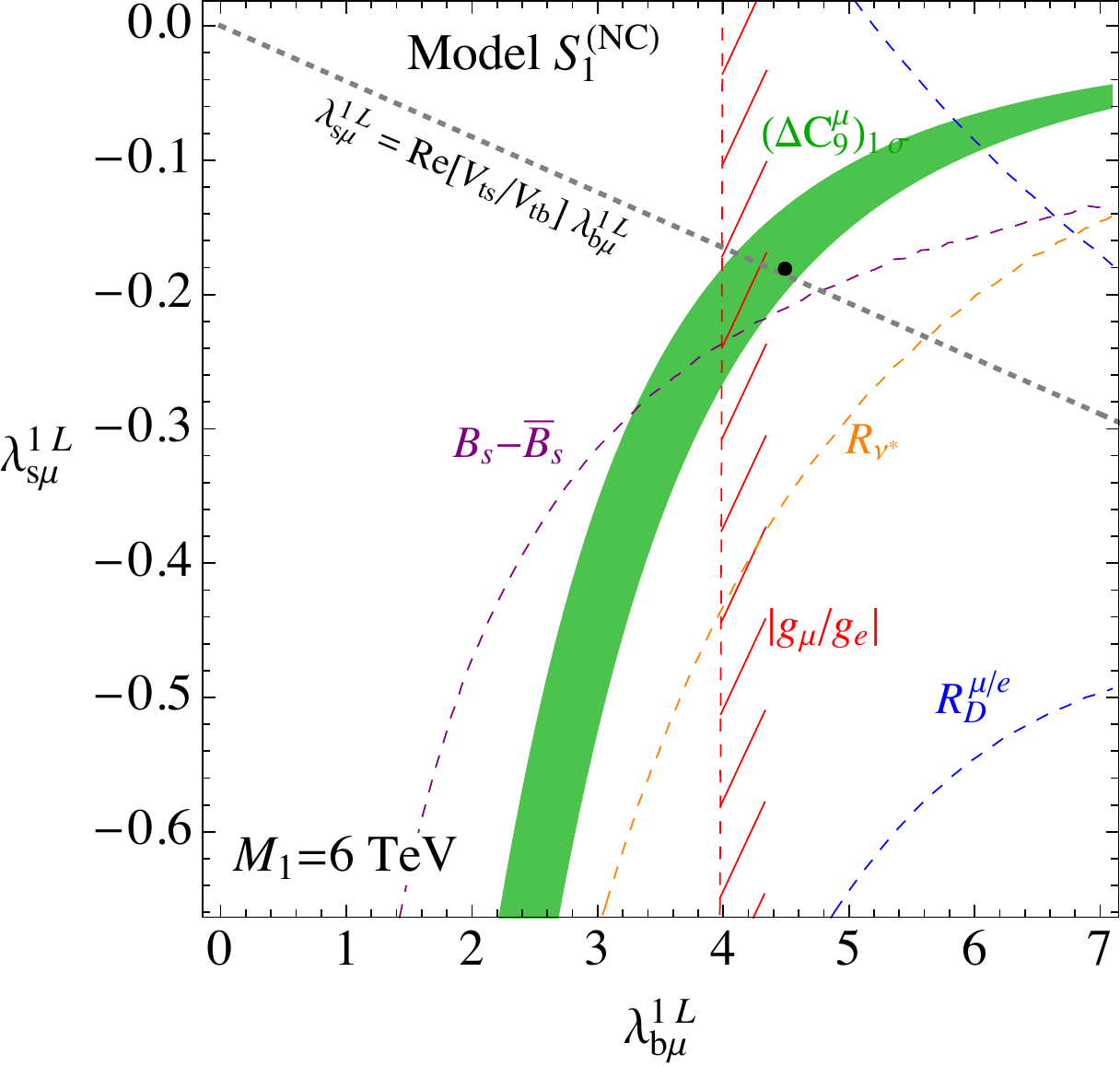}
\caption{\small Left: 95\% CL limits in the plane of $\lambda^{1L}_{b\mu} - M_1$, for $\lambda^{1L}_{s\mu} = - \Re[V_{cb}/V_{cs}] \lambda^{1L}_{b\mu}$.
Right: 95\% CL limits from individual observables in the plane of $\lambda^{1L}_{b\mu} - \lambda^{1L}_{s\mu}$, fixing $M_1 = 6 \TeV$. In both plots the green region is the $1\sigma$ favourite one from $\Delta C_9^\mu$. }\label{fig:S1NCbounds}
\end{figure}

One may attempt to fit neutral-current anomalies in $b\to s \ell \ell$ from the one-loop contributions from $S_1$. This scenario has been considered for the first time in \cite{Bauer:2015knc}. Significant contributions to $\Delta C _9 ^\mu $ may only come from the two muon couplings $\lambda ^{1L}_{b\mu}$ and $\lambda ^{1L}_{s\mu}$, whereas the universal contribution is always negligible ($C_9 ^u\approx 0$), c.f. Eqs.~(\ref{eq:Lbsll},\ref{eq:C9num}).

$B_s$-mixing, c.f. Eq.~\eqref{eq:DeltaF2LQ}, and $B \to K^{(*)} \nu \nu$, c.f. Eq.~\eqref{eq:BoundBKnunuFin}, put strong constraints on the product of the two couplings. Thanks to the different scaling of these observables and $\Delta C_9^\mu$ on the leptoquark couplings, the limits can be avoided by a suitably large leptoquark mass, as can be seen in Fig.~\ref{fig:S1NCbounds} (left). For $M_1 \gtrsim 3\,\text{TeV}$ it is possible to avoid these limits while having couplings still in the perturbative range (see also \cite{Cai:2017wry,Azatov:2018kzb}).
Nevertheless, even while marginally evading the $B_s$-mixing constraint, the $\Delta C _9 ^\mu $ deviation remains in $\approx 2 \sigma$ tension with the bound on $\lambda ^{1L}_{b\mu}$ arising from the LFU limit from $\tau$ decays, $|g_\mu /g_e|$ (see App.~\ref{app:tauLFU}). The situation is illustrated in the right panel of Fig.~\ref{fig:S1NCbounds}. This is slightly exacerbated by a $\sim 1\sigma$ deviation in the opposite direction measured in $|g_\mu /g_e|$.

We thus conclude that the $S_1$ leptoquark is not able to fit neutral-current anomalies while remaining completely consistent with all other constraints. The situation regarding NC anomalies is not modified significantly by letting also the other couplings vary in the fit. This issue could be avoided by allowing a mild cancellation by tuning a further contribution to this observable, possibly arising from some other state. 
Fixing \mbox{$M_1 = 6 \TeV$} we find the best-fit point for:
\be
	S_1^{~(NC)} (M_1 = 6\TeV)|_{\text{best-fit}} : \quad
	\lambda^{1L}_{b\mu} \approx 4.5, \quad
	\lambda^{1L}_{s\mu} \approx -0.18.
\ee

%---------------------------------------------------------------------------------------
\subsubsection{Addressing $(g-2)_{\mu,e}$}

Thanks to the $m_t$ enhancement of the left-right contribution, c.f. Eq.~\eqref{eq:TRbetaalfa}, 
$S_1$ is a good candidate to address the observed deviation in the muon anomalous magnetic moment. The leading contribution is given numerically by (see Eq.~\eqref{eq:amu_num}):
\be
	\Delta a_\mu \approx 8.23 \times 10^{-7} \frac{\lambda^{1L}_{b\mu} \lambda^{1R}_{t\mu}}{M_1^2/1\TeV^2} (1 + 0.53 \log \frac{M_1^2}{1\TeV^2})~.
\ee
The observed deviation can thus be addressed for small couplings, and no other observable is influenced significantly.
Analogously, it is possible to address the (smaller and less significant) deviation in the electron magnetic moment, see Table~\ref{tab:obs2}.

A combined explanations of both deviations with a single mediator was thought not to be viable, due to the very strong constraint from $\mu \to e \gamma$, see e.g. Refs.~\cite{Crivellin:2018qmi}.
More recently, in the updated versions of \cite{Bigaran:2020jil,Dorsner:2020aaz} it was realized that a possible way out is to align the $S_1$ leptoquark-muon couplings to the top quark, while the leptoquark-electron ones to the charm.\footnote{We thank the authors of \cite{Bigaran:2020jil} and \cite{Dorsner:2020aaz} for pointing this out to us.}
In our formalism this can be achieved aligning the $S_1$ couplings as
$ \lambda^{1L}_{i e} = V_{c i} \lambda^{1L}_e$ and $\lambda^{1L}_{i \mu} = V_{t i} \lambda^{1L}_\mu$.
In this way one has (c.f. Eqs.(\ref{eq:TRbetaalfa},\ref{eq:muega_num},\ref{eq:amu_num})):
\be\begin{split}
	\Delta a_e &\approx 1.9 \times 10^{-10} \frac{\lambda^{1L}_{e} \lambda^{1R}_{ce}}{M_1^2/1\TeV^2} (1 + 0.09 \log \frac{M_1^2}{1\TeV^2})	~, \\
	\Delta a_\mu &\approx 8.23 \times 10^{-7} \frac{\lambda^{1L}_{\mu} \lambda^{1R}_{t\mu}}{M_1^2/1\TeV^2} (1 + 0.53 \log \frac{M_1^2}{1\TeV^2})	~, \\
	\Br(\mu \to e \gamma) &= 0~.
\end{split}\ee
On the one hand, the strong limit from $\mu \to e \gamma$ implies that the alignment described above must be held with high accuracy. On the other hand, radiative corrections to leptoquark couplings from SM Yukawas are expected to induce deviations from it. For this reason we will not investigate this direction further.

%---------------------------------------------------------------------------------------
\subsubsection{Addressing CC and $(g-2)_\mu$}
\label{sec:S1CCamu}

\begin{figure}[t]
\centering
\includegraphics[height=6.5cm]{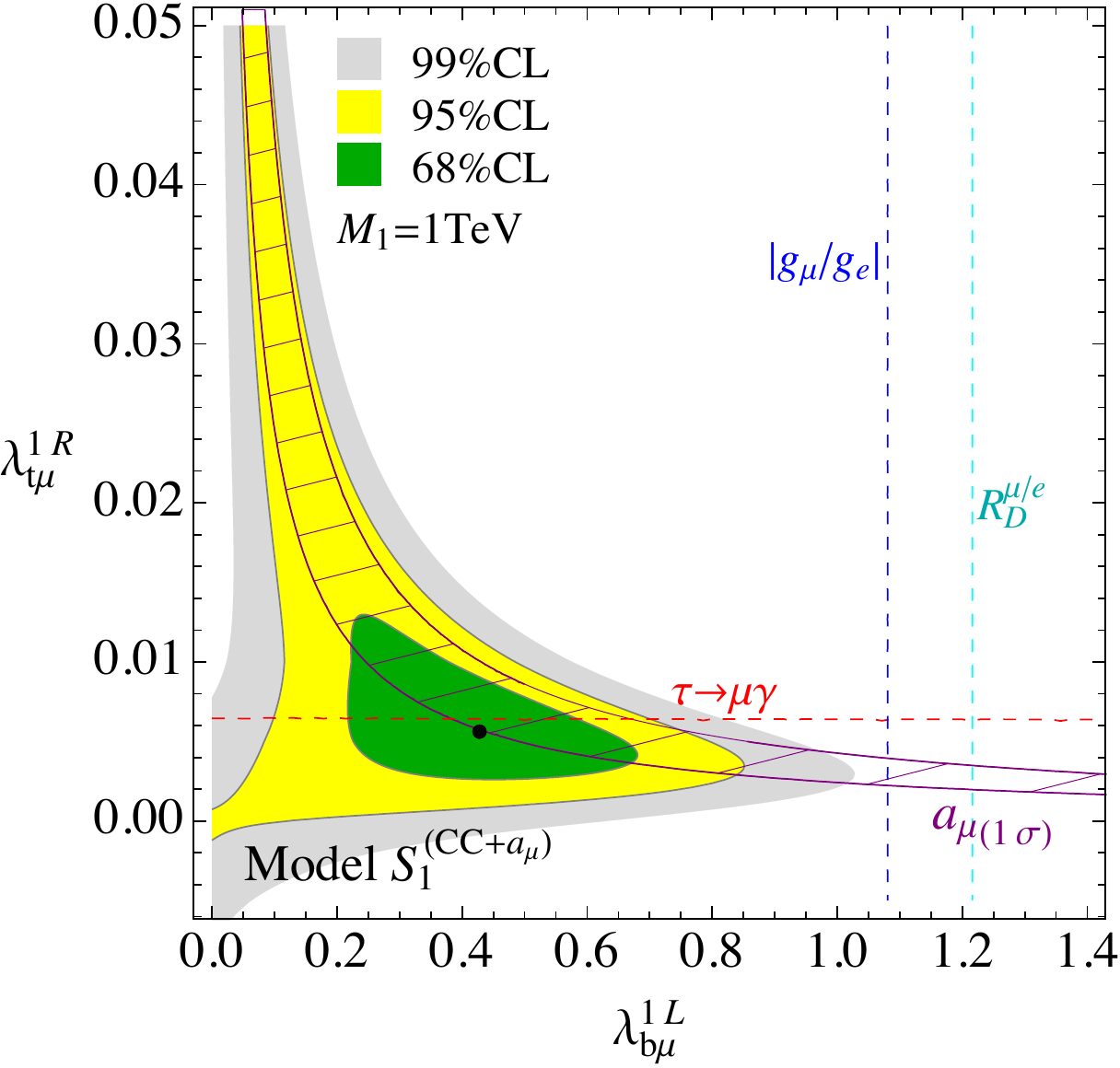}~
\includegraphics[height=6.5cm]{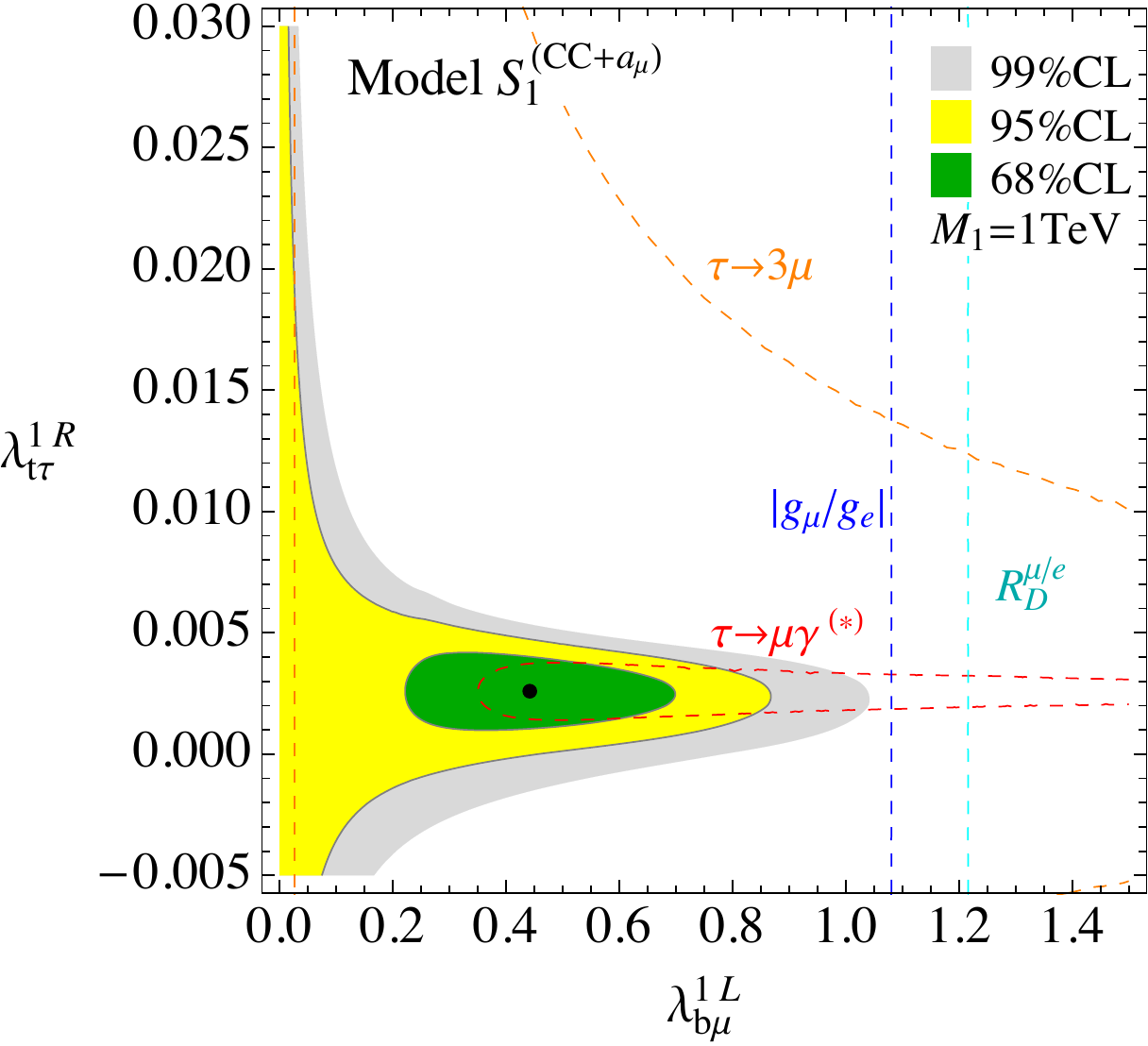} \\
\includegraphics[height=6.5cm]{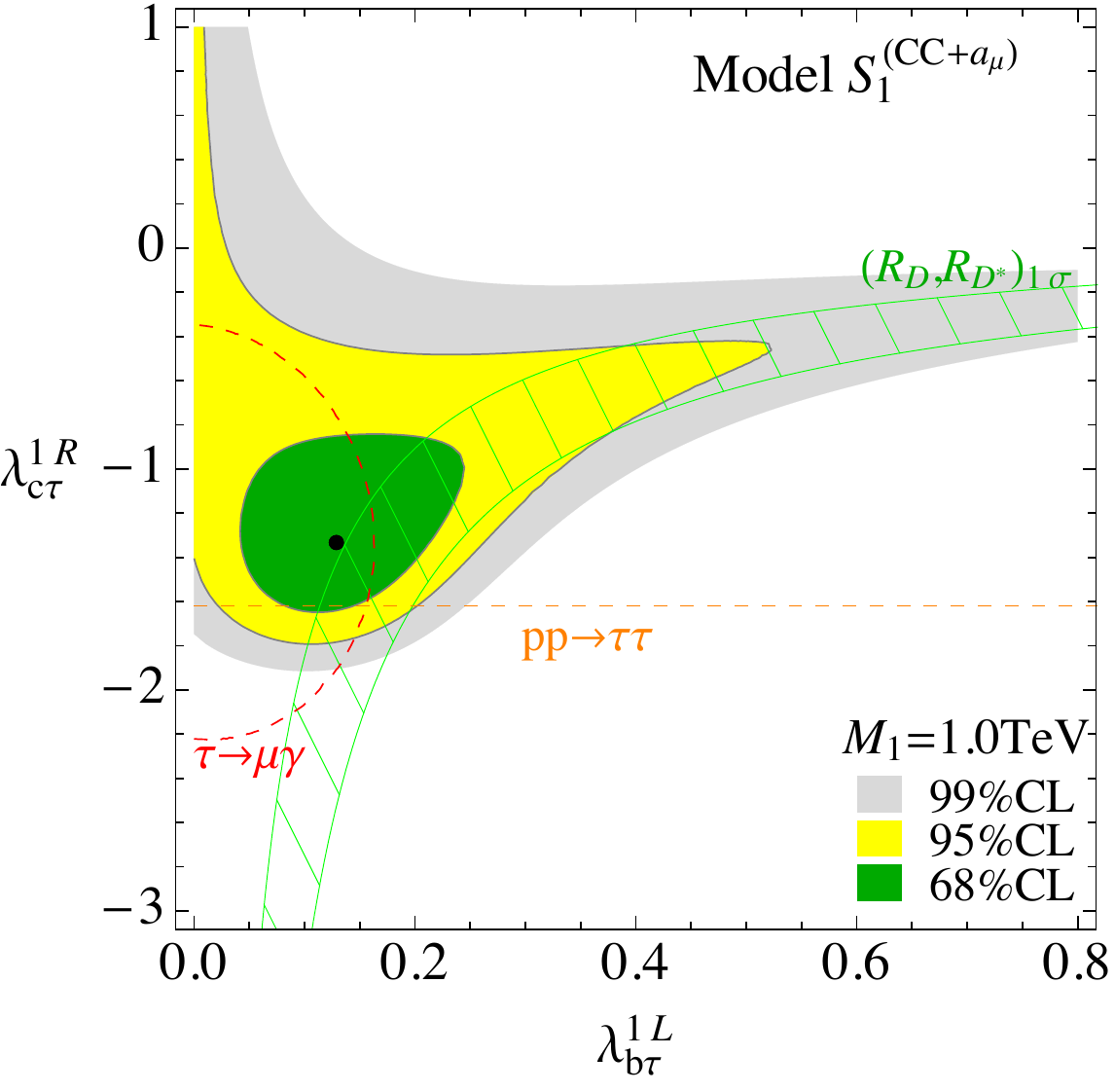} ~
\includegraphics[height=6.5cm]{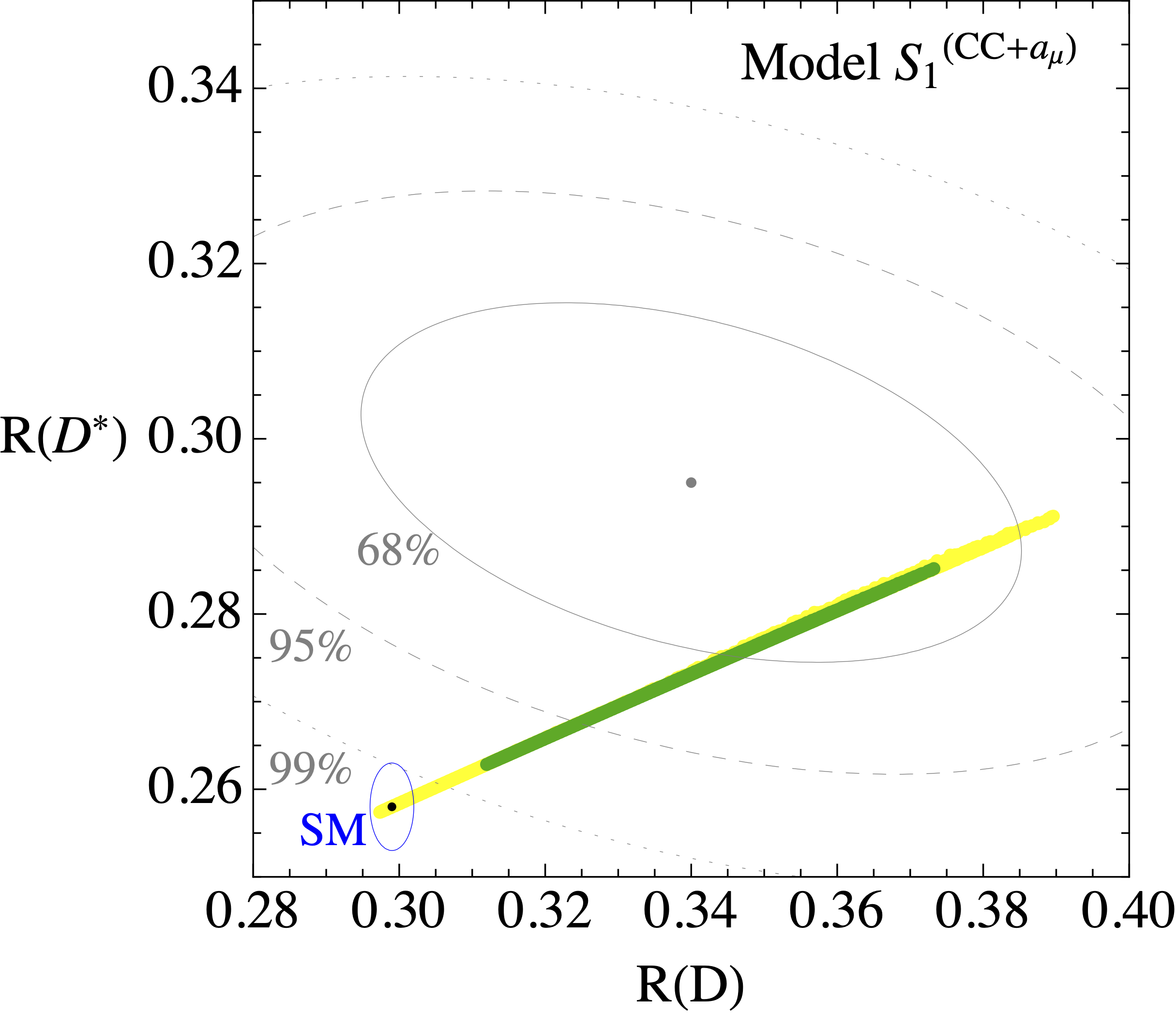}
\caption{\small Result from the fit in the $S_1$ model aimed to addressing the CC and $(g-2)_\mu$ anomalies. We show the preferred regions in the planes of two couplings, where those not shown are profiled over. The dashed lines show, for illustrative purposes, $2\sigma$ limits from individual observables where the other couplings are fixed at the best-fit point (black dot).
In the lower-right panel we show where the preferred region is mapped in the plane of $R(D)-R(D^*)$.}\label{fig:S1CCamu}
\end{figure}

Following the previous Subsections, the next natural step is to attempt to fit both $R(D^{(*)})$ and $a_\mu$ anomalies with $S_1$.
The relevant couplings are $\lambda^{1L}_{b\tau}$, $\lambda^{1L}_{b\mu}$, $\lambda^{1R}_{t\tau}$, $\lambda^{1R}_{c\tau}$, $\lambda^{1R}_{t\mu}$.

This setup is not simply the combination of those discussed previously.
Indeed, due to the  $\lambda^{1L}_{b\tau}$ and $\lambda^{1R}_{c\tau}$ couplings on the one hand (required to fit the charged-current $B$-anomaly), and $\lambda^{1R}_{t\mu}$, $\lambda^{1L}_{b\mu}$ on the other hand (necessary to fit $a_\mu$), sizeable contributions to $\tau \to \mu \gamma$ are generated at one-loop, see Eq.~(\ref{eq:taumuga_num}).
The values of $\lambda^{1L}_{b\tau, b\mu}$ and $\lambda^{1R}_{c\tau, t\mu}$ required to fit $R(D^{(*)})$ and $(g-2)_\mu$ would induce a too large contribution to this LFV decay.
However, we find that the large contribution to $\tau \to \mu \gamma$ arising from the product of $\lambda^{1L}_{b\mu} \lambda^{1R}_{c\tau}$ can be mostly cancelled by the $\lambda^{1L}_{b\mu} \lambda^{1R}_{t\tau}$ term, for $\lambda^{1R}_{t \tau} \sim 0.003$, corresponding to a small tuning of approximately one part in 5.
Such a small value does not affect any other observable in our fit. It should be noted that the only effect of this coupling is to tune this observable.

We show the preferred region in parameters space, obtained from our fit, in Fig.~\ref{fig:S1CCamu}. In the upper two panels we show the fit in the $(\lambda^{1L}_{b\mu}, \lambda^{1R}_{t\mu})$ and $(\lambda^{1L}_{b\mu}, \lambda^{1R}_{t\tau})$ planes, including the preferred region from the global fit as well as the individual 95\%CL limits from single observables, to help illustrating the physics behind the analysis. It can be noted that the observed value of the muon magnetic moment can be reproduced, and that $\tau \to \mu \gamma$ requires a non-zero value of $\lambda^{1R}_{t\tau}$, as discussed above. Regarding the top-right panel, the single-observable constraint from $\tau \to \mu \gamma$ (red dashed line) is shown by imposing that $a_\mu$ is fixed to the value we get at the best-fit point.
In the lower-left panel we show the fit in the couplings contributing to $R(D^{(*)})$, $(\lambda^{1L}_{b\tau}, \lambda^{1R}_{c\tau})$, and how this preferred region is mapped in the plane of $R(D) - R(D^{*})$ (lower-right panel).
Comparing to the allowed region in the same plane in the model studied in Fig.~\ref{fig:S1CC}, we see that $\tau \to \mu \gamma$ strongly reduced the allowed region, while still not preventing a good fit of the charged-current $B$-anomalies. Due to this reduction of the allowed $(\lambda^{1L}_{b\tau},\lambda^{1R}_{c\tau})$ parameter space, specifically with smaller values of $\lambda^{1L}_{b\tau}$ , the points in the $R(D^*)-R(D)$ plane line up more closely in a line than what is observed in Fig.~\ref{fig:S1CC}.

We conclude that the $R(D^{(*)})$ and $(g-2)_\mu$ anomalies can be addressed by the $S_1$ leptoquark, for perturbative couplings and TeV-scale mass. We find the best-fit point, with $M_1 = 1\TeV$, for
\be
	\left. S_1~^{(CC + a_\mu)}  \right|_{\text{best-fit}}: \quad
	\begin{array}{l l l} \lambda^{1L}_{b\tau} \approx 0.13, & 
	\lambda^{1L}_{b\mu} \approx 0.44, \\[0.2cm]
	 \lambda^{1R}_{t\tau} \approx 0.0026, &
	\lambda^{1R}_{c\tau} \approx -1.33, &
	\lambda^{1R}_{t\mu} \approx 0.0051. 
	\end{array}
\ee

%---------------------------------------------------------------------------------------
\subsection{Single-leptoquark $S_3$}
\label{sec:S3CCNC}

\begin{figure}[t]
\centering
\includegraphics[width=0.44\hsize]{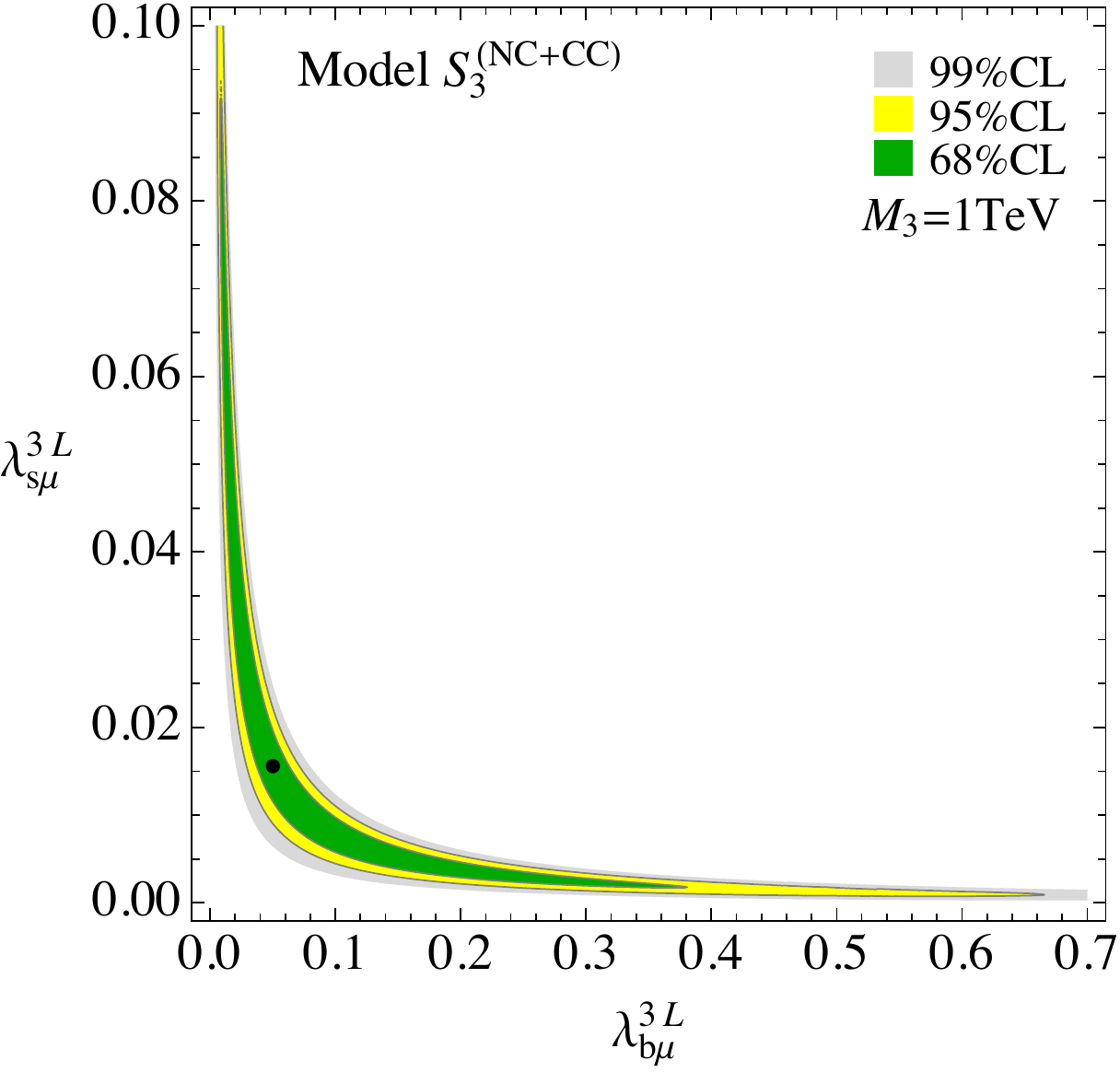}~
\includegraphics[width=0.44\hsize]{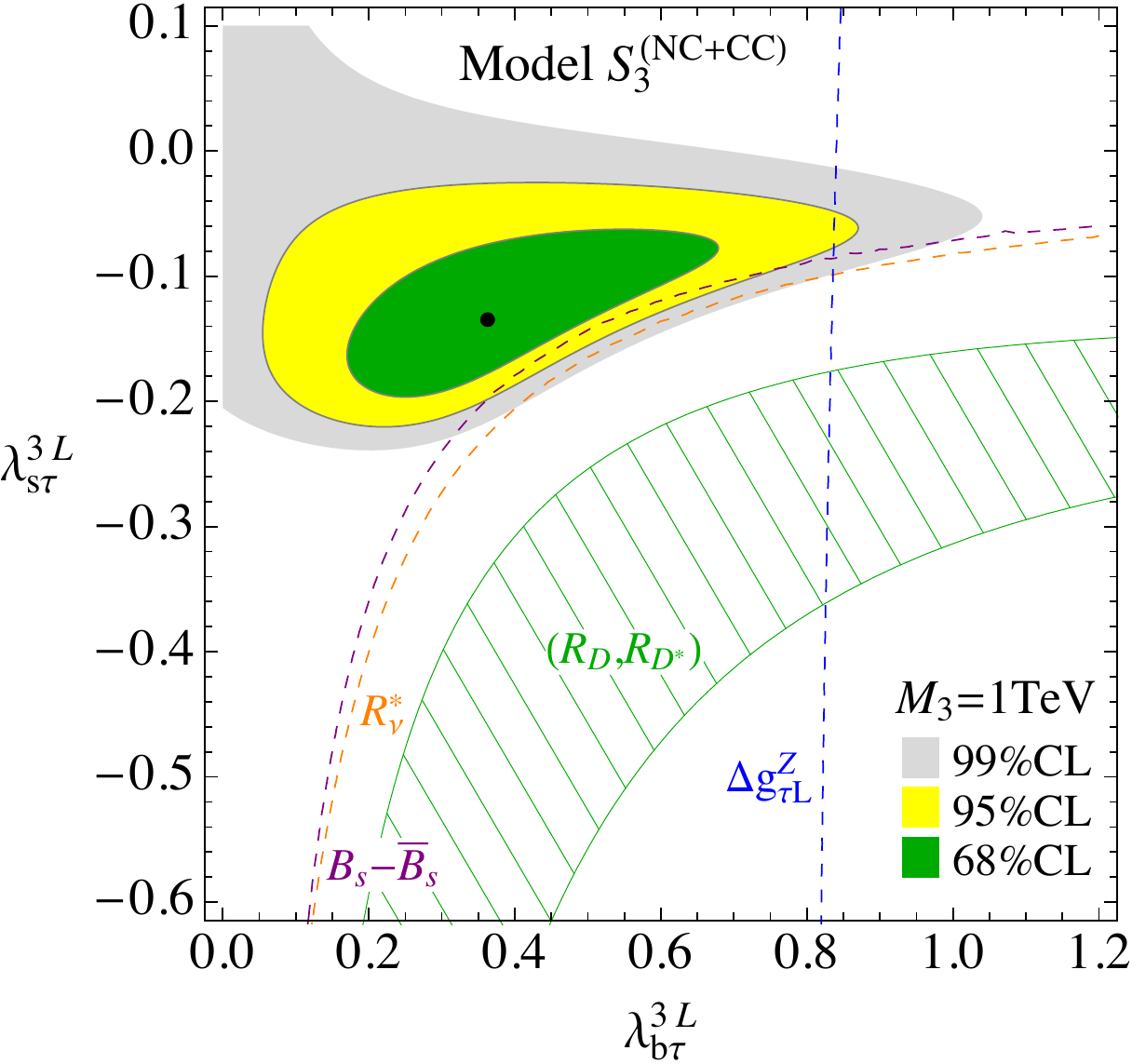} \\
\includegraphics[width=0.42\hsize]{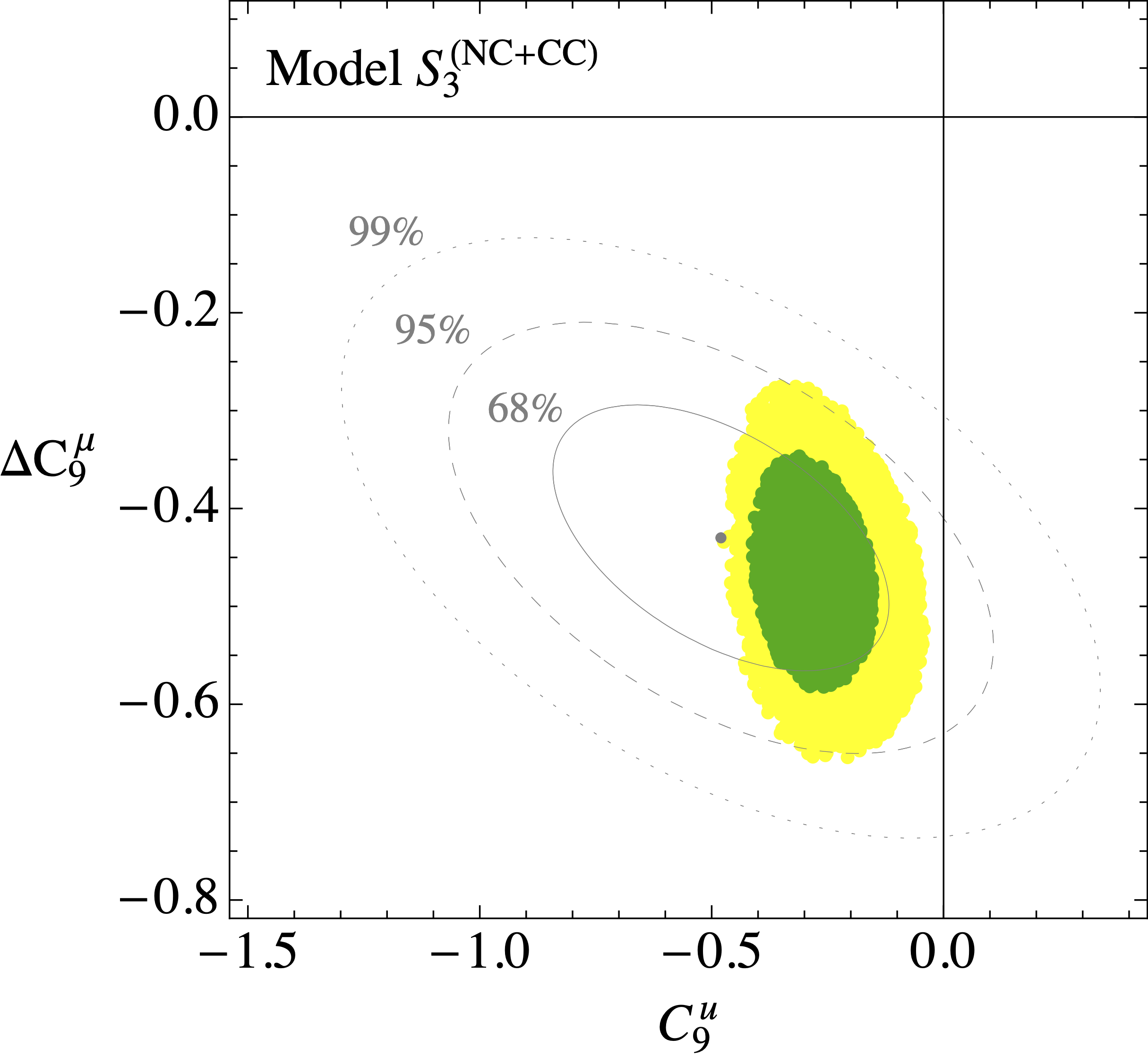} ~
\includegraphics[width=0.45\hsize]{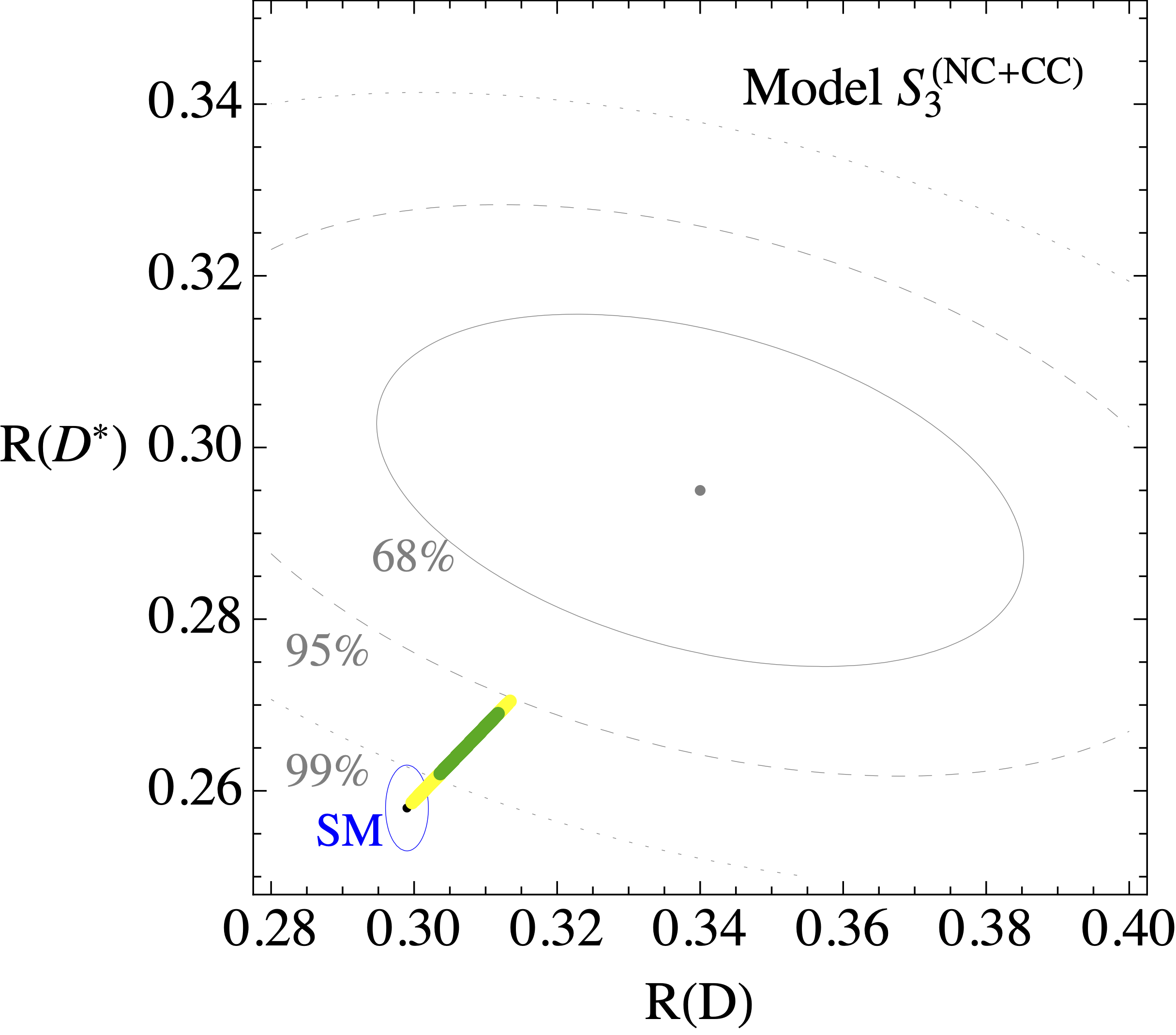} \\
\caption{\small Result from the fit in the $S_3$ model. In the upper panels we show the preferred regions in the planes of two couplings, where the two not shown are profiled over.
The dashed lines show, for illustrative purposes, $2\sigma$ limits from individual observables where the other couplings are fixed at the best-fit point (black dot).
In the lower panels we show where the preferred region is mapped in the planes of the neutral and charged-current anomalies.}\label{fig:S3CCNC}
\end{figure}

We move on to examine $S_3$, and we attempt directly  a combined explanation of charged and neutral current anomalies. It is well known that $S_3$ provides a simple and good explanation for the deviations observed in $b\to s \ell \ell$, thanks to its tree-level contribution to the partonic process. The couplings required are $\lambda ^{3L}_{b\mu}$-$\lambda ^{3L}_{s\mu}$, with small enough values that other observables do not pose relevant constraints.
The leading contribution to both $R(D^{(*)})$ and $C_9 ^u$, instead, arises via the $\lambda ^{3L}_{b\tau}$-$\lambda ^{3L}_{s\tau}$ couplings. For concreteness we fix $M_3=1\,\text{TeV}$, but the fit would be very similar for a slightly larger mass.

Our results can be seen in Fig.~\ref{fig:S3CCNC}. As expected, the model is successful in fitting $\Delta C_9^\mu$.
The couplings to the tau allow to also fit $C_9 ^u$, while charged-current anomalies cannot be reproduced. The main limiting observables are $B_s$-mixing and $B \to K^{(*)} \nu\nu$, as can be seen from the top-right panel.

The best-fit point, for $M_3 = 1\TeV$, is found for
\be
	\left. S_3\,^{(CC + NC)} \right|_{\text{best-fit}}: \quad
	\lambda^{3L}_{b\tau} \approx 0.36, \quad
	\lambda^{3L}_{s\tau} \approx -0.13, \quad
	\lambda^{3L}_{b\mu} \approx 0.050, \quad
	\lambda^{3L}_{s\mu} \approx 0.015.
\ee

%---------------------------------------------------------------------------------------
\subsection{$S_1 + S_3$ with LH couplings only}
\label{sec:S1S3LH}

\begin{figure}[t]
\centering
\includegraphics[width=0.45\hsize]{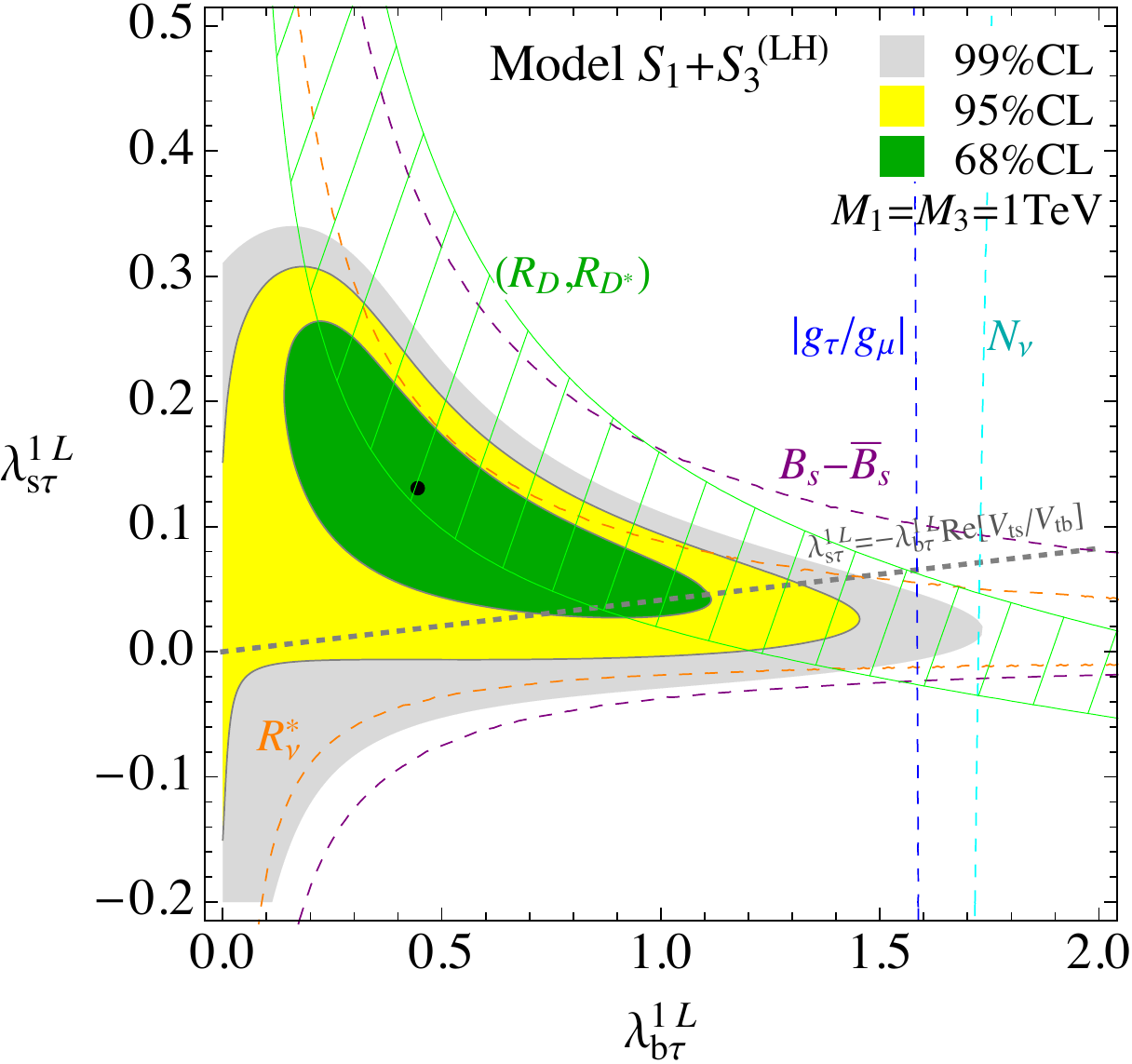}~
\includegraphics[width=0.45\hsize]{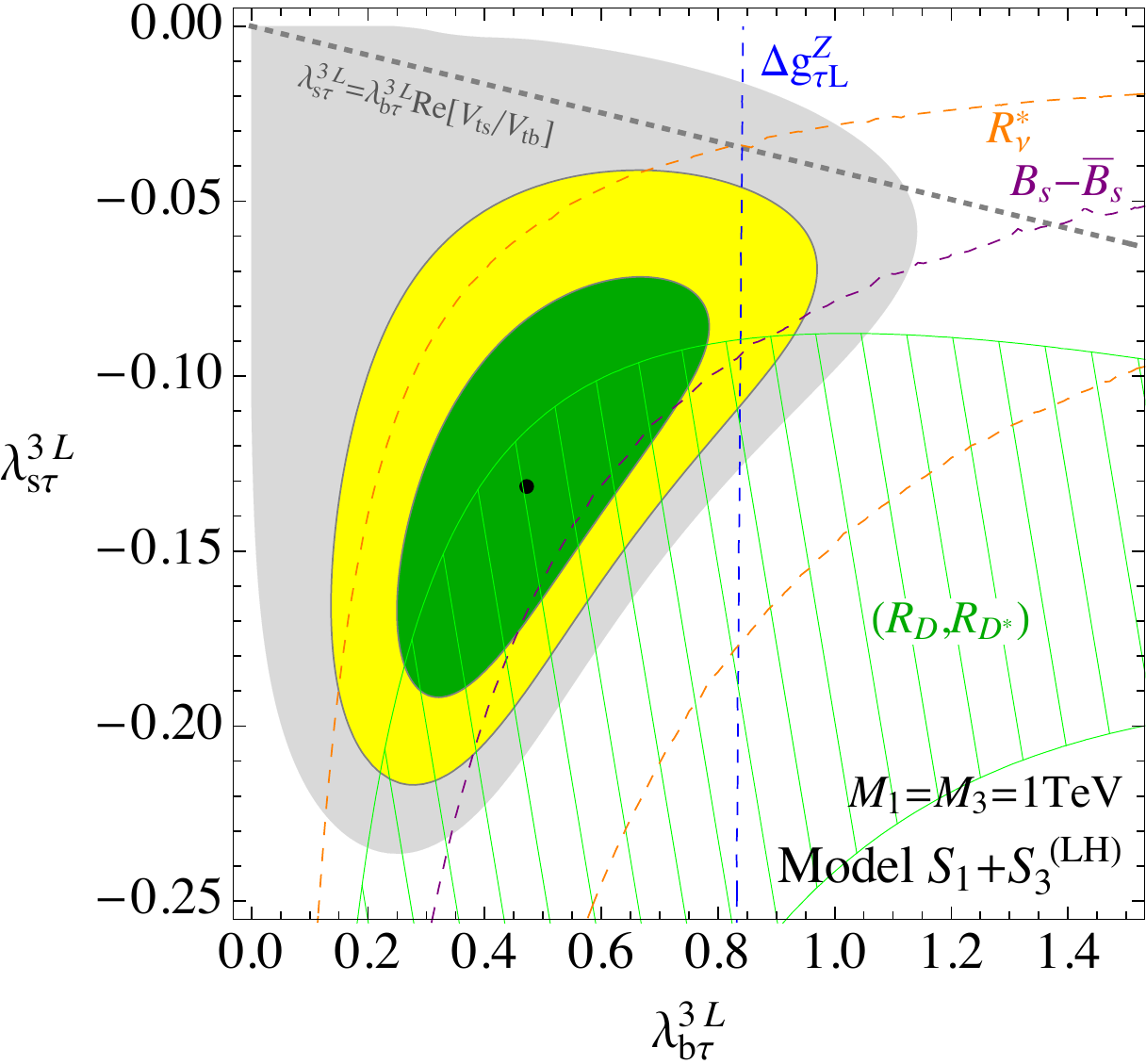} \\[5pt]
\includegraphics[width=0.43\hsize]{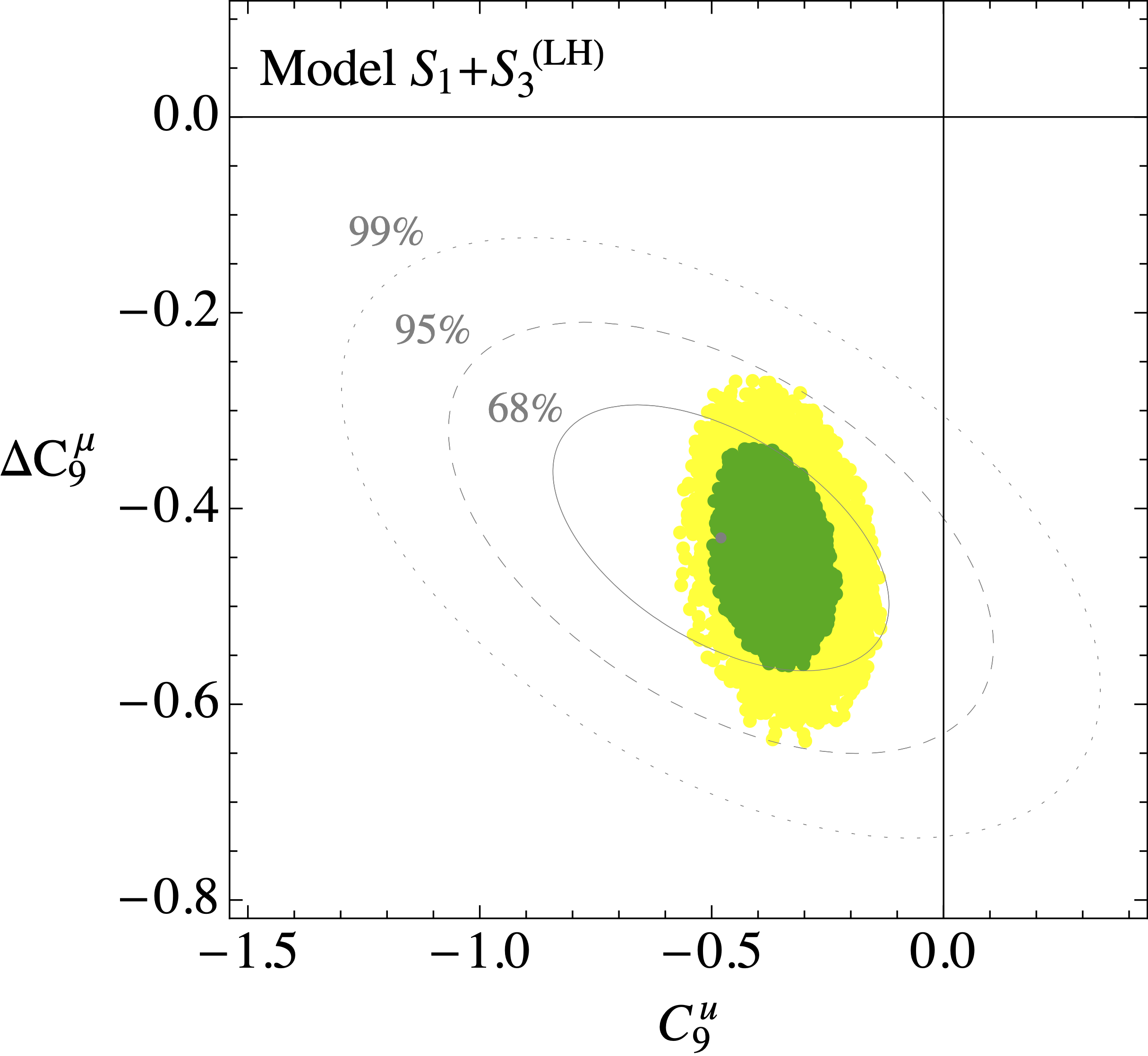} ~
\includegraphics[width=0.46\hsize]{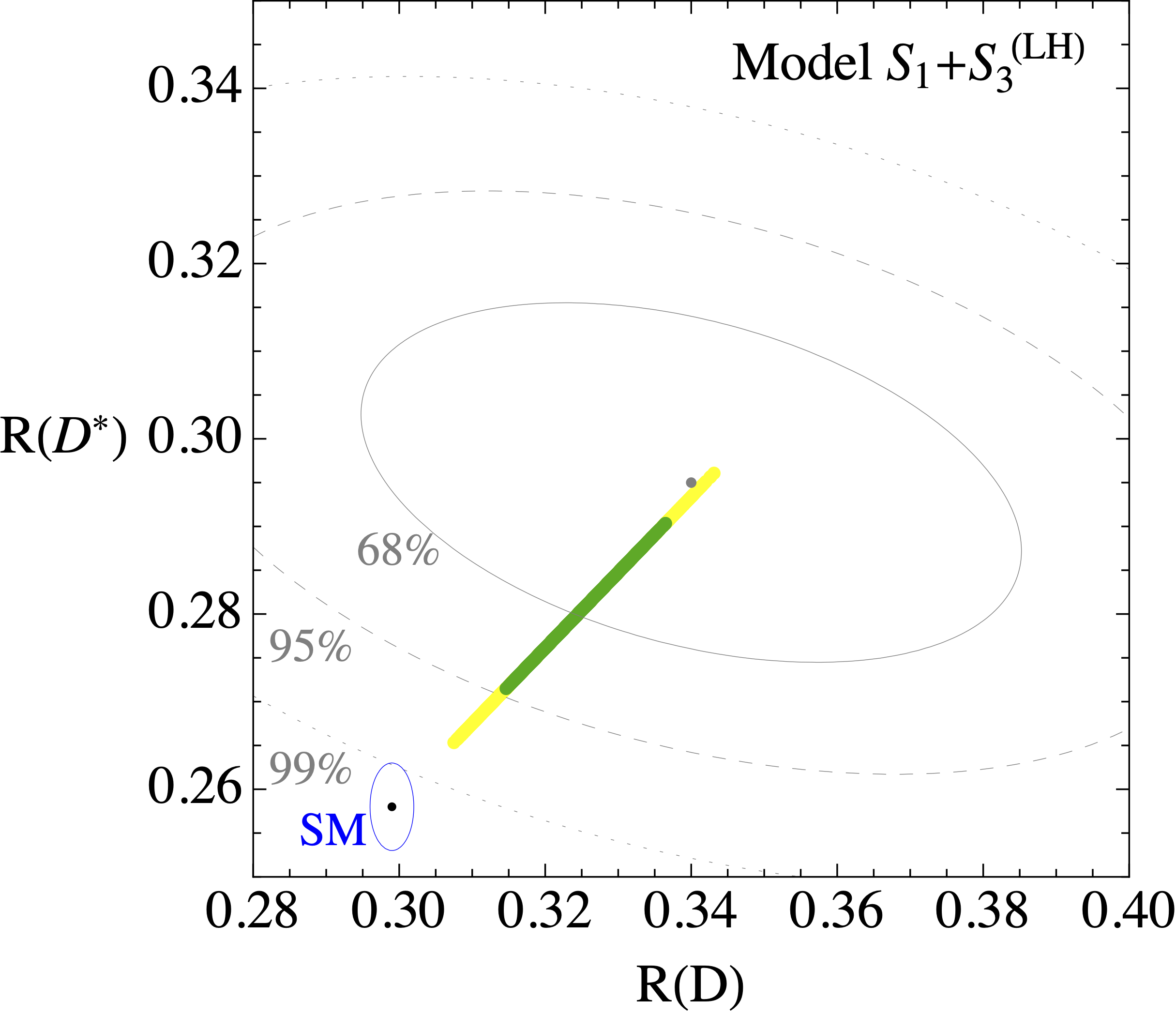} \\
\caption{\small Result from the fit in the $S_1 + S_3 \, ^{(\text{LH})}$ model, with only left-handed couplings. In the upper panels we show the preferred regions in the planes of two couplings, where the two not shown are profiled over.
The dashed lines show, for illustrative purposes, $2\sigma$ limits from individual observables where the other couplings are fixed at the best-fit point (black dot).
In the lower panels we show where the preferred region is mapped in the planes of the neutral and charged-current anomalies.}\label{fig:S1S3LH}
\end{figure}

Models involving $S_1$ and $S_3$ with left-handed couplings have been first considered in \cite{Crivellin:2017zlb,Buttazzo:2017ixm,Marzocca:2018wcf}. In particular, in \cite{Buttazzo:2017ixm,Marzocca:2018wcf} it was shown how this setup could fit both charged- and neutral-current anomalies with couplings compatible with a minimally broken $\U(2)^5$ flavor symmetry, albeit with a tension between $R(D^{(*)})$ and the $B_s$-mixing constraint. Since then, new experimental updates on $R(D^{(*)})$ pushed the preferred region closer to the SM, thus also alleviating the tension with meson mixing. Here, we update the fit for this scenario, without assuming a priori a specific flavor structure for the relevant couplings.

The relevant couplings are $\lambda^{1L}_{[bs]\tau}$, $\lambda^{3L}_{[bs]\mu}$, and $\lambda^{3L}_{[bs]\tau}$. A first qualitative understanding of the model can be obtained by noticing the main roles of the various couplings with regard to the anomalies:
\be
\lambda^{3L}_{[b,s]\mu}\to \Delta C_9 ^\mu,\qquad \lambda ^{3L}_{[b,s]\tau}\to C_9 ^u,\qquad (\lambda^{1L}_{[b,s]\tau}, \lambda^{3L}_{[b,s]\tau}) \to R(D^{(*)}).
\ee
In this model, the relative deviation in $R(D)$ and $R(D^*)$ from the respective SM values is predicted to be exactly the same, since it is only due to the same left-handed vector-vector operator generated in the SM. The contribution is given by the combination in Eq.~\eqref{eq:CVL}.

The most salient features of the fit are summarized in Fig.~\ref{fig:S1S3LH}.
In the top two panels we show the preferred regions in the $\lambda^{1L}_{b\tau} - \lambda^{1L}_{s\tau}$ and $\lambda^{3L}_{b\tau} - \lambda^{3L}_{s\tau}$ planes, together with the single-observable $2\sigma$ limits obtained fixing the other couplings to the global best-fit value. The favoured region in the $\lambda^{3L}_{b\mu} - \lambda^{3L}_{s\mu}$ plane is very similar to the one of model $S_3^{\,(CC + NC)}$ (Fig.~\ref{fig:S3CCNC} top-left), thus we do not show it again. 
The constraint from $B \to K^{(*)} \nu \nu$ is avoided thanks to a slight cancellation between the tree-level contributions of the two leptoquarks \cite{Buttazzo:2017ixm}, see Eq.~\eqref{eq:EFTRnu}. 
There is a (small) leftover tension in the $R(D^{(*)})$ fit, due to constraints from $B_s$-mixing. It should be noted that this tension grows with larger LQ masses (thus larger required couplings) since the deviation in $R(D^{(*)})$ scales as $\lambda^2/M^2$ while the contribution to meson mixing goes as $\lambda^4/M^2$.

We also point out that the parameter-region preferred by the fit is compatible with the relations between couplings predicted by a minimally-broken $\U(2)^5$ flavor symmetry, $\lambda_{s \alpha} = c_{\U(2)} V_{ts} \lambda_{b \alpha}$, with $c_{\U(2)}$ an $\mathcal{O}(1)$ complex parameter, see e.g. \cite{Buttazzo:2017ixm} and references therein. The case with $|c_{\U(2)}| = 1$ is shown with grey dashed lines in the upper panels.

In the lower two panels we show how the preferred regions in parameter space maps into the anomalous $B$-decay observables. As can be seen, it is possible to reproduce them within $1\sigma$.
The best-fit point, for $M_1 = M_3 = 1\TeV$, is found for
\be
	\left. \begin{array}{c} S_1 + S_3 ~^{(LH)} \\  \end{array} \right|_{\text{best-fit}}: \quad
	\begin{array}{l l l l}
		\lambda^{3L}_{b\tau} \approx 0.47, &
		\lambda^{3L}_{s\tau} \approx -0.13, &
		\lambda^{3L}_{b\mu} \approx 0.056, &
		\lambda^{3L}_{s\mu} \approx 0.014, \\[0.2cm]
		\lambda^{1L}_{b\tau} \approx 0.45, &
		\lambda^{1L}_{s\tau} \approx 0.13.
	\end{array}
\ee

%---------------------------------------------------------------------------------------
\subsection{$S_1 + S_3$  addressing CC, NC, and $(g-2)_\mu$} 
\label{sec:S1S3amu}

\begin{figure}[p]
\centering
\includegraphics[height=6.5cm]{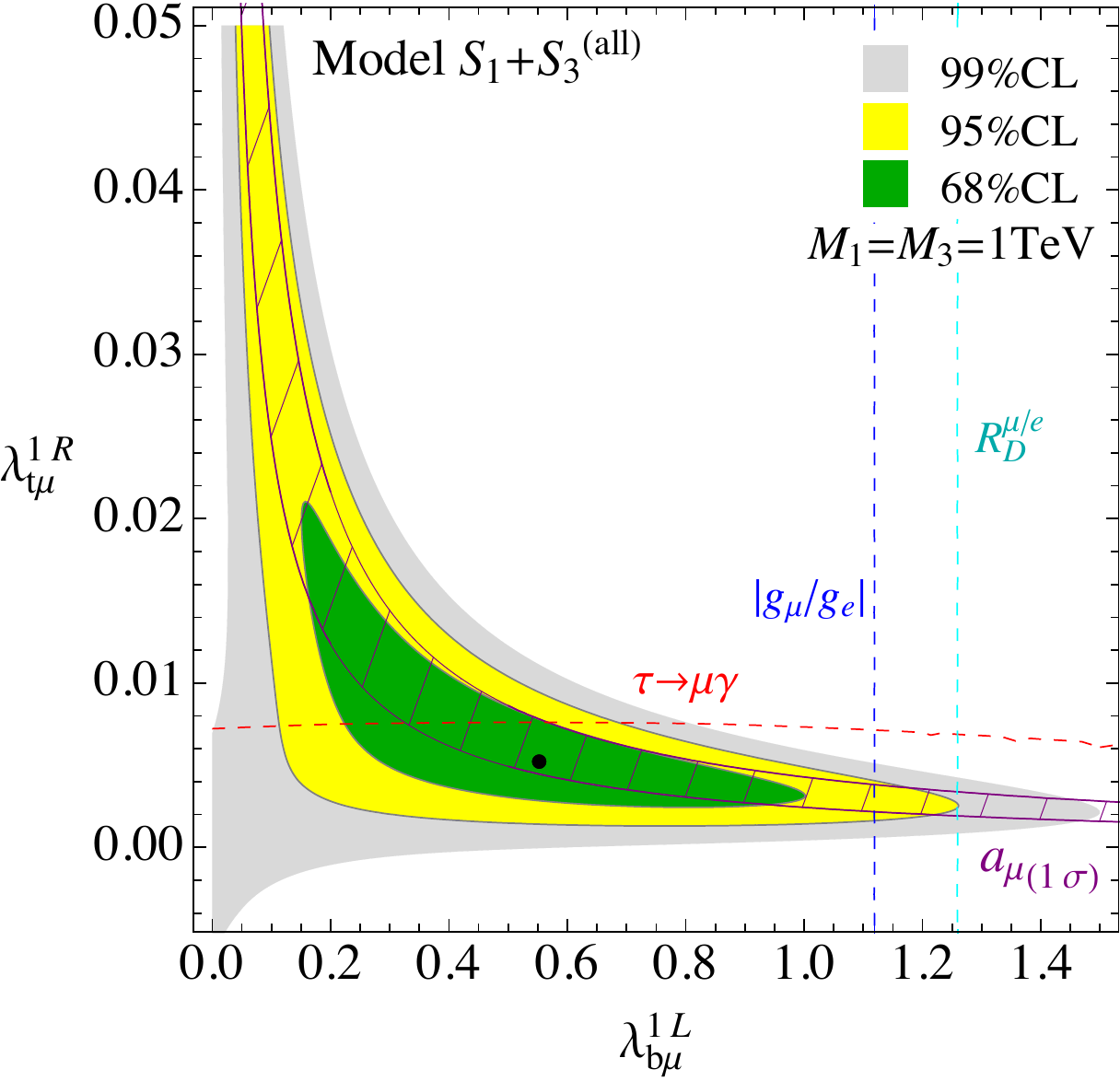}~
\includegraphics[height=6.5cm]{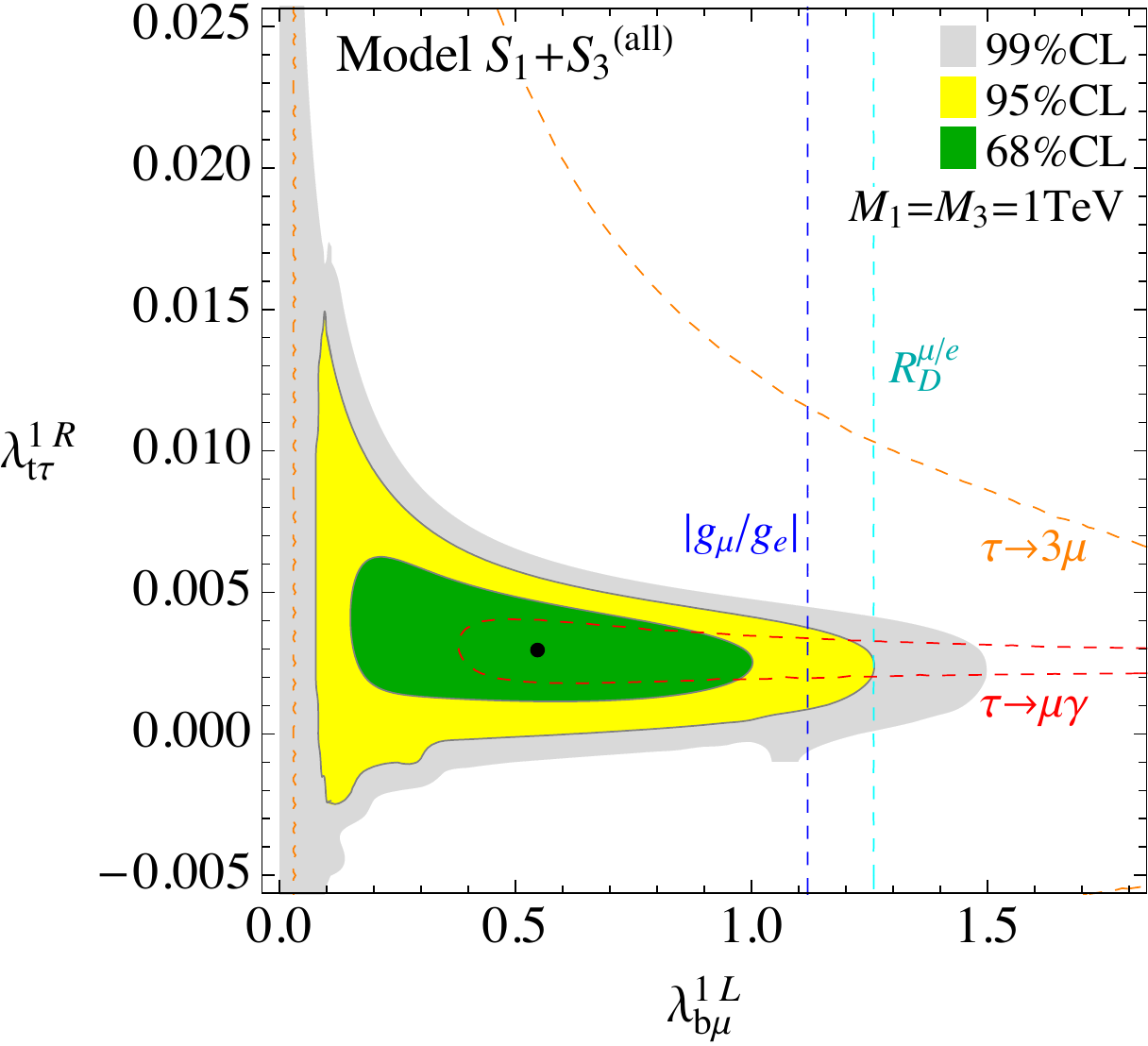} \\[5pt]
\includegraphics[height=6cm]{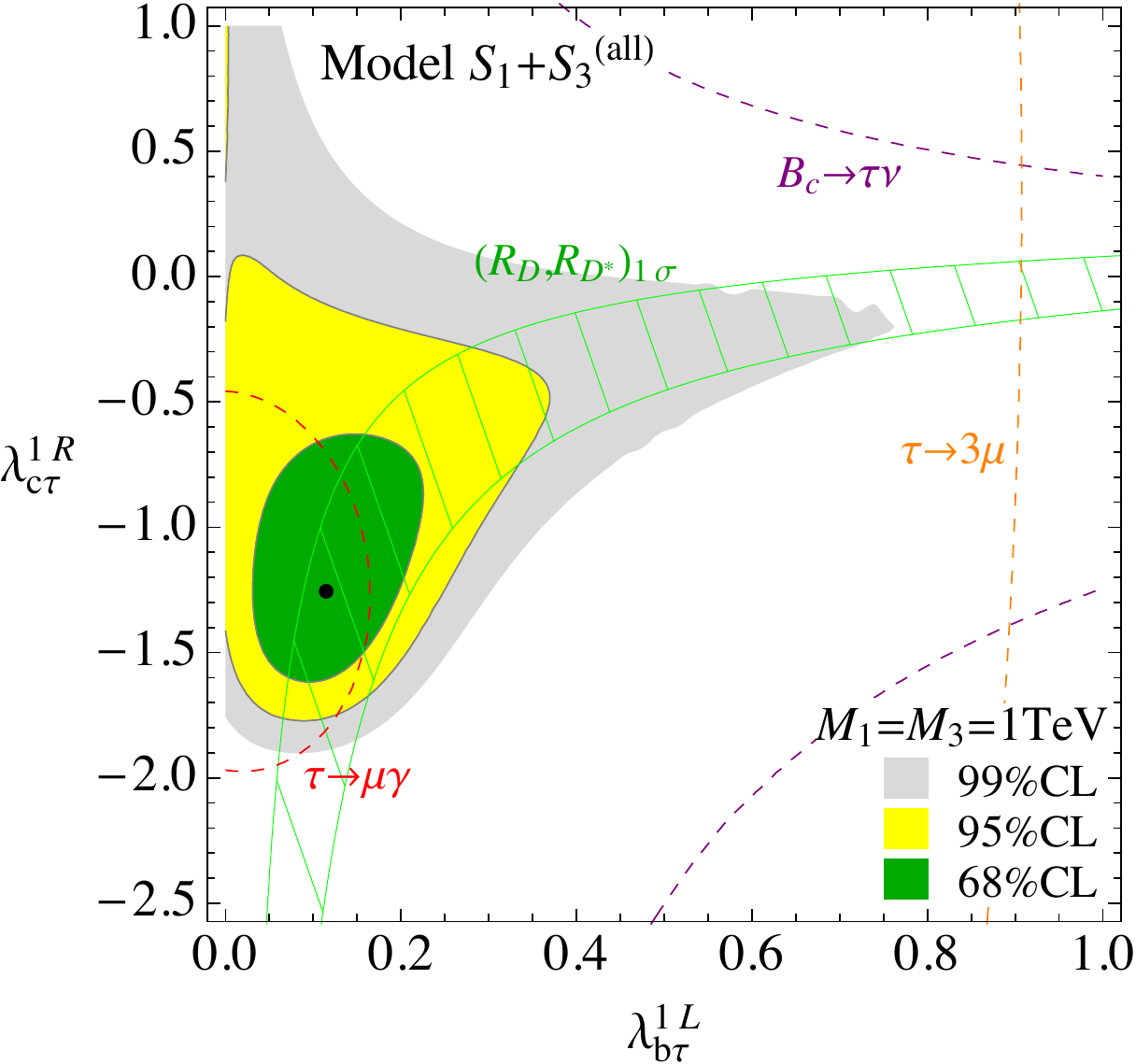}~
\includegraphics[height=6cm]{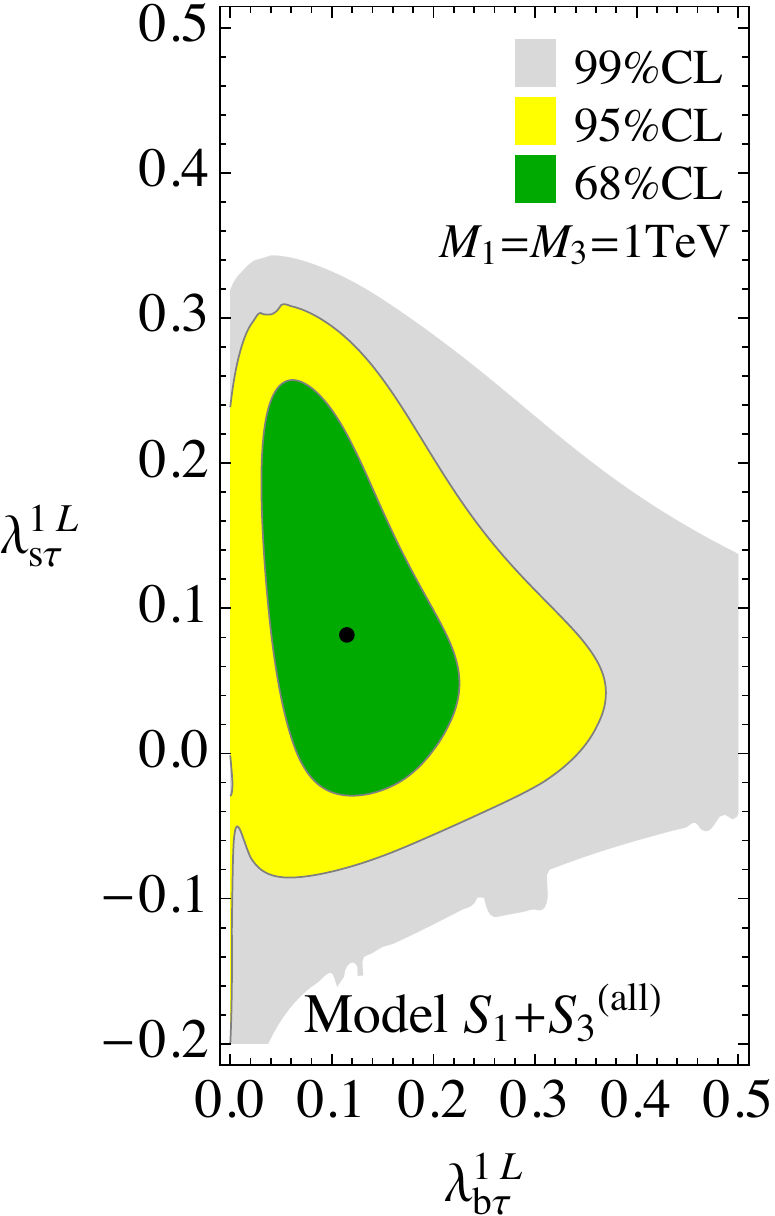} ~
\includegraphics[height=6cm]{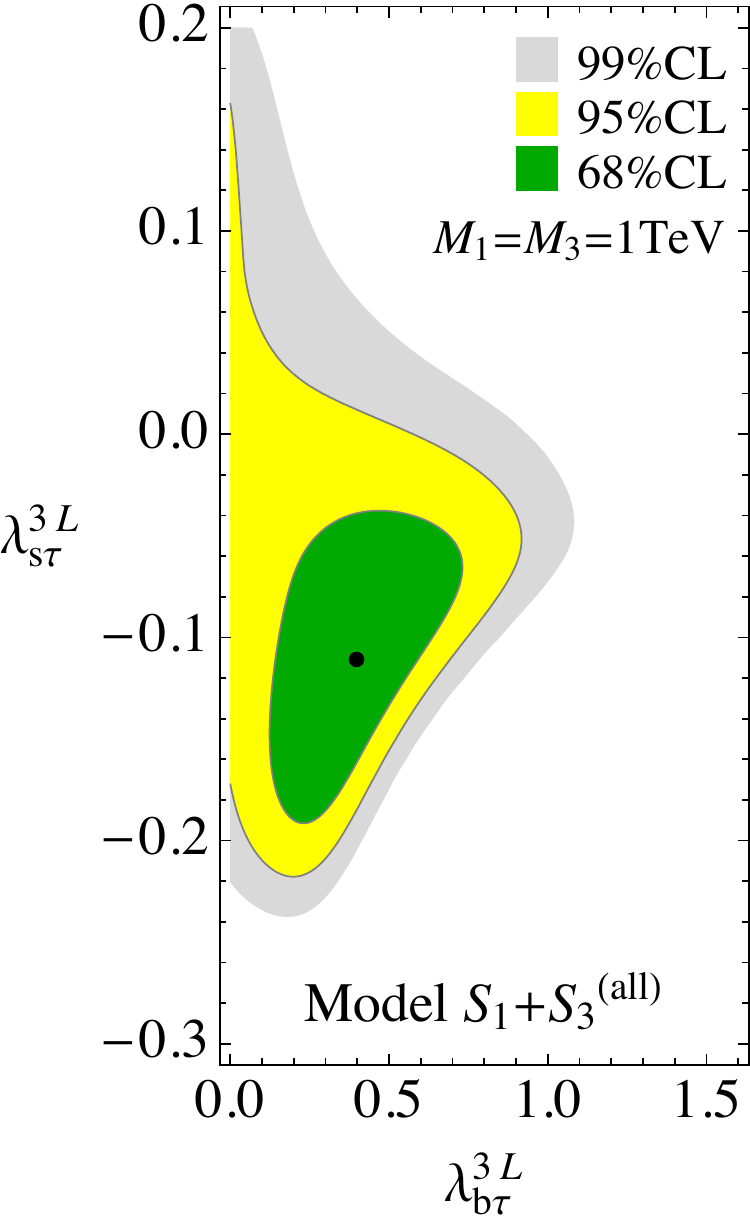} \\[5pt]
\includegraphics[height=6cm]{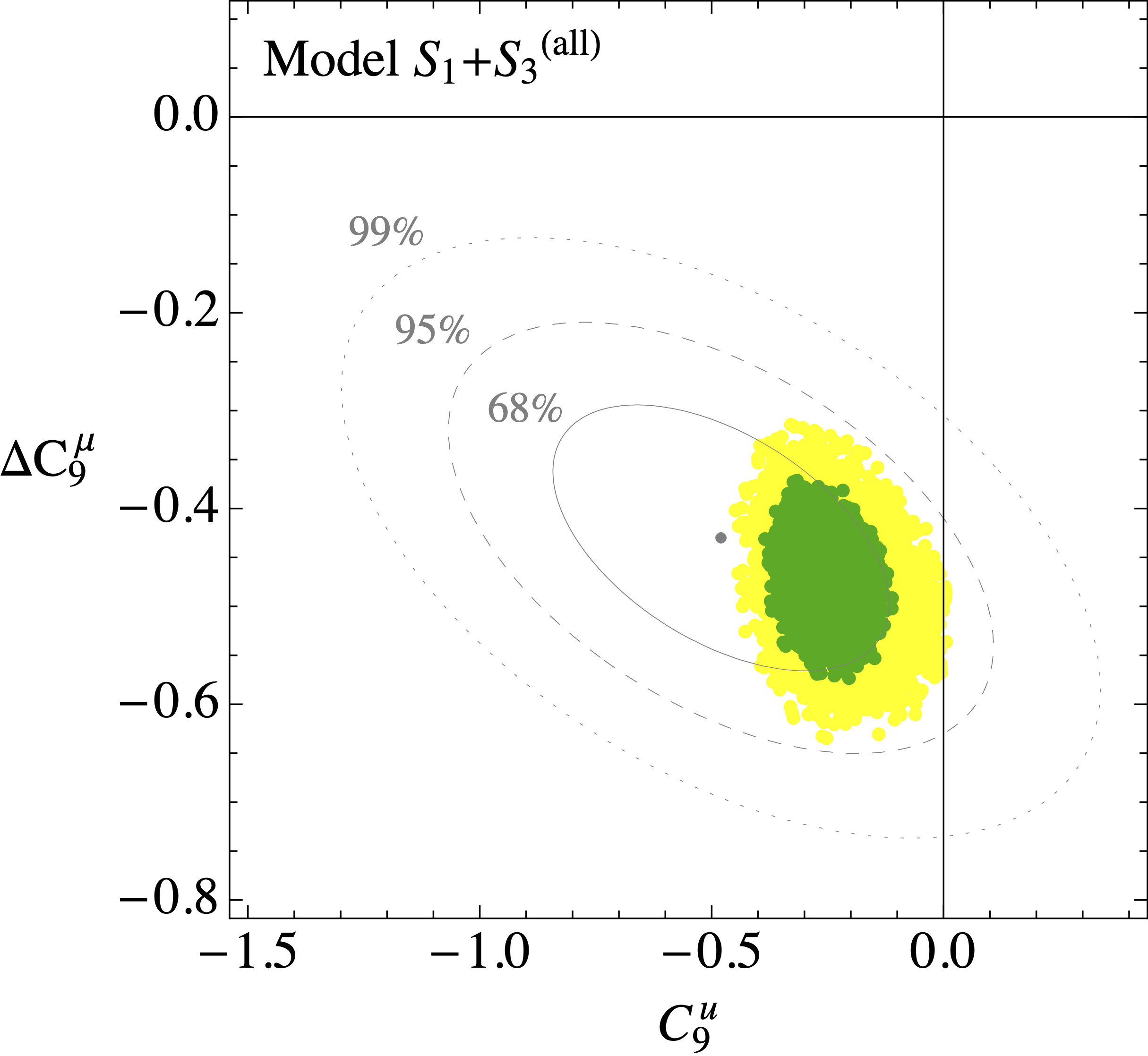} ~
\includegraphics[height=6cm]{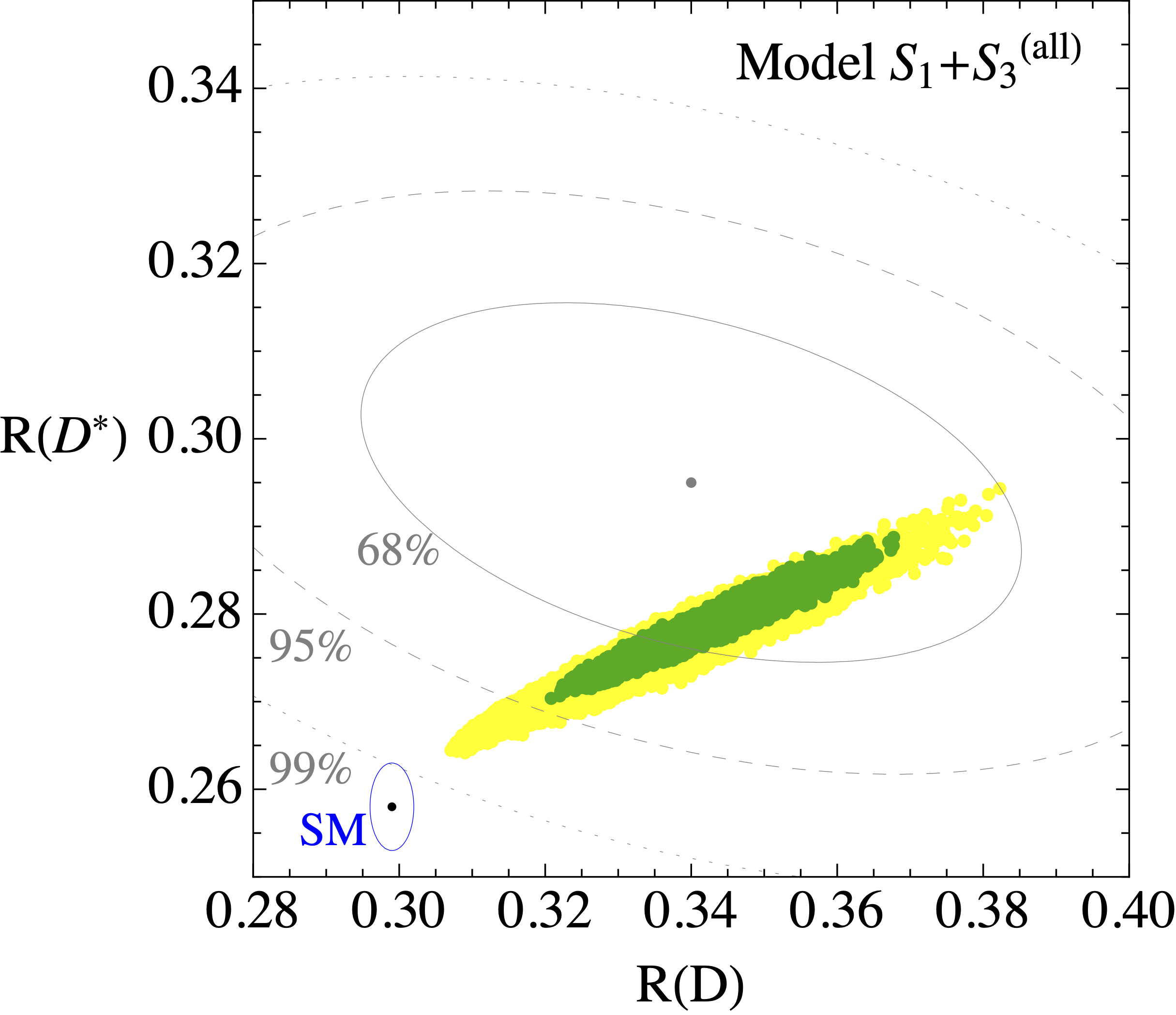} \\
\caption{\small Result from the fit in the $S_1 + S_3 \,^{(\text{all})}$ model, aimed at addressing all anomalies (see description in the text).}\label{fig:S1S3IV}
\end{figure}

From the previous Sections it is clear that in order to address all anomalies, both $S_1$ and $S_3$ leptoquarks are required. NC anomalies are addressed only by $S_3$, the muon anomalous magnetic moment only by $S_1$, while $R(D^{(*)})$ receives sizeable contributions from both.
For our most general analysis we keep ten active couplings:
\be
	\lambda ^{3L}_{[b,s]\tau}, \quad
	\lambda^{3L}_{[b,s]\mu}, \quad
	\lambda^{1L}_{[b,s]\tau}, \quad
	\lambda^{1L}_{b\mu}, \quad
	\lambda^{1R}_{[t,c]\tau}, \quad
	\lambda^{1R}_{t\mu}.
\ee
The results of our fit are shown in Fig.~\ref{fig:S1S3IV}.

In the first row of Fig.~\ref{fig:S1S3IV} we show the preferred regions for the couplings relevant for the $a_\mu$ fit. The situation is very similar to what already discussed for model $S_1^{\,(CC+a_\mu)}$, Sec.~\ref{sec:S1CCamu}.

The couplings relevant for the $R(D^{(*)})$ fit are shown in the second row. They show a behavior very similar to the one already seen in the models $S_1^{\,(CC+a_\mu)}$ and $S_1 + S_3^{\,(LH)}$. The main contribution is due to the scalar+tensor operators generated via the $\lambda^{1R}_{c\tau} \lambda^{1L}_{b\tau}$ couplings, but a sizeable contribution, which helps to improve the fit with respect to model $S_1^{\,(CC)}$, is induced via the left-handed couplings $\lambda^{1L}_{[b,s]\tau}$ and $\lambda^{3L}_{[b,s]\tau}$, analogously to what we saw in model $S_1 + S_3^{\,(LH)}$. Contrary to that case, however, here the preferred region avoids any tension with both $B_s$-mixing and $B \to K^{(*)} \nu \nu$.

We do not show in Fig.~\ref{fig:S1S3IV} the preferred values for $\lambda^{3L}_{[s,b]\mu}$, which are necessary to fit $\Delta C_9^\mu$, since they are analogous to what we saw for model $S_3^{\,(CC+NC)}$ (see Fig.~\ref{fig:S3CCNC} top-left).

We conclude that all the anomalies in $R(D^{(*)})$, $b \to s \mu\mu$, and $(g-2)_\mu$, can be completely addressed in this model, for perturbative couplings and TeV-scale leptoquark masses.
The best-fit point, for $M_1 = M_3 = 1\TeV$, is found for
\be
	\!\! \left.  S_1 + S_3~^{\rm (all)}  \right|_{\text{best-fit}}:
	\begin{array}{l l l l}
		\lambda^{3L}_{b\tau} \approx 0.40, &
		\lambda^{3L}_{s\tau} \approx -0.11, &
		\lambda^{3L}_{b\mu} \approx 0.31, &
		\lambda^{3L}_{s\mu} \approx 0.0024, \\[0.2cm]
		\lambda^{1L}_{b\tau} \approx 0.11, &
		\lambda^{1L}_{s\tau} \approx 0.082, &
		\lambda^{1L}_{b\mu} \approx 0.55, \\[0.2cm]
		\lambda^{1R}_{t\tau} \approx 0.0029, &
		\lambda^{1R}_{c\tau} \approx -1.26, &
		\lambda^{1R}_{t\mu} \approx 0.0052.
	\end{array}
\ee

%---------------------------------------------------------------------------------------
\subsection{Leptoquark potential couplings}
\label{sec:S1S3Higgs}

In this Section we study available constraints for the potential couplings of leptoquark with the Higgs boson in the third and fourth lines of Eq.\eqref{eq:S1S3Model}.
There are four such couplings: $\lambda_{H1}$, $\lambda_{H3}$, $\lambda_{\epsilon H3}$, and $\lambda_{H13}$.
All contribute only at one-loop level in the matching to SMEFT operators, therefore possible phenomenological effects are suppressed both by a loop factor and by the LQ mass scale.
We focus on effects of these couplings which are independent on the LQ couplings to fermions. We thus need precisely measured quantities in the bosonic sector of the SM.

\begin{table}[t]
\begin{center}
\begin{tabular}{ | c | c l | c |}
\hline
\cellcolor[gray]{0.92} {\bf Observable} & \cellcolor[gray]{0.92} {\bf Measurement} & \cellcolor[gray]{0.92} & \cellcolor[gray]{0.92} {\bf Reference} \\
\hline
\hline
$S$ & $0.04 \pm 0.08$ & & \cite{Haller:2018nnx}\\
\hline
$T$ & $ 0.08 \pm 0.07$  &($\rho_{S, T} = 0.92$) & \cite{Haller:2018nnx} \\
 \hline                                       
$\kappa_{g}$ & $1.00 \pm 0.06$ & &  \cite{Aad:2019mbh}  \\
 \hline                                       
$\kappa_{\gamma}$ & $1.03 \pm 0.07$ & ($\rho_{\gamma, g} = -0.44$) &  \cite{Aad:2019mbh}  \\
 \hline                                       
$\sigma/\sigma_{\SM}(Z\gamma)$ & $2.0^{+1.0}_{-0.9}$ & (ATLAS) &  \cite{Aad:2020plj}  \\
 \hline  
$\sigma/\sigma_{\SM}(Z\gamma)$ & $< 3.9$ @ 95\% CL  & (CMS) &  \cite{Sirunyan:2018tbk}  \\
 \hline                                       
\end{tabular}
\caption{ Bosonic observables for the LQ potential couplings. \label{tab:obsBosonic}}
\end{center} 
\end{table}

Obvious candidates are the gauge-boson oblique corrections measured at LEP \cite{Barbieri:2004qk}: $\hat S$, $\hat T$, $Y$, $W$, as well as the analogous effect for QCD, $Z$. All these parameters are measured at the per-mille level, and are able to constrain multi-TeV scale physics. Given the expressions in the Warsaw basis of \cite{Wells:2015uba} and our one-loop matching of the SMEFT to the LQ model, we obtain (see App.~\ref{app:EWPO} for details)
\be\begin{split}
	\hat{S} & = \frac{\alpha}{4 s_W^2} S = - \frac{g^2 N_c v^2 Y_{S_3}}{48 \pi^2} \frac{\lambda_{\epsilon H3}}{M_3^2} \approx - 5.4 \times 10^{-5} \lambda_{\epsilon H3} / m^2~,\\
	\hat{T} & = \alpha T = \frac{N_c v^2 \lambda_{\epsilon H3}^2}{48 \pi^2 M_3^2}  + \frac{N_c v^2 }{16 \pi^2} |\lambda_{H13}|^2 \frac{M_1^4 - M_3^4 - 2 M_1^2 M_3^2 \log M_1^2/M_3^2}{(M_1^2 - M_3^2)^3} = \\
		&\approx 3.8 \times 10^{-4} \lambda^2_{\epsilon H3} / m^2 + 3.8 \times 10^{-4} |\lambda_{H13}|^2 / m^2~,
	\label{eq:STnumappr}
\end{split}\ee
where in the numerical expressions for simplicity we fixed $M_1 = M_3 = m \TeV$. The contributions to $Y$, $W$, and $Z$ are instead at, or below, the $10^{-6}$ level and thus completely negligible given the present experimental precision.
The constraints on $S$ and $T$ from \cite{Haller:2018nnx} are reported in Table~\ref{tab:obsBosonic}.
The contribution to the $T$ parameter from the $\lambda_{H13}$ coupling has been also studied in \cite{Dorsner:2019itg}, albeit not in the EFT approach. We checked that we agree once the EFT limit is taken into account.

\begin{figure}[t]
\centering
\includegraphics[width=0.45\hsize]{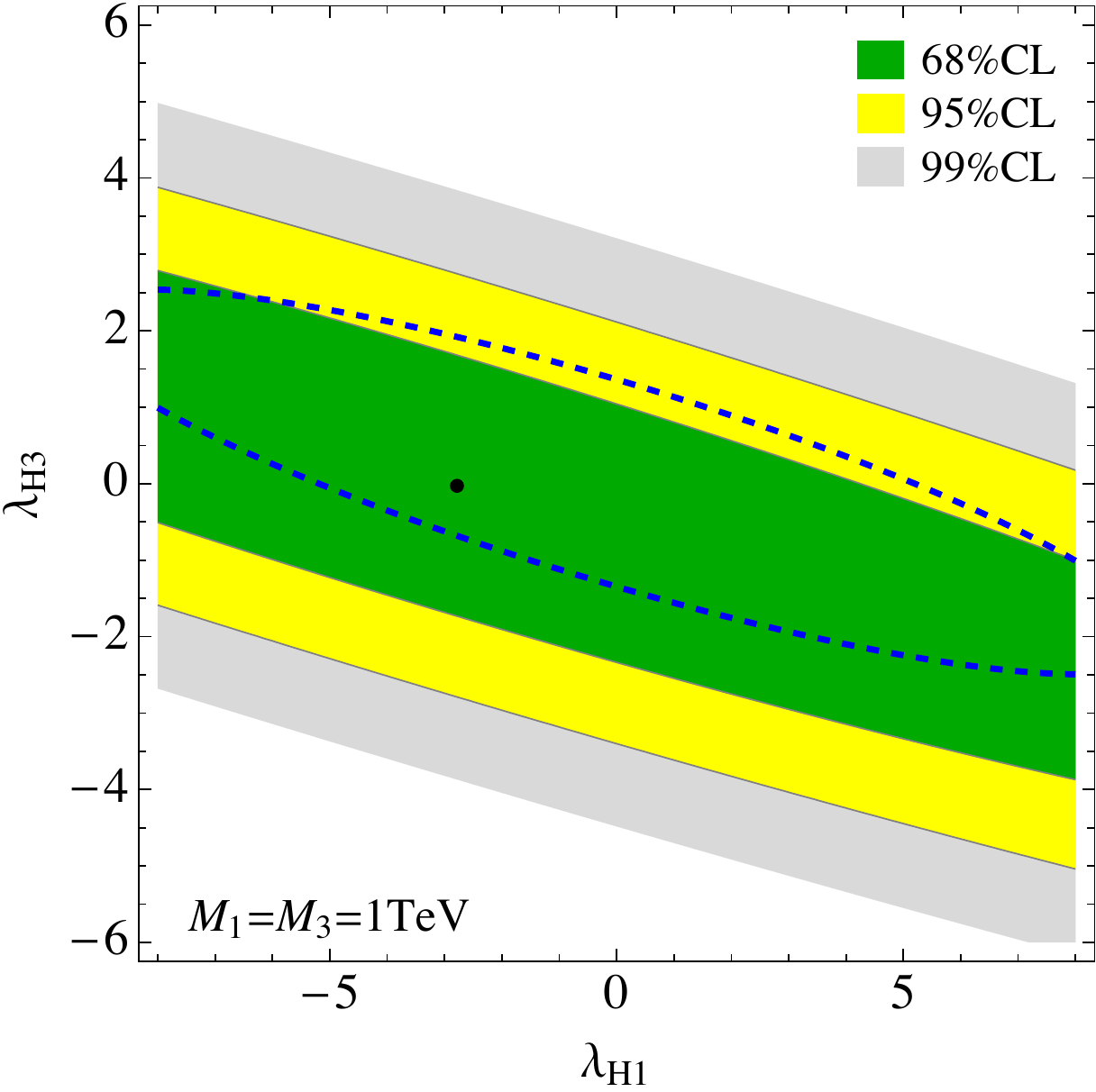}~
\includegraphics[width=0.45\hsize]{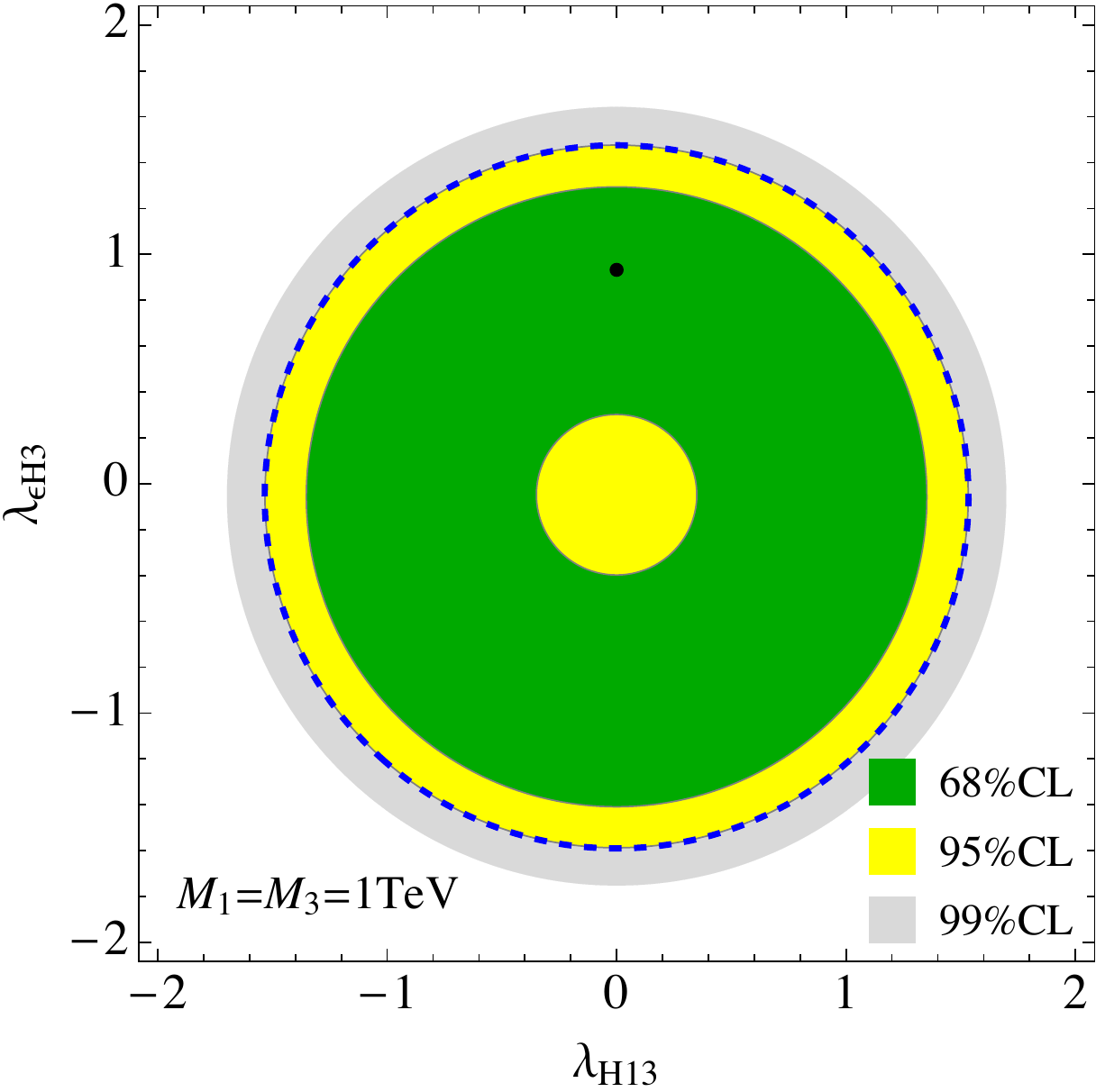} 
\caption{\small Limits on LQ potential couplings from oblique corrections and Higgs measurements. In each panel, the other two couplings have been marginalised. The black point represents the best-fit point while the dashed blue contours are the prospects for 95\%CL limits after HL-LHC.}\label{fig:potential}
\end{figure}

The LQ couplings to the Higgs also generate at one-loop contributions to $h g g$, $h \gamma\gamma$, and $h Z \gamma$ couplings. Since these are also loop-generated in the SM, the percent-level precision presently available for the Higgs couplings to photons and gluons couplings allows to probe heavy new physics. Loop contributions to other couplings, which arise at tree-level in the SM, are instead too small to have a sizeable impact. We thus consider the combined fit of Higgs couplings in the $\kappa$-framework where only $\kappa_\gamma$ and $\kappa_g$ are left free, and a constraint on $\sigma/\sigma_{\SM}(Z\gamma) = \kappa_{g}^2 \kappa_{Z\gamma}^2$, which is however still not precisely measured, see Table~\ref{tab:obsBosonic}. The approximate contributions to these parameters in our model are given by (details in App.~\ref{app:EWPO})
\be\begin{split}
	\kappa_{g} - 1 &= - (3.51 \lambda_{H3} + 1.17 \lambda_{H1} ) \times 10^{-2}  / m^2 ~, \\
	\kappa_{\gamma} - 1 &= -(2.32 \lambda_{H3} + 0.66 \lambda_{\epsilon H3}  - 0.11 \lambda_{H1} ) \times 10^{-2}  / m^2 ~, \\
	\kappa_{Z\gamma} - 1 &= -(1.89 \lambda_{H3} + 0.23 \lambda_{\epsilon H3}  - 0.033 \lambda_{H1} ) \times 10^{-2} / m^2 ~.
	\label{eq:HCouplNum}
\end{split}\ee

Analogously to what presented above for flavour observables, we combine Higgs couplings and oblique constraints in a global likelihood. From this we find the maximum likelihood point and construct the 68, 95, and 99\% CL regions in planes of two couplings, where the other two are marginalised. The results in the ($\lambda_{H1}$,$\lambda_{H3}$) and ($\lambda_{H13}$,$\lambda_{\epsilon H3}$) planes are shown in Fig.~\ref{fig:potential} for $M_1 = M_3 = 1 \TeV$.
We observe that a limit of about $1.5$ can be put on both $\lambda_{H13}$ and $\lambda_{\epsilon H3}$ (right panel). This comes mainly from the contribution to the $\hat{T}$ parameter, Eq.~\eqref{eq:STnumappr}, which is quadratic in the two couplings and thus allows to constrain both at the same time. The $\lambda_{H1}$ and $\lambda_{H3}$ couplings, instead, are constrained mainly from their contribution to the $h\gamma\gamma$ and $h g g$ couplings, Eq.~\eqref{eq:HCouplNum}. We see that with present experimental accuracy the limits are still rather weak, and there is an approximate flat direction which doesn't allow to put any relevant bound on $\lambda_{H1}$. 

This situation will marginally improve with the more precise Higgs measurements from HL-LHC \cite{Cepeda:2019klc}. The future expected 95\%CL contours are shown as dashed blue lines. This however has no appreciable effect on the limits shown in the right panel, since those are dominated by the constraint on the $T$ parameter, which will instead improve substantially from measurements on the $Z$ pole at FCC-ee. A more detailed analysis of FCC prospects are however beyond the scope of this paper.

%----------------------------------------------------
\subsection{Comparing with literature}

In recent months the $S_1 + S_3$ model at one-loop accuracy has been studied in Refs.~\cite{Crivellin:2019dwb,Saad:2020ihm} for what regards the flavour anomalies, while Ref.~\cite{Crivellin:2020ukd} studied electroweak and Higgs limits on the leptoquark-Higgs couplings. Given the similarity of the goals with out work, we discuss in this Section the main differences.
The most important lies in the approach used to calculate radiative leptoquark contributions to observables. While previous works employed direct computations of leptoquark loop contributions to the desired low-energy amplitudes, in this work we use an EFT approach, whereby the only model-dependent part of the computation is the one-loop matching to the SMEFT. As argued in the introduction, we believe such an approach has several advantages, the most important being the automatic inclusion of all new physics effects to all observables at leading order in $1 / M_{\rm LQ}^2$ expansion and to one-loop accuracy: there is indeed no need to simplify the computation neglecting given terms or couplings, for example all electroweak corrections are included automatically in our computation.\footnote{While in the approximate semi-analytical expressions we might neglect some sub-leading terms, in order to simplify the presentation, all contributions are kept in the numerical analysis.}

This approach thus allow us to study a larger set of observables than the ones considered in previous studies. One example is $D^0 - \bar{D}^0$ mixing, which was not considered in \cite{Crivellin:2019dwb,Saad:2020ihm}. Considering the benchmark points selected by both works, we find that all of them are excluded (by a large amount) by $D$-meson mixing, due to large $\lambda^{1(3)L}_{23}$ couplings. In the benchmarks of \cite{Saad:2020ihm} we also find a large tension in $\tau \to \mu \gamma$, while we are in agreement with the analytic expressions for the decay.
We also find some tension in the leptonic decay $D_s \to \tau \nu$ for several benchmarks points of both studies.

We should also point out that the Higgs-$S_3$ interaction proportional to $\lambda_{\epsilon H 3}$, c.f. Eq.~\eqref{eq:S1S3Model}, was not included in the analysis of Ref.~\cite{Crivellin:2020ukd}.

%%%%%%%%%%%%%%%%%%%%%%%%%%%%%%%%%
\section{Prospects}
\label{sec:prospects}

In this Section, we discuss the implications of future Belle II measurements of
\begin{itemize}
	\item LFV $B$ decays induced at parton level by $b \to s \tau \mu$ (see App.~\ref{app:BLFVnLFV});
	\item $B$ decays induced at parton level by $b \to s \tau \tau$ (see App.~\ref{app:BLFVnLFV}).
\end{itemize}
These processes, in fact, are particularly interesting for leptoquark scenarios aiming at addressing both neutral and charged-current $B$-anomalies. Both are induced at tree-level by $S_3$ and, by $\SU(2)_L$ relations, the $b \to c \tau \nu_\tau$ transition, tree-level in the SM, is related to the FCNC transition $b \to s \tau \tau$. Also, LFV is a natural consequence of leptoquark couplings once also the coupling to muons is considered, as required by neutral-current anomalies.
While the LFV $B$-meson decays are already included in the global fits described in the previous Sections, the current bounds on  $b \to s \tau \tau$ observables are of the order of $\sim 10^{-3}$ and thus too weak to set constraints on the model parameters. However, Belle II, with $50$ab$^{-1}$ of luminosity, will strongly improve the sensitivity, in particular for the branching fraction of the semileptonic decays. On the other hand, the Upgrade II of LHCb will set competitive bounds on the leptonic decay $B_s \to \tau \tau$.
 The relevant future expected limits at $95\%$ C.L. for Belle II \cite{Kou:2018nap} and LHCb \cite{Bediaga:2018lhg}  are summarised in Table~\ref{tab:BelleIIprosp}.
 
 \begin{table}[t]
\begin{center}
\begin{tabular}{ | c | c | c | c|}
\hline
\textbf{Observable} & \textbf{Present limit} & \textbf{Belle II} $(5) 50$ab$^{-1}$ & \textbf{LHCb} Up.-II\\
\hline
\hline
\cellcolor[gray]{0.92} $b\to s \tau \mu$ observables & \cellcolor[gray]{0.92} & \cellcolor[gray]{0.92} & \cellcolor[gray]{0.92} \\
\hline
Br$(B^+ \to K^+ \tau^\pm \mu^\mp)$ & $< 3.3 (5.4) \times 10^{-5}$ \cite{Lees:2012zz,Aaij:2020mqb} &  $3.9 \times 10^{-6}$ & $\mathcal{O}(10^{-6})$ \\
\hline
Br$(B_s \to \tau^\pm \mu^\mp)$ & $<4.2 \times 10^{-5}$ \cite{Aaij:2019okb}&  $\sim 4 \times 10^{-6}$ & $\sim 1 \times 10^{-5}$ \\
\hline   
\hline
\cellcolor[gray]{0.92} $b\to s \tau \tau$ observables & \cellcolor[gray]{0.92} & \cellcolor[gray]{0.92} & \cellcolor[gray]{0.92} \\
\hline
Br$(B^+ \to K^+ \tau^+ \tau^-)$ & $< 2.8 \times 10^{-3}$ \cite{TheBaBar:2016xwe} & $(7.7) \, 2.4 \times 10^{-5}$ & - \\
\hline
Br$(B_s \to  \tau^+ \tau^-)$ & $< 6.8 \times 10^{-3}$ \cite{Aaij:2017xqt} & $(9.7) 3 \times 10^{-4}$ & $5 \times 10^{-4}$ \\
\hline
 \end{tabular}
\caption{Future Belle II and LHCb sensitivities, at $95\%$ C.L., for $b \to s \tau \mu$ and $b \to s \tau \tau$ observables. \label{tab:BelleIIprosp}}
\end{center} 
\end{table}

\begin{figure}[t]
\centering
\includegraphics[height=7cm]{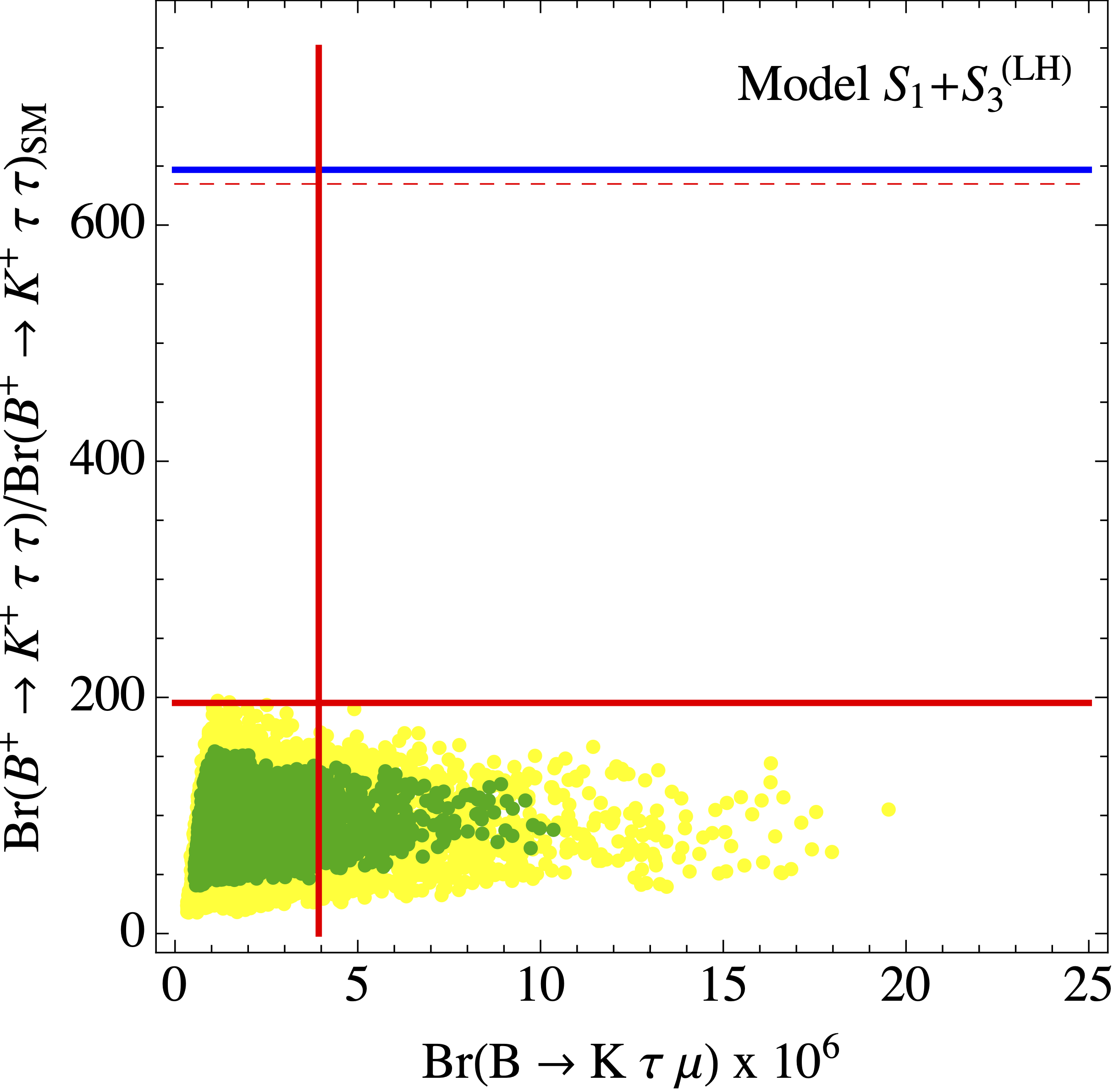}\quad
\includegraphics[height=7cm]{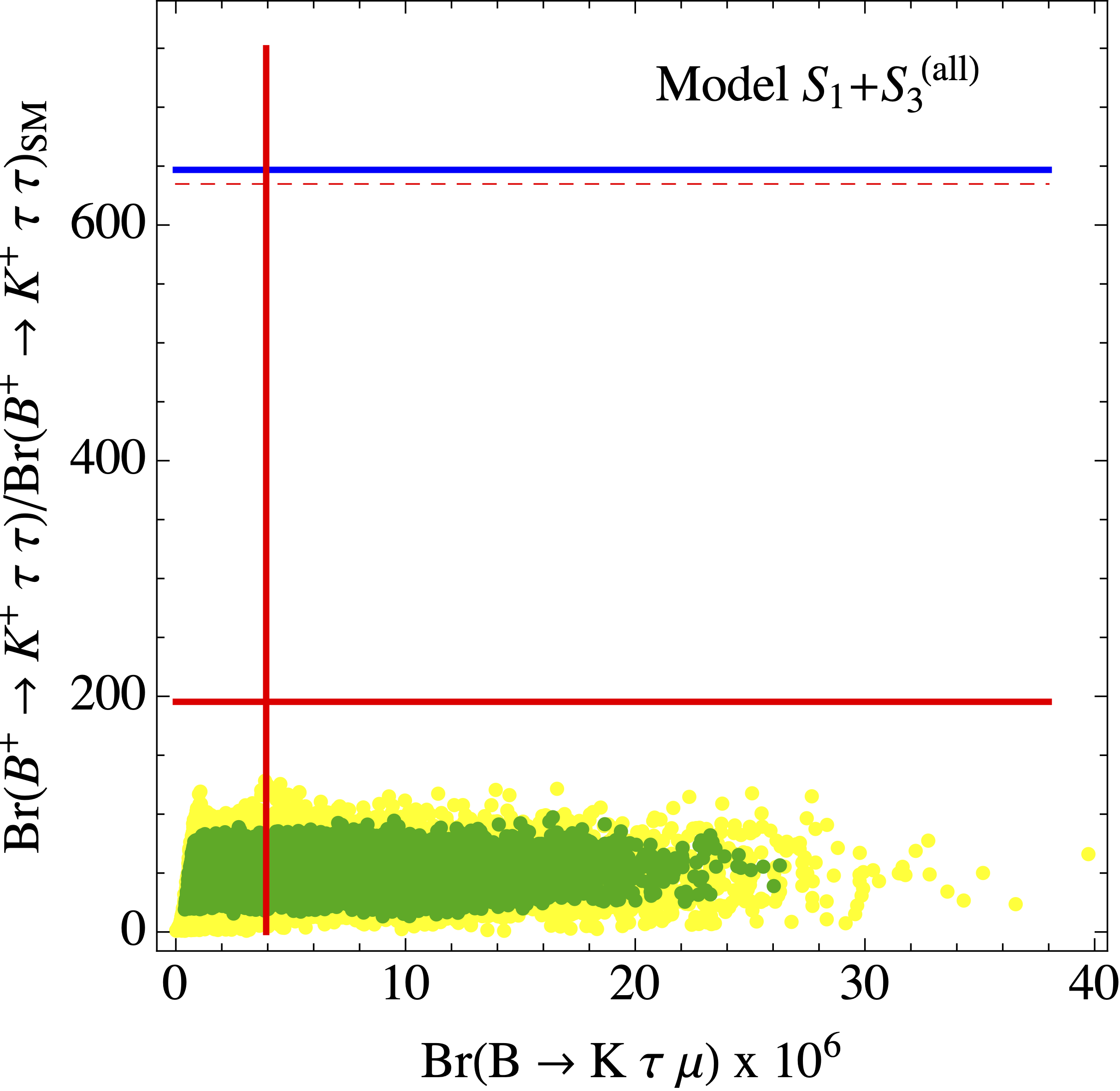}\\[10pt]
\caption{\small Results from the fit in the $S_1 + S_3^{\, \rm (LH)}$ model (left panel) and in the $S_1 + S_3^{\, \rm (all)}$ (right panel). The red solid (dashed) lines correspond to the $50$ab$^{-1}$ ($5$ab$^{-1}$) Belle II future bounds, at 95\% C.L. . The blue solid line is the prospected bound for the LHCb Upgrade II, on Br$(B_s\to  \tau \tau)/$Br$(B_s\to  \tau \tau)_{\rm SM}$.}\label{fig:BelleIIprosp}
\end{figure}

In Fig.~\ref{fig:BelleIIprosp} we show how the preferred parameter-space regions for the models $S_1 + S_3^{\, (LH)}$ (left) and $S_1 + S_3^{\,(all)}$ (right) map in the plane of the branching fractions of the LFV decay $B^+ \to K^+ \tau \mu$ and the decay $B^+ \to K^+ \tau \tau$ (normalised to the SM value).\footnote{It should be noted that at tree-level in our model this ratio is the same for all decays involving the $b \to s \tau \tau$ transition, e.g. $B_s \to \tau \tau$ (see App.~\ref{app:BLFVnLFV}).}
The red horizontal lines correspond to the Belle II future bounds at $95\%$ C.L. on Br$(B^+ \to K^+ \tau^+ \tau^-)$ at $5$ab$^{-1}$ (dashed lines) and $50$ab$^{-1}$ (solid lines), while the vertical ones represent the Belle II $50$ab$^{-1}$ prospect for Br$(B^+ \to K^+ \tau \mu)$. One can see that, in both scenarios, the predictions for both the non-LFV and LFV semileptonic $B$ decay into $\tau$ are in the ballpark of the future Belle II sensitivity at $50$ab$^{-1}$, while the expected bounds at $5$ab$^{-1}$ are still too weak to set significant constraints on the models. Furthermore, one can notice that the future measurements of $b\to s \tau \tau$ observables are constraining more strongly the parameter space of the $S_1 + S_3~^{\rm (LH)}$ model than the one of the $S_1 + S_3~^{\rm (all)}$ model. For the leptonic decay $B_s \to  \tau^+ \tau^-$ at Belle II only the prospect at luminosity of $5$ab$^{-1}$ is available; it is not shown in the plots since it is weaker with respect to the semileptonic decays and correspond to a horizontal line at $\sim 1250$. On the other hand, for the Upgrade II of LHCb, the prospected bound on Br$(B_s \to  \tau^+ \tau^-)$ (blue horizontal lines) is stronger and leads to constraints similar to the ones that we obtain from the $B^+ \to K^+ \tau^+\tau^-$ decay measured at $5$ab$^{-1}$ Belle II. In order to evaluate the constraining power of future Br$(B_s \to \tau \mu)$ measurements, in Fig.~\ref{fig:BelleIIprosp}, one could keep in mind that in our model we have Br$(B_s \to \tau \mu)/$Br$(B^+\to K^+ \tau\mu) \approx 0.89$, at tree-level.

%%%%%%%%%%%%%%%%%%%%%%%%%%%%%%%%%
\section{Conclusions}
\label{sec:conclusions}

\begin{figure}[t]
\centering
\includegraphics[height=13cm]{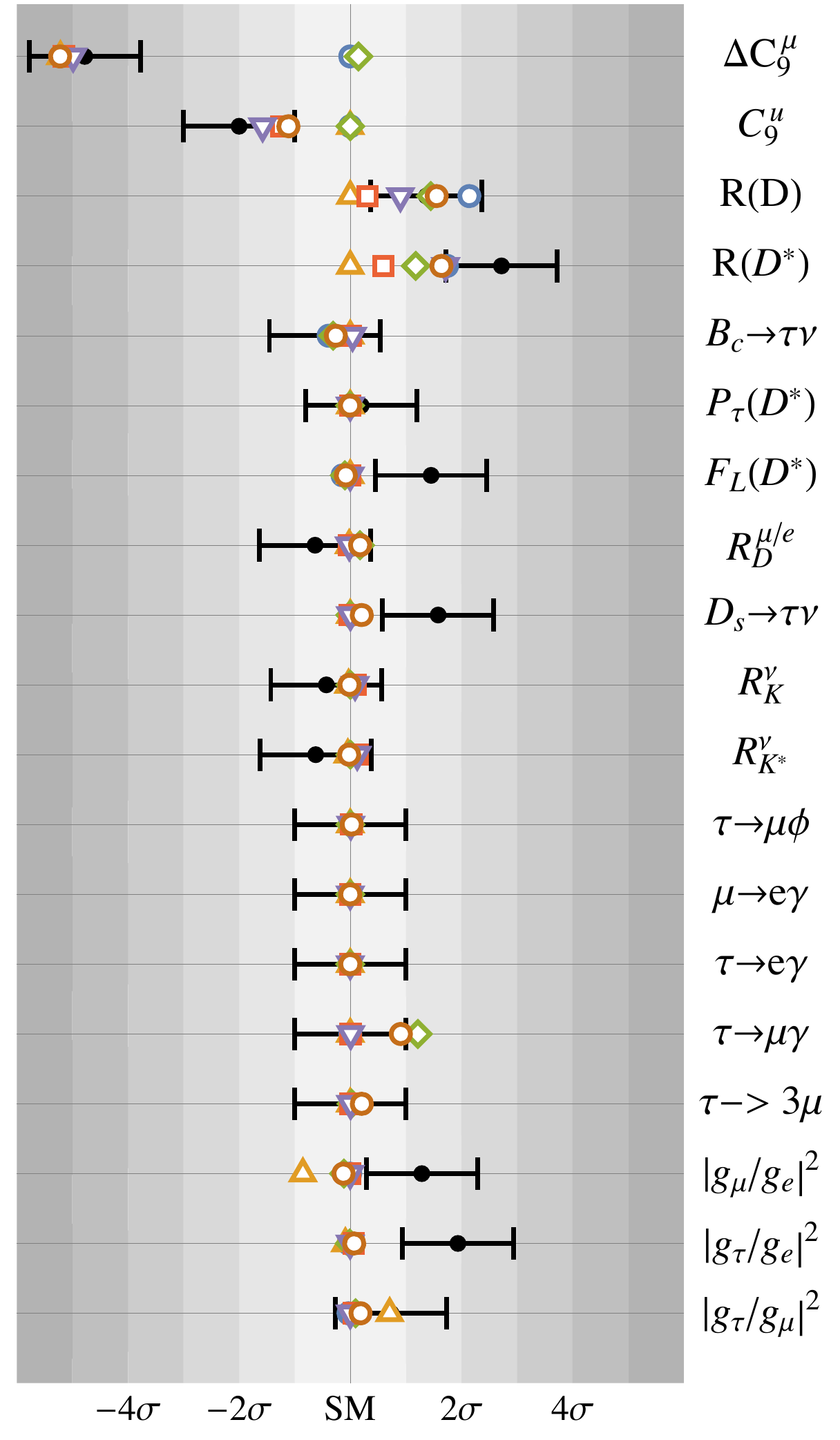}\quad
\includegraphics[height=13cm]{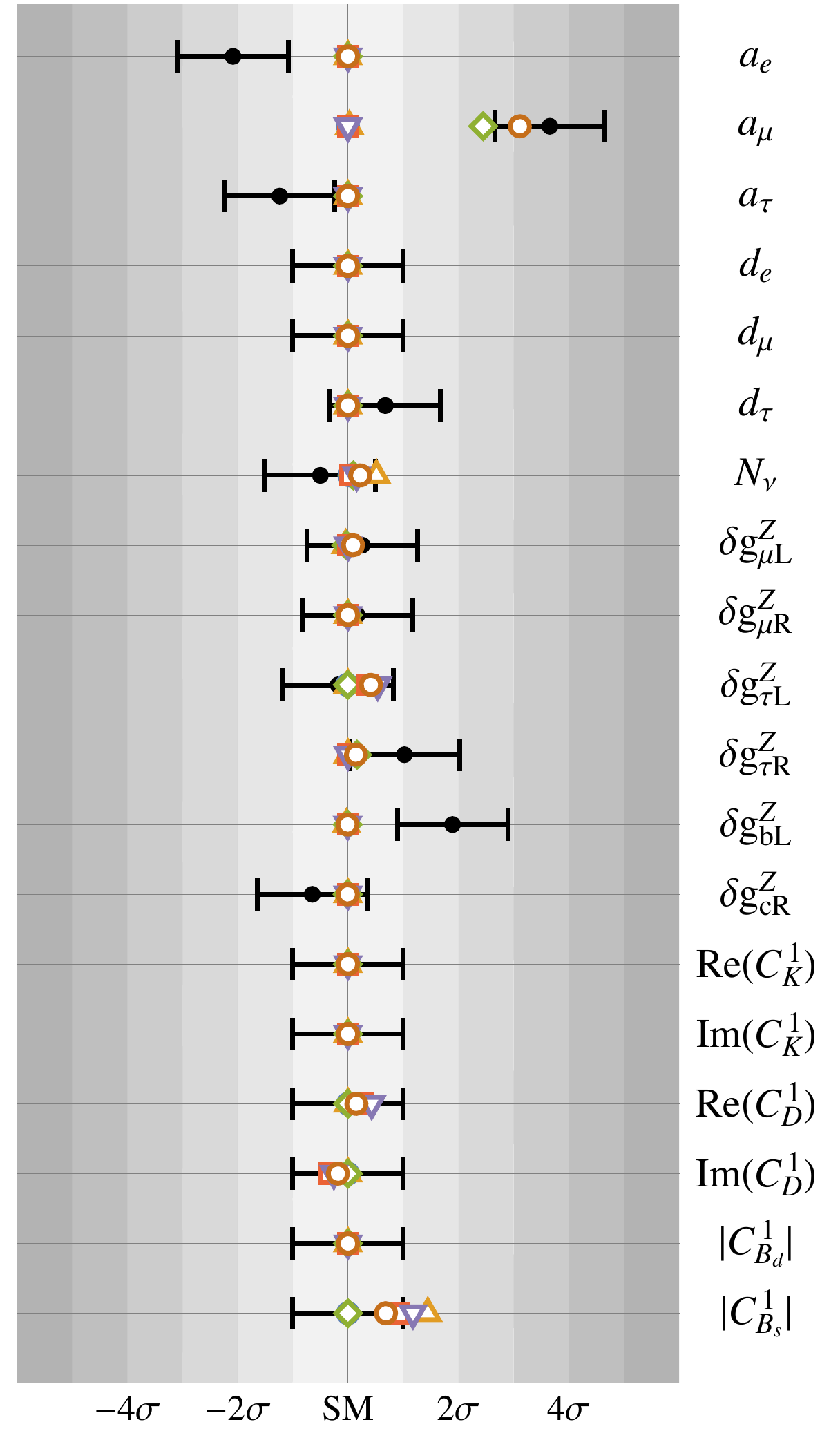}\\[10pt]
\includegraphics[width=8cm]{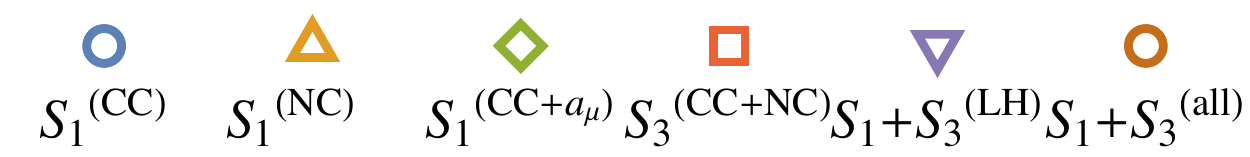}
\caption{\small For the best-fit points in each model studied in this work, we show the relative deviations from the Standard Model prediction in all observables, in terms of number of sigmas given by the experimental precision in that observable. The black intervals represent the experimental measurements}\label{fig:summary}
\end{figure}

In this work we examined in detail and at one-loop accuracy the phenomenology of Standard Model
extensions involving the two leptoquarks $S_{1}$ and $S_{3}$, motivated
by the experimental discrepancies observed in $B$-meson decays and
in the muon anomalous magnetic moment $(g-2)_{\mu}$.

To this aim, we performed global fits for several benchmark models to a comprehensive list of
flavor and electroweak precision observables, each computed at one-loop
accuracy, leveraging on our previous work \cite{Gherardi:2020det}. For each
model, we identify best-fit regions and major sources of tension,
when present, and also provide prospects for $B$-decays to $\tau \tau$ and $\tau \mu$, in the experimental scope of Belle II and LHCb.

It is found that models involving only the $S_1$ leptoquark can consistently address $R(D^{(*)})$ and $(g-2)_\mu$ anomalies, while a fully-satisfactory solution for $b\to s\mu\mu$ anomalies is prevented by the combination of constraints from $B_s$-mixing and LFU in $\tau$ decays.
Conversely, the $S_3$ leptoquark when taken alone can only address neutral-current $B$-meson anomalies.
A model with both $S_{1}$ and $S_{3}$, and only left-handed couplings for $S_{1}$, can address both $B$-anomalies but not the muon magnetic moment.
Finally, allowing for right handed $S_{1}$ couplings makes it possible
to fit also $(g-2)_{\mu}$.
Concerning the prospects for both the LF conserving branching fraction $\text{Br}(B\to K\tau\tau)$ and the LFV one $\text{Br}(B\to K\tau\mu)$, they are found to be in the ballpark of the future expected sensitivity of Belle-II and LHCb.

A quick glance summary of the various models is provided by Fig.~\ref{fig:summary} where we show, for the best-fit point of each model, the deviations from the SM prediction of each of the most relevant observables studied in the global fit. The black dots and intervals represent the experimentally preferred values and uncertainties, and for each observable we normalize the $x$-axis to the corresponding uncertainty (i.e. we count the number of standard deviations).
Detailed informations for each model can be found in Subsections~\ref{sec:S1}-\ref{sec:S1S3amu}.
A separate analysis is provided in Sec.~\ref{sec:S1S3Higgs} for Higgs physics observables and electroweak oblique corrections, which put constraints on leptoquark-Higgs couplings.

To conclude, we find that the combination of $S_{1}$ and $S_{3}$
provides a good combined explanation of several experimental anomalies:
charged and neutral-current $B$-meson anomalies as well as the muon magnetic moment. Their mass is necessarily close to the $ 1\TeV$ scale, particularly to address charged-current anomalies $R(D^{(*)})$, and is thus in the region that could still show some signals at HL-LHC, if they are light enough, but that will definitely be tested at future hadron colliders.

In the next few years, several experiments are expected to provide concluding answers as to the nature of all these puzzles.
While at this time it is still very possible that some, or all, of these will turn out to be only statistical fluctuations and will be shown to be compatible with SM predictions, the possibility that even only one will instead be confirmed is real.
Such an event would have profound and revolutionary implications for our understanding of Nature at the smallest scales. The scalar leptoquarks considered here are very good candidates for combined explanations and could thus be the heralds of a new physics sector lying at the TeV scale.

%%%%%%%%%%%%%%%%%
\subsection*{Acknowledgements}
%%%%%%%%%%%%%%%%%

We thank Francesco Saturnino for email exchanges, which led us to improve the analysis. DM acknowledges support by the INFN grant SESAMO, MIUR grant PRIN\_2017L5W2PT, as well as partial support by the European Research Council (ERC) under the European Union’s Horizon 2020 research and innovation programme, grant agreement 833280 (FLAY). EV has been partially supported by the DFG Cluster of Excellence 2094 ORIGINS, the Collaborative Research Center SFB1258 and the BMBF grant 05H18WOCA1 and thanks the Munich Institute for Astro- and Particle Physics (MIAPP) for hospitality.

%%%%%%%%%%%%%%%%%%%%%%%%%%%%%%%%%%%%%
%%%%%%%%%%%%%%%%%%%%%%%%%%%%%%%%%%%%%
%%%%%%%%%%%%%%%%%%%%%%%%%%%%%%%%%%%%%

\appendix 

%%%%%%%%%%%%%%%%%%%%%%%%%%%%%%%%%
%%%%%%%%%%%%%%%%%%%%%%%%%%%%%%%%%

%%%%%%%%%%%%%%%%%%%%%%%%%%%%%%%%%
\section{Analysis of observables and pseudo-observables}
\label{app:obs}

We collect in this Appendix the details for all observables considered in this study. Whenever feasible and relevant, we emphasize the model independent steps of the calculations.
In the derivations below, we employ several results available from the literature. Of particular relevance to us are the following references:
\begin{enumerate}

	\item The complete one-loop matching equations of the $S_1 + S_3$ model onto the SMEFT \cite{Gherardi:2020det}.
	\item The SMEFT one-loop renormalization group equations \cite{Jenkins:2013zja,Jenkins:2013wua,Alonso:2013hga}.
	\item The one-loop matching equations of SMEFT onto LEFT \cite{Dekens:2019ept}.
	\item The LEFT one-loop renormalization group equations \cite{Jenkins:2017dyc}.
\end{enumerate}	

We consider up to dimension-six operators in the effective theories. Operator bases are taken from Refs. \cite{Grzadkowski:2010es} and \cite{Jenkins:2017jig}, for SMEFT and LEFT respectively, and reported in Appendix~\ref{sec:op_bases} for ease of reference.

Some remarks about notation and conventions: 
\begin{itemize}

\item The sign convention for gauge couplings follows from Eq. \eqref{eq:g_convention}. This is the opposite convention to the one employed in our previous work \cite{Gherardi:2020det}, but it agrees with all other references mentioned here. We make this sign change for ease of comparison with these references.
\item SM Yukawa couplings are defined by:
\be
	\LL_{\rm Yuk} = -(y_E)_{\alpha\beta} \bar \ell_\alpha e_\beta H -(y_U)_{ij} \bar q_i u_j \widetilde H - (y_D)_{ij} \bar q_i d_j  H + \text{h.c.},
\ee
where $\widetilde H = i \sigma_2 H^*$. This is the same convention used in Ref. \cite{Grzadkowski:2010es} and in our work \cite{Gherardi:2020det}. All other references mentioned in this section use instead the notation: 
\begin{NoHyper} 
$y_{\text{\cite{Jenkins:2013zja,Jenkins:2013wua,Alonso:2013hga,Jenkins:2017dyc,Jenkins:2017jig,Dekens:2019ept}}} = y_{\text{this work}}^\dagger.$
\end{NoHyper}
\item Any other notation/convention agrees with those employed in our work \cite{Gherardi:2020det}.
\end{itemize}

Finally, we give, for convenience, our numerical expressions in terms of:
\be
m_{1,3} \equiv M_{1,3} / (1 \TeV).
\ee

%%----------------------------------------------------------------------------------

\subsection{Observables for $b\to s \ell\ell$}
\label{app:bsll}

One of the goals of the present analysis is to provide an explanation for the hints of non LFU in the neutral-current semileptonic decay of $B$-meson into $K^{(*)}$. We study here the observables related to $b\to s \ell^+ \ell^-$, focusing on those for which it is possible to obtain a robust SM prediction, i.e. the LFU ratios $R_K$ and $R_{K^*}$ in several $q^2$ bins, as well as the leptonic decay $B_s \to \mu^+ \mu^-$. Their SM predictions and experimental measurements are reported in the first five lines of Table~\ref{tab:obs}.

The standard notation for the effective operators relevant to these processes is
\be
	\mathcal{H}_{\rm eff} \supset - \frac{4 G_F}{\sqrt{2}} \frac{\alpha}{4 \pi} V_{ts}^* V_{tb} \sum_i \mathcal{C}_i \OO_i~,
	\label{eq:LeffDF1}
\ee
where the $\mathcal{C}_i$ are evaluated at $m_b$ scale and the operators $\OO_i$ are defined as
\be\begin{array}{ll}
	\OO_9^{(\prime) \ell\ell } = (\bar s \gamma_\alpha P_{L (R)} b) (\bar \ell \gamma^\alpha \ell)~, \quad &
	\OO_{10}^{(\prime) \ell\ell } = (\bar s \gamma_\alpha P_{L (R)} b) (\bar \ell \gamma^\alpha \gamma_5 \ell)~, \\
	\OO_S^{(\prime) \ell\ell } = m_b (\bar s  P_{R (L)} b) (\bar \ell \ell)~, \quad &
	\OO_P^{(\prime) \ell\ell } = m_b (\bar s  P_{R (L)} b) (\bar \ell \gamma_5 \ell)~, \\
	\OO_7^{(\prime)} = \frac{m_b}{e} (\bar s  \sigma_{\alpha\beta} P_{R (L)} b) F^{\alpha\beta}~.
	\label{eq:Opbsll}
\end{array}\ee
The expressions for the Wilson coefficients of the above operators in terms of the LEFT ones are 
\be\begin{array}{l l}
	\mathcal{C}_{9(10)}^{sb\ell\ell} = \frac{\NN_{sb}^{-1}}{2} \left( [L_{de}^{V,LR}]_{sb\ell\ell} \pm [L_{ed}^{V,LL}]_{\ell\ell sb} \right), ~ &
	\mathcal{C}_{9(10)}^{\prime sb\ell\ell} = \frac{\NN_{sb}^{-1}}{2} \left( [L_{ed}^{V,RR}]_{\ell\ell sb} \pm [L_{ed}^{V,LR}]_{\ell\ell sb} \right), \\
	\mathcal{C}_{S(P)}^{sb\ell\ell} = \frac{\NN_{sb}^{-1}}{2 m_b} \left( [L_{ed}^{S,RR}]_{\ell\ell sb} \pm [L_{ed}^{S,RL}]^*_{\ell\ell sb} \right),  ~ &
	\mathcal{C}_{S(P)}^{\prime sb\ell\ell} = \frac{\NN_{sb}^{-1}}{2 m_b} \left( [L_{ed}^{S,RL}]_{\ell\ell sb} \pm [L_{ed}^{S,RR}]^*_{\ell\ell sb} \right), \\
	\mathcal{C}_7 = \NN_{sb}^{-1} \frac{e}{m_b} [L_{d\gamma}]_{sb}~, ~ &
	\mathcal{C}_7^\prime = \NN_{sb}^{-1} \frac{e}{m_b} [L_{d\gamma}]^*_{bs}~,
	\label{eq:C910bsll}
\end{array}\ee
where $\NN_{sb} = \frac{4 G_F}{\sqrt{2}} \frac{\alpha}{4 \pi} V_{ts}^* V_{tb}$.

Experimental measurements of the observables taken into account set constraints on the $\mathcal{C}_{9(10)}^{sb\ell\ell} $ coefficients of $\OO_{9(10)}^{(\prime) \ell\ell }$. The only tree-level contributions come from $[\OO_{ed}^{V,LL}]_{\ell\ell sb}$. One-loop corrections might be important in a precision analysis, in particular in regions of the parameter space in which the tree-level term turns out to be small. The leading contributions are the ones proportional to the squared top Yukawa and to the large logs associated to the electromagnetic RG in the LEFT, but there might be relevant effects also from terms in which the product of four LQ couplings enters. Therefore, an approximated expression for these semileptonic Wilson coefficients, at $m_b$ scale, in terms of the model parameters is
\begin{align}
	[L_{ed}^{V,LL}]_{\ell\ell sb}(m_b) \approx& ~ \frac{\lambda^{3L*}_{s \ell} \lambda^{3L}_{b\ell}}{M_3^2} % \\
	- \frac{ (\laLdag{1}\laL{1})_{\ell\ell} (\laLst{1}\laLT{1})_{sb} }{64 \pi^2 M_{1}^{2}} \nonumber \\
	& + \frac{y_t^2}{32\pi^2} 2 \eta_{tt}^{s i} \eta_{tt}^{j b} \frac{\lambda^{1L *}_{i \ell} \lambda^{1L}_{j \ell}}{2 M_1^2} 
		 + \frac{\alpha}{6\pi}  \frac{ \lambda^{3L*}_{s \tau} \lambda^{3L}_{b \tau}}{M_3^2}  \log \frac{m_{\tau}^2}{M_3^2} \, ,\nonumber\\	 
	[L_{de}^{V,LR}]_{sb\ell\ell}(m_b) \approx& ~ - \frac{ (\laLst{1} \laLT{1})_{sb} (\laRdag \laR)_{\ell\ell} }{64 \pi^2 M_{1}^{2}} + \frac{y_t^2}{64\pi^2} \eta^{sb}_{tt}  \frac{\lambda_{t \ell}^{1R\,*} \lambda^{1R}_{t \ell}}{M_1^2} \left( \log \frac{M_1^2}{m_t^2} - \frac{3}{2} \right) + \nonumber \\
	& + \frac{\alpha}{6\pi} \frac{\lambda^{3L*}_{s \tau} \lambda^{3L}_{b \tau}}{M_3^2} \log \frac{m_\tau^2}{M_3^2} \, ,\nonumber\\
	  [L_{X}]_{sb\ell\ell}(m_b) \approx & 0 \quad \text{for all the other }L_{X} \, ,
	  \label{eq:Lbsll}
\end{align}
where $\eta^{ij}_{kl} \equiv V_{ki}^* V_{lj}$ and for $S_3$ we kept only the tree-level contribution and the electromagnetic universal and vector-like loop correction.

Therefore, the coefficients in Eq.~(\ref{eq:C910bsll}) are all vanishing at first approximation, apart from $\mathcal{C}_{9(10)}^{sb\ell\ell}$ which can be obtained, using Eq.~(\ref{eq:Lbsll}), as
\be
	\mathcal{C}_{9(10)}^{sb\ell\ell} = \frac{\NN_{sb}^{-1}}{2} \left( [L_{de}^{V,LR}]_{sb\ell\ell}(m_b) \pm [L_{ed}^{V,LL}]_{\ell\ell sb}(m_b) \right)~.
\ee

The relevant observables for these operators are the LFU ratios $R_K$ \cite{Aaij:2014ora,Aaij:2019wad,Abdesselam:2019lab} and $R_{K^*}$ \cite{Aaij:2017vbb,Abdesselam:2019wac}, the branching ratio of $B_s \to \mu^+ \mu^-$ \cite{Aaij:2017vad,Aaboud:2018mst,Sirunyan:2019xdu}, the angular observable $P_5^\prime$ in $B \to K^* \mu^+ \mu^-$ \cite{Aaij:2020nrf}, and branching ratios in other $b \to s \mu\mu$ transitions \cite{Aaij:2014pli,Aaij:2015nea,Aaij:2015esa,Aaij:2016flj}.
In our analysis, we use the results from the global fit of $b\to s \ell \ell$ observables done in~\cite{Aebischer:2019mlg}. In particular, with negligible $S_1$ couplings to right-handed muon $\lambda^{1R}_{i\mu} \approx 0$, the fit relevant for us is the one with a LFU-violating contribution to muons along the left-handed combination $\Delta\mathcal{C}_{9}^{sb\mu\mu} \equiv \mathcal{C}_{9}^{sb\mu\mu}- \mathcal{C}_{9}^{\text{univ}} \approx -\mathcal{C}_{10}^{sb\mu\mu} $, and a flavor-universal contribution along the vector-like direction $\mathcal{C}_{9}^{\text{univ}}$:
\be\begin{split}
\mathcal{C}_{9}^{\text{univ}}&\equiv \mathcal{C}_{9}^{sbee} = -0.48\pm 0.24, \\
\Delta\mathcal{C}_{9}^{sb\mu\mu}&\equiv \mathcal{C}_{9}^{sb\mu\mu}- \mathcal{C}_{9}^{\text{univ}}= -\mathcal{C}_{10}^{sb\mu\mu}= -0.43\pm 0.09 ,
\end{split}\ee
with a correlation of $\rho \approx -0.5$.\footnote{We use the results updated in April 2020 from P. Stangl's slides at \href{https://conference.ippp.dur.ac.uk/event/876/timetable/}{\tt https://conference.ippp.dur.ac.uk/event/876/timetable/}.}
The numerical expressions we get in our setup are, assuming real LQ couplings:
\be\begin{split}
\mathcal{C}_{9}^{\text{univ}}&\approx 0.25  \frac{\sum_\alpha \lambda^{3L}_{s\alpha}\lambda^{3L}_{b\alpha}}{|V_{ts}|  m_3^2} (1 + 0.079 \log m_3^2 ) - 0.013 \frac{\sum_\alpha \lambda^{1L}_{s\alpha}\lambda^{1L}_{b\alpha}}{m_1^2} + \ldots ,  \\
\Delta\mathcal{C}_{9}^{sb\mu\mu}&\approx - 23.7 \frac{\lambda^{3L}_{b\mu} \lambda^{3L}_{s\mu}}{|V_{ts}| m_3^2}  - 0.80 \frac{\lambda^{1L}_{b\mu} \lambda^{1L}_{s\mu} }{m_1^2} + 0.94 \frac{(\sum_i  \lambda^{1L}_{i\mu} \lambda^{1L}_{i\mu}) (\sum_\alpha \lambda^{1L}_{s\alpha} \lambda^{1L}_{b\alpha}) }{m_1^2} + \ldots .
	\label{eq:C9num}
\end{split}\ee

We report for completeness the LFU ratios $R_{K}$ and $R_{K^*}$ dependence on these Wilson coefficients:
\be\begin{split}
	R_{K^*}([0.045,1.1] \GeV^2) &\approx 0.94 + 0.11 \Re \Delta C_9^{sb\mu\mu} - 0.13 \Re \Delta C_{10}^{sb\mu\mu} + \ldots~, \\
	R_{K^*}([1.1,6] \GeV^2) &\approx 1.00 + 0.22 \Re \Delta C_9^{sb\mu\mu} - 0.25 \Re \Delta C_{10}^{sb\mu\mu}  + \ldots ~, \\
	R_{K}([1.1,6] \GeV^2) &\approx 1.00 + 0.24 \Re \Delta C_9^{sb\mu\mu} - 0.24 \Re \Delta C_{10}^{sb\mu\mu}  + \ldots ~, \\
\end{split}\ee
where the SM prediction is known up to $\mathcal{O}(1\%)$ electromagnetic corrections \cite{Bordone:2016gaq} and we neglected subleading contributions due to quadratic terms or imaginary parts of the Wilson coefficients

%%-------------------------------------------------------
\subsection{Observables for $b\to c \tau\nu$}
\label{app:bctaunu}

In this Section, we analyze the effects of the $S_{1,3}$ model on the observables related to $b\to c\ell \nu$, in order to account for the hints of non LFU in the charged-current semileptonic $B$ decays.
In the following, we will at first assume that the NP contribution to the $\tau\nu_\tau$ channel is the dominant one and all the others might be neglected. Indeed, as shown in the next Section, LFU in light leptons has been checked at the percent level, this limits the size of the possible effect in the muon channel in $R(D^{(*)})$ to be sub-leading.

The relevant four-fermion effective Hamiltonian, at the $\mu = m_b$ scale is
\be
	\mathcal{H}_{\rm eff} \supset \frac{4 G_F}{\sqrt{2}}  V_{cb} \sum_i \mathcal{C}_i \OO_i~,
	\label{eq:LeffDF1}
\ee
where the $\mathcal{C}_i$ are evaluated at $m_b$ scale and the operators $\OO_i$ are defined as
\begin{align}
&\OO_{V_L}  =(\overline{c}\gamma^{\mu}P_{L}b)(\overline{\tau}\gamma_{\mu}P_{L}\nu)\, , & \OO_{V_R}=  (\overline{c}\gamma^{\mu}P_{R}b)(\overline{\tau}\gamma_{\mu}P_{L}\nu)\, , \nonumber\\
&\OO_{S_L}  =(\overline{c}P_{L}b)(\overline{\tau}P_{L}\nu)\, , & \OO_{S_R}=  (\overline{c}P_{R}b)(\overline{\tau}P_{L}\nu)\, ,\\
& \OO_{T}  =(\overline{c}\sigma^{\mu\nu}P_{L}b)(\overline{\tau}\sigma_{\mu\nu}P_{L}\nu)\, . \nonumber
\label{eq:Opbctaunu}
\end{align}
They coincide with the $\nu_\tau\tau bc$ matrix elements of the hermitian conjugates of the LEFT $\OO_{\nu e du}$ operators. The expressions for the Wilson coefficients in terms of the LEFT ones are given by
\begin{align}
	C_{V_L} & = - \frac{1}{2 \sqrt{2} G_F V_{cb}}[L_{\nu edu}^{V,LL}(m_b)]_{\nu_\tau \tau bc}^{*}\, , &
	C_{V_R} & = - \frac{1}{2 \sqrt{2} G_F V_{cb}}[L_{\nu edu}^{V,LR}(m_b)]_{\nu_\tau \tau bc}^{*}\, , \nonumber \\
	C_{S_L} & = - \frac{1}{2 \sqrt{2} G_F V_{cb}}[L_{\nu edu}^{S,RR}(m_b)]_{\nu_\tau \tau bc}^{*}\, , & 
	C_{S_R} & = - \frac{1}{2 \sqrt{2} G_F V_{cb}}[L_{\nu edu}^{S,RL}(m_b)]_{\nu_\tau \tau bc}^{*}\, ,\\
	C_{T} & = - \frac{1}{2 \sqrt{2} G_F V_{cb}}[L_{\nu edu}^{T,RR}(m_b)]_{\nu_\tau \tau bc}^{*}\, . \nonumber
\end{align}

At tree-level, in our model, non vanishing contributions come only from $\OO^{V,LL}_{\nu e du}$, $\OO^{S,RR}_{\nu e du}$ and $\OO^{T,RR}_{\nu e du}$, which is to say $\OO_{V_L}$, $\OO_{S_L}$ and $\OO_{T}$.
Important one-loop corrections are due to QCD RG effects on the Wilson coefficients of the scalar and tensor operators, particularly below the electroweak scale  \cite{Gonzalez-Alonso:2017iyc} (while $L^{V,LL}_{\nu e du}$ is not affected by this running), and by electroweak and Yukawa RGE from the electroweak to the leptoquark scale.
In our approach we also include finite corrections arising from the one-loop matching between the LEFT and SMEFT  \cite{Dekens:2019ept}, as well as in the SMEFT to LQ matching. These effects have a comparable impact to the RGE, as we will show below. Going from LEFT coefficients at the $m_b$ scale up to SMEFT ones at $1\TeV$ we get:
\begin{align}
	[ L_{\nu edu}^{V,LL}(m_{b}) ] {}_{\nu_\tau \tau bc} &\approx - 2 V_{cb}^* [C_{H \ell}^{(3)}(1 \TeV) ]_{33} - 2 V_{ci}^* [C_{H q}^{(3)}(1 \TeV)]_{3i} + \nonumber \\
	& + 2.1 V^*_{c i} [ C_{lq}^{(3)}(1\TeV) ]_{333i}  - 0.056 V_{ci}^* [ C_{lq}^{(1)}(1\TeV) ]_{333i} + \\
	&  - 0.16 V_{cb}^* V_{t j} V_{tb}^* [ C_{lq}^{(3)}(1\TeV) ]_{33j3}  - \sum_{i \neq 3 } 0.14 V_{cb}^* V_{t j} V^*_{t i} [ C_{lq}^{(3)}(1\TeV) ]_{33ji} ,\nonumber \\
	[ L_{\nu edu}^{V,LR}(m_{b}) ] {}_{\nu_\tau \tau bc} &= - [C_{H u d}(1 \TeV)]_{23} ~,\\
	[ L_{\nu edu}^{S,RL}(m_{b}) ] {}_{\nu_\tau \tau bc} &= -1.67 V_{ci}^* [C_{l e d q}(1 \TeV)]_{333i} ~, \\
	\left( \begin{array}{c}
	{}	[ L_{\nu edu}^{S,RR}(m_{b}) ] {}_{\nu_\tau \tau bc} \\
	{}	[ L_{\nu edu}^{T,RR}(m_{b}) ] {}_{\nu_\tau \tau bc}
	\end{array}\right) 
	&\approx
	\left( \begin{array}{c c}
		1.48 & 0 \\
		0 & 0.88
	\end{array}\right)_{\rm QCD}
	\left( \begin{array}{c}
	{}	[  L_{\nu edu}^{S,RR}(m_{t}) ] {}_{\nu_\tau \tau bc} \\
	{}	[  L_{\nu edu}^{T,RR}(m_{t}) ] {}_{\nu_\tau \tau bc}
	\end{array}\right) \approx \nonumber \\
	&\approx
	\left( \begin{array}{c c}
		1.68 & - 0.22 \\
		-0.0032 & 0.85
	\end{array}\right)
	\left( \begin{array}{c}
	{}	[   C_{\ell e q u}^{(1)}(1\TeV) ] {}_{3 3 3 2} \\
	{}	[   C_{\ell e q u}^{(3)}(1\TeV) ] {}_{3 3 3 2}
	\end{array}\right)~,
\end{align}
where in the last expression we neglected small CKM and loop-suppressed terms proportional to $ \sum_{i=1,2} V_{ti} [   C_{\ell e q u}^{(1,3)}(1\TeV) ] {}_{3 3 i 2} $, which are however kept in the numerical analysis.

The tree-level matching between SMEFT and the LQ theory is given in Eq.~\eqref{eq:EFTS13treematch}, which sets the relation $[C_{\ell e q u}^{(1)}]^{(0)}_{3 3 3 2} = - 4 [C_{\ell e q u}^{(3)}]^{(0)}_{3 3 3 2}$. This gets modified by finite one-loop contributions to $[C_{\ell e q u}^{(1)}(M_{\rm LQ})]_{3 3 3 2} \approx - 3.68 [C_{\ell e q u}^{(3)}(M_{\rm LQ})]_{3 3 3 2}$, see \cite{Gherardi:2020det}.

Considering a NP scale around $\sim $TeV, an approximate expression of the Wilson coefficients at $m_b$ scale as a function of the LQ couplings, is
\begin{align}
&C_{V_L}  \approx  \frac{v^2}{2} \sum_j \left(1.09 \frac{\lambda_{b\tau}^{1L}\lambda_{j\tau}^{1L*} }{2M_{1}^{2}} - 1.02 \frac{\lambda_{b\tau}^{3L}\lambda_{j\tau}^{3L*} }{2M_{3}^{2}} \right) \frac{V_{cj}}{V_{cb}} , & C_{V_R}  \approx0, \nonumber \\
&C_{S_L}  \approx - 1.63 \frac{v^2}{2V_{cb}} \frac{\lambda_{b\tau}^{1L}\lambda_{c\tau}^{1R*}}{2M_{1}^{2}}, & C_{S_R}  \approx0, \label{eq:Cbctaunu} \\
&C_{T}  \approx 0.88 \frac{v^2}{2V_{cb}} \frac{\lambda_{b\tau}^{1L}\lambda_{c\tau}^{1R*}}{8M_{1}^{2}}. \nonumber
\end{align}
In terms of low-energy coefficients at the $m_b$ scale, we obtain the relation $C_{S_L} \approx - 7.43 C_{T}$.

The observables mostly relevant to the $b\to c \tau \nu$ transition, for which a robust SM prediction is possible, are
\be
	R(D),\quad R(D^*),\quad P_\tau,\quad F_L(D^*),\quad \text {Br}(B_c \to \tau \nu)~.
\ee
The approximate dependence on the non-vanishing EFT coefficients in Eq.~(\ref{eq:Cbctaunu}), valid up to order $\mathcal O (\alpha _s) $ and $\mathcal O (\Lambda _{\text{QCD}}/m_c)$, are \cite{Iguro:2018vqb}:
\begin{align} 
	\dfrac{R(D)}{R(D)_\text{SM}} &= 
		1 + \Re[ 2 C_{V_L} + 1.49 C_{S_L}^* + 1.14 C_T^* ] + \mathcal{O}(C^2)~, \\
	\dfrac{R(D^*)}{R(D^*)_\text{SM}} &=
		1 + \Re[ 2 C_{V_L} - 0.11 C_{S_L}^* - 5.12 C_T^* ] + \mathcal{O}(C^2)~, \\
	\dfrac{P_\tau(D)}{P_\tau(D)_\text{SM}} &=
		\left( \dfrac{R(D)}{R(D)_\text{SM}}\right)^{-1} \left(
		1 + \Re[ 2 C_{V_L} + 4.65 C_{S_L}^* - 1.18 C_T^* ] + \mathcal{O}(C^2) \right) ~, \\
	\dfrac{P_\tau(D^*)}{P_\tau(D^*)_\text{SM}} &= 
		\left( \dfrac{R(D^*)}{R(D^*)_\text{SM}}\right)^{-1} \left( 
		1 + \Re[ 2 C_{V_L} + 0.22 C_{S_L}^* - 3.37 C_T^* ] + \mathcal{O}(C^2) \right) ~, \\
	\dfrac{F_L ^{D^*}}{[F_L ^{D^*}]_\text{SM}} &=
		\left( \dfrac{R(D^*)}{R(D^*)_\text{SM}}\right)^{-1} \left(
		1 + \Re[ 2 C_{V_L} - 0.24 C_{S_L}^* - 4.37 C_T^* ] + \mathcal{O}(C^2) \right) ~, \\
	\dfrac{\text{Br}(B_c^+\to \tau ^+ \nu)}{\text{Br}(B_c^+\to \tau ^+ \nu)_\text{SM}} &=
		1 + 2 \Re[ C_{V_L} - 4.33 C_{S_L} ] + \mathcal{O}(C^2)~.
\end{align}
In the numerical analysis we use the complete expressions from \cite{Iguro:2018vqb}, including quadratic terms. The global avarage of $R(D)$ and $R(D^*)$ \cite{Lees:2012xj,Lees:2013uzd,Aaij:2015yra,Huschle:2015rga,Sato:2016svk,Hirose:2016wfn,Hirose:2017dxl,Aaij:2017uff,Aaij:2017deq,Siddi:2018avt,Belle:2019rba} is taken from \cite{Amhis:2016xyh} (Spring 2019 update), the measurement of $P_\tau(D^{*})$ from \cite{Hirose:2017dxl}, $F_L^{D^*}$ from \cite{Abdesselam:2019wbt},  and $\text{Br}(B_c^+\to \tau ^+ \nu)$ from \cite{Akeroyd:2017mhr}. They are summarised in Table~\ref{tab:obs} together with the SM predictions.

Approximate numerical expressions for $R(D)$, $R(D^*)$, and $\text{Br}(B_c^+\to \tau ^+ \nu)$ assuming real LQ couplings:
\begin{align} 
	\frac{R(D)}{R(D)_\text{SM}} &\approx 
		1 + 2 \Re[C_{V_L}] - 0.79 \frac{ \lambda^{1L}_{b\tau} \lambda^{1R}_{c\tau} }{m_1^2} (1 + 0.05 \log m_1^2) +  0.36 \frac{ (\lambda^{1L}_{b\tau} \lambda^{1R}_{c\tau})^2 }{m_1^4} (1 + 0.08 \log m_1^2) , \\
	\dfrac{R(D^*)}{R(D^*)_\text{SM}} &\approx
		1 + 2 \Re[C_{V_L}] - {0.34} \frac{ \lambda^{1L}_{b\tau} \lambda^{1R}_{c\tau} }{m_1^2} (1 + {0.02} \log m_1^2) + 0.1 \frac{ (\lambda^{1L}_{b\tau} \lambda^{1R}_{c\tau})^2 }{m_1^4} (1 + 0.01 \log m_1^2) , 
\end{align}
\be
	\dfrac{\text{Br}(B_c^+\to \tau ^+ \nu)}{\text{Br}(B_c^+\to \tau ^+ \nu)_\text{SM}} \approx
	1 + 2 \Re[C_{V_L}]  {+ 5.1} \frac{ \lambda^{1L}_{b\tau} \lambda^{1R}_{c\tau} }{m_1^2} (1 + 0.04 \log m_1^2) +  {6.5} \frac{( \lambda^{1L}_{b\tau} \lambda^{1R}_{c\tau} )^2 }{m_1^4} (1 + 0.08 \log m_1^2) ,
\ee
where
\be
	2 C_{V_L} \approx {0.77} \frac{ \lambda^{1L}_{b\tau}(V \lambda^{1L*})_{c \tau} }{V_{cs} m_1^2} - {0.72} \frac{ \lambda^{3L}_{b\tau}(V \lambda^{3L*})_{c \tau} }{V_{cs} m_3^2}~.
	\label{eq:CVL}
\ee

\medskip

%%--------------------------------------------------
\subsection{LFU in $b \to c \mu(e) \nu$}
\label{app:RDmue}

Relaxing the hypothesis that NP is coupled only to the third generation, we can have also corrections to $b\to c \nu e$ and $b\to c \nu\mu$ interactions. Lepton Flavour Universality between muons and electrons in these charged-current transitions has been checked experimentally at the percent level \cite{Aubert:2008yv,Glattauer:2015teq}. Since these processes are also used to extract the CKM element $V_{cb}$, if also new physics affects them a careful analysis should be performed in general, see e.g.~\cite{Jung:2018lfu}. 
In our case, since we assume negligible leptoquark couplings to electrons, we avoid this issue by using only the LFU ratio
\be
	R_D^{\mu/e} \equiv \frac{\Br(B \to D \mu \nu)}{\Br(B \to D e \nu)}.
\ee
to constrain new physics. Since this ratio is insensitive to $V_{cb}$, and a global fit for this SM parameter gets contributions from a large number of observables, we argue that the extraction of the CKM element wouldn't change in a sizeable way by removing this single observable.
This ratio has been measured by both Babar \cite{Aubert:2008yv} and Belle \cite{Glattauer:2015teq} collaborations:
\begin{equation*}
	\left. R_D^{\mu/e} \right|_{\rm Babar} = 0.950 \pm 0.021 \pm 0.053 ~, \quad
	\left. R_D^{\mu/e} \right|_{\rm Belle} = 0.995 \pm 0.022 \pm 0.039 ~,
\end{equation*}
where the first uncertainty is statistical and the second systematic and in deriving this for Babar we took into account the correlation between systematic uncertainty in the muon and electron channels. Combining these two results, we get
\be
	\left. R_D^{\mu/e} \right|_{\rm comb} = 0.978 \pm 0.035~.
\ee
The decay rate into light leptons, as function of the same operators as in Eq.~\eqref{eq:Opbctaunu} (but with $\ell = \mu, e$ instead of the tau) is given by \cite{Azatov:2018knx}
\be\begin{split}
	\Br(B \to D \ell \nu) \approx& 10^{-3} \left( 23.3 (1 + C^\ell_{V_{L+R}})^2  + 1.0 (1 + C^\ell_{V_{L+R}}) (C^\ell_T + 2 C^\ell_{S_{L+R}} ) + \right. \\
	& \qquad \left.  + 33.5 (C^\ell_{S_{L+R}})^2 + 3.5 (C^\ell_T)^2 \right),
\end{split}\ee
where $C_{X_{L+R}} \equiv C_{X_L} + C_{X_R}$.
Assuming negligible couplings to electrons, the leading contributions to the LFU ratio are given by
\be
	R_D^{\mu/e} = 1 + 2 \Re[C^\mu_{V_L}] - {0.047} \frac{ \lambda^{1L}_{b\mu} \lambda^{1R}_{c\mu} }{m_1^2} + {0.50} \frac{ |\lambda^{1L}_{b\mu} \lambda^{1R}_{c\mu}|^2 }{m_1^4} +  \ldots~,
\ee
where the expression for $2 C^\mu_{V_L}$ is the same as in Eq.~\eqref{eq:CVL} with $\tau \to \mu$.
%

%%--------------------------------------------------
\subsection{$D_s \to \tau \nu$}
\label{app:Ds}

In the LQ model the leading contributions to this leptonic decay arise at the tree-level. Given the limited present experimental accuracy, and the fact that the tree-level contribution is expected to be the dominant one in these modes, we limit ourselves to this. 

The branching ratio of the $D_s\to \tau \nu$ decay can be expressed as a function of the LEFT coefficients ${L}^{V,LL}_{\nu e d u} $ and ${L}^{S,RR}_{\nu e d u} $, evaluated at the charm quark mass scale,
\be \begin{split}
 \text{Br}(D_s\to \tau \nu) &= \dfrac{\tau_{D_s} f^2_{D_s} m_{D_s}}{64\pi} m_\tau^2\left(1-\frac{m_\tau^2}{m_{D_s}^2}\right)^2 \times \\
 & \qquad \times \sum_\alpha \Big| \dfrac{\delta _{\alpha 3}}{{\Lambda_{\text{SM}}^{cs}}^2} - [{L}^{V,LL}_{\nu e d u} ]_{\nu_\alpha \tau s c}  -  [{L}^{S,RR}_{\nu e d u} ]_{\nu_\alpha \tau s c}\dfrac{m_{D_s}^2}{m_\tau (\overline{m_s}+\overline{m_c})}\Big|^2 ~,
\end{split}\ee
where $1 / \Lambda_{\text{SM}}^{cs \, 2} =2\sqrt{2} G_F V_{cs}$, $f_{D_s}=0.25$GeV~\cite{Aoki:2016frl}, $m_{D_s}=1.968$GeV, $\tau_{D_s}=5\times 10^{-13} s$~\cite{Tanabashi:2018oca}, $\overline {m}_c = 1.27\,\text{GeV}$ and $\overline {m}_s = 93\,\text{MeV}$.

The tree-level matching to the LQ model, augmented by the QCD RG evolution, gives
\begin{align}
	[ {L}^{V,LL}_{\nu e d u} (m_c) ]_{\nu_\alpha \tau s c}&=\left( - \frac{\lambda^{1L*}_{s \alpha} \lambda^{1L}_{k \tau}}{2 M_1^2} +  \frac{\lambda^{3L *}_{s \alpha} \lambda^{3L}_{k\tau}}{2 M_3^2}\right)  V^*_{ ck } \, , \\ 
	[ {L}^{S,RR}_{\nu e d u}  (m_c)]_{\nu_\alpha \tau s c}&= \left(\frac{\alpha_{s}(m_{c})}{\alpha_{s}(\mu_{W})}\right)^{\frac{12}{23}} \left(\frac{\alpha_{s}(\mu_{W})}{\alpha_{s}(M)}\right)^{\frac{12}{21}} \frac{\lambda^{1R}_{c\tau} \lambda_{s \alpha}^{1L*}}{2 M_1^2} \, .
\end{align} 
The dominant contribution arises from the interference with the SM, thus from the $\tau$ channel:
\be\begin{split}
	\frac{ \Br(D_s^+\to \tau^+ \nu_\tau)}{ \Br(D_s^+\to \tau^+ \nu_\tau)_{\SM}}\approx 1 + 2 \times 10^{-2} \Re & \left[ 1.5 \frac{\lambda^{1L *}_{s\tau} \lambda^{1L }_{i\tau} V_{c i}^* / V_{c s}^*}{m_1^2} - 1.5 \frac{\lambda^{3L *}_{s\tau} \lambda^{3L }_{i\tau} V_{c i}^* / V_{c s}^*}{m_3^2} + \right. \\
	& \left. - 4.6 \frac{\lambda^{1L *}_{s\tau} \lambda^{1R }_{c\tau} }{m_1^2} \right] ~,
\end{split}\ee
where $\Br(D_s^+\to \tau^+ \nu_\tau)_{\SM} = (5.169 \pm 0.004) \times 10^{-2}$ \cite{Aoki:2016frl}.
The experimental measurement for the $D_s^+\to \tau^+ \nu_\tau$ branching ratio is~\cite{Tanabashi:2018oca}
\be
	\dfrac{\text{Br}(D_s\to \tau \nu)}{\text{Br}(D_s\to \tau \nu)_{\text{SM}}} = 1.060 \pm 0.044~.
\ee

%----------------------------------------------------------------------------------
\subsection{$B\to K^{(*)} \nu\nu$, $K^+ \to \pi^+\nu\nu$, and $K_L \to \pi^0\nu\nu$}
\label{app:bsnunu}

The relevant parton level processes $d_i \to d_j \nu_\alpha \nu_\beta$ are described by the effective four-fermion Lagrangian
\be
	\LL_{\rm eff}^{d_i d_j\nu\nu} \supset \frac{4 G_F}{\sqrt{2}} \frac{\alpha}{4 \pi} V_{ti}^* V_{tj} \sum_i C_i \OO_i~,
	\label{eq:LeffDF1bsnunu}
\ee 
with the following $\Delta F=1$ operators
\begin{align}
\OO_L^{ij\alpha\beta} &=\left(\bar{d}_i \gamma_\mu P_L d_j \right) \, \left(\bar\nu_\alpha \gamma^\mu (1-\gamma_5) \nu_\beta \right) \, ,
 \quad \OO_R^{ij\alpha\beta} = \left(\bar{d}_i \gamma_\mu P_R d_j \right) \, \left(\bar\nu_\alpha \gamma^\mu (1-\gamma_5) \nu_\beta \right) \, .
\label{eq:Opbsnunu}
\end{align}
The relations between the Wilson coefficients of the operators above and the LEFT ones are given by
\be
	[C_L]_{ij\alpha\beta} = N_{ij} \sum_{\alpha\beta}[L_{\nu d}^{V,LL}]_{\alpha\beta ij} \, , \qquad
	[C_R]_{ij\alpha\beta} = N_{ij} \sum_{\alpha\beta}[L_{\nu d}^{V,LR}]_{\alpha\beta ij} \, ,
\label{eq:Cbsnunu}
\ee
where
\be
	N_{ij} = \dfrac{1}{G_F}  \dfrac{\pi}{\sqrt{2}\alpha V_{ti}^* V_{tj}}~.
\ee
In our model, at tree-level only $C_L$ is not vanishing. While at loop level a contribution to $C_R$ is generated, it is suppressed by small down-quarks Yukawa couplings, and thus it is completely negligible.
The leading contributions to the Wilson coefficients at $m_b$ scale, in terms of the UV parameters, are
\begin{align}
	&[C_L]_{ij\alpha\beta}=&N_{ij}  \Big[ \frac{\lambda^{1L*}_{i \alpha} \lambda^{1L}_{j \beta}}{2 M_1^2} +  \frac{\lambda^{3L *}_{i \alpha} \lambda^{3L}_{j\beta}}{2 M_3^2}+\frac{1}{16 \pi^2}\frac{1}{12} \frac{m_t^2}{v^2} \Big[24 V^*_{t i} V_{tj} |V_{tb}|^2 \left(   \frac{\lambda^{3L *}_{b\alpha } \lambda^{3L}_{b\beta }}{M_3^2}\right)+ \nonumber\\
	&& -3(3+2 \log (M^2/m_t^2)) \left(\left(  \frac{\lambda^{1L*}_{i\alpha } \lambda^{1L}_{k\beta }}{2 M_1^2} +  \frac{\lambda^{3L *}_{i\alpha } \lambda^{3L}_{k\beta }}{2 M_3^2}\right)V_{t j} V^*_{t k} + \right.  \nonumber \\
	&&  \left. +\left(  \frac{\lambda^{1L*}_{k\alpha } \lambda^{1L}_{j\beta }}{2 M_1^2} +  \frac{\lambda^{3L *}_{k\alpha } \lambda^{3L}_{j\beta }}{2 M_3^2}\right)V_{t k} V^*_{t i}\right)\Big] \Big]+ \ldots , \nonumber \\
	&[C_R]_{ij\alpha \beta} \approx 0 , & \label{eq:EFTRnu}
\end{align}
where $M\sim M_1 ,\, M_3$ is the matching scale to the UV model.

For $B \to K^{(*)} \nu \nu$ it is customary to define the ratio with the (clean) SM prediction,
\be
R_K^\nu=\frac{\text{Br}(B \to K \nu \nu)}{\text{Br}_{SM}(B \to K \nu \nu)} \, , \quad R_{K^*}^\nu=\frac{\text{Br}(B \to K^* \nu \nu)}{\text{Br}_{SM}(B \to K^* \nu \nu)} \, .
\ee
The experimental limits at $95\%$ C.L. are~\cite{Grygier:2017tzo}
\be
R_K^\nu< 4.65 \, , \quad R_{K^*}^\nu<3.22 \, .
\label{eq:RKnuBound}
\ee
These ratios can be expressed in terms of 2 real parameters $\epsilon >0$ and $\eta \in [-1/2,1/2]$ \cite{Buras:2014fpa}:
\be
R_K^\nu=\sum_{\alpha \beta} \frac{1}{3}(1-2\eta_{\alpha \beta})\epsilon_{\alpha \beta}^2 \,, \quad R_{K^*}^\nu=\sum_{\alpha \beta}\frac{1}{3}(1+\kappa_\eta \eta_{\alpha \beta}) \epsilon_{\alpha \beta}^2~,
\ee
with
\be\begin{split}
	\epsilon_{\alpha \beta} &=\dfrac{\sqrt{|C_L^{\SM, sb}\delta_{\alpha \beta}+[C_L]_{sb\alpha \beta}|^2+|[C_R]_{sb\alpha \beta}|^2}}{ |C_L^{\SM, sb}|} ~, \\
	\eta_{\alpha \beta} &= -\dfrac{\text{Re}\left[ \left( C_L^{\SM, sb}\delta_{\alpha \beta} +[C_L]_{sb\alpha \beta}\right) [C_R]_{sb\alpha \beta}^*\right]}{|C_L^{\SM, sb}\delta_{\alpha \beta}+[C_L]_{sb\alpha \beta}|^2+|[C_R]_{sb\alpha \beta}|^2} ~ .
\end{split}\ee

The $\kappa_\eta$ parameter depends on form factors and its numerical value in [0,$q^2_{max}$] is $1.34\pm 0.04$, where $q_{max}$ is the kinematic limit of 22.9 GeV for $K$ and 19.2 GeV for $K^*$. At the leading order, in our model, $C_R=0$ and therefore $ \eta_{\alpha \beta}=0 $ and $R_K^\nu=R_{K^*}^\nu=\epsilon^2 $.
The SM Wilson coefficient is given by
\be
C_L^{\SM, sb}= - X_t/s_W^2 \, , \, \, \text{with} \, \, X_t=1.469\pm 0.017 \, .
\ee
The bounds of Eq.~(\ref{eq:RKnuBound}) can probe a region of the parameter space in which the $B \to K^{(*)} \nu \nu$ cross section is dominated by the squared BSM amplitude. Assuming real couplings we can provide approximate numerical expressions:
\be\begin{split}
	R_K^\nu &\approx R_{K^*}^\nu \approx 1 + 1.37 \frac{\sum_{\alpha}  \lambda^{1L}_{s \alpha} \lambda^{1L}_{b \alpha}}{|V_{ts}| m_1^2} + 1.25 \frac{\sum_{\alpha} \lambda^{3L}_{s \alpha} \lambda^{3L}_{b \alpha}}{|V_{ts}| m_3^2} + \\
		& + 1.42 \frac{\sum_{\alpha \beta} (\lambda^{1L}_{s \alpha})^2 (\lambda^{1L}_{b \beta})^2 }{|V_{ts}|^2 m_1^4}  + 1.16 \frac{\sum_{\alpha \beta} (\lambda^{3L}_{s \alpha})^2 (\lambda^{3L}_{b \beta})^2 }{|V_{ts}|^2 m_3^4}  
		+ 2.57 \frac{\sum_{\alpha} \lambda^{3L}_{s \alpha} \lambda^{3L}_{b \alpha} \lambda^{1L}_{s \alpha} \lambda^{1L}_{b \alpha}}{|V_{ts}|^2 m_1^2 m_3^2} ~,
	\label{eq:BoundBKnunuFin}
\end{split}\ee
where $m_i=M_i/$TeV.

Let us now discuss Kaon decays. In the Standard Model, the coefficients of this four-fermion interaction are~\cite{Buras:1998raa}
\be
	[C_{L}^{\SM, ds}]_{\alpha\beta} = -\frac{1}{s_W^2}\left(  \, X_t+ \frac{V_{cs} V_{cd}^*}{V_{ts} V_{td}^*} \, X_c^\alpha \right)\delta_{\alpha\beta}  \, ,
\ee
with $X_t=1.48$, $X_c^e=X_c^\mu=1.053\times 10^{-3}$, and $X_c^\tau=0.711\times 10^{-3}$.

Since $C_R \approx 0$ in our model, the branching ratios for the $K^+ \to \pi^+ \nu \nu$ and $K_L \to \pi^0 \nu \nu$ decays can be expressed as 
\begin{align}
&\text{Br}(K^+ \to \pi^+ \nu \nu)= \text{Br}(K^+ \to \pi^+ \nu_e \nu_e)_{\SM}\sum_{\alpha,\beta=1,2}\Big| \delta_{\alpha\beta}+\frac{[C_L]_{ds\alpha\beta}}{[C_{L}^{\SM, ds}]_{11}}\Big|^2+ \nonumber\\
&+\text{Br}(K^+ \to \pi^+ \nu_\tau \nu_\tau)_{\SM} 
	\left[ \sum_{\alpha=1,2}\left(\Big|\frac{[C^{ sd}_L]_{\alpha 3}}{[C_{L}^{\SM, ds}]_{33}}\Big|^2+\Big|\frac{[C^{ sd}_L]_{ 3\alpha}}{[C_{L}^{\SM, ds}]_{33}}\Big|^2\right) +\Big| 1+\frac{[C^{ sd}_L]_{33}}{[C_{L}^{\SM, ds}]_{33}}\Big|^2 \right] , \\
&\text{Br}(K_L \to \pi^0 \nu \nu)=  \text{Br}(K_L \to \pi^0 \nu \nu)_{\SM} \, \frac{1}{3} \,
	\Bigg[\sum_{\alpha,\beta=1,2}
		\left( \delta_{\alpha\beta}+\frac{\text{Im}[C_L]_{ds\alpha\beta}}{\text{Im}[C_{L}^{\SM, ds}]_{11}} \right)^2+ \nonumber \\
&+\sum_{\alpha=1,2}\left(\left(\frac{\text{Im}[C^{ sd}_L]_{\alpha 3}}{\text{Im}[C_{L}^{\SM, ds}]_{33}}\right)^2+\left(\frac{\text{Im}[C^{ sd}_L]_{ 3\alpha}}{\text{Im}[C_{L}^{\SM, ds}]_{33}}\right)^2\right) +\left( 1+\frac{\text{Im}[C^{ sd}_L]_{33}}{\text{Im}[C_{L}^{\SM, ds}]_{33}}\right)^2 \Bigg] ,
\label{eq:BrKpinunu}
\end{align}
where
\be\begin{split}
	\text{Br}(K^+ \to \pi^+ \nu_e \nu_e)_{\SM} &=3.06 \times 10^{-11} , \\
	\text{Br}(K^+ \to \pi^+ \nu_\tau \nu_\tau)_{\SM} &= 2.52 \times 10^{-11} ,\\
	\text{Br}(K_L \to \pi^0 \nu \nu)_{\SM} &= 3.4 \times 10^{-11} .
\end{split}\ee
The experimental bounds are \cite{Tanabashi:2018oca,CortinaGil:2020vlo}\footnote{NA62 results on $K^+ \to \pi^+ \nu\nu$ have been updated in a \href{https://indico.cern.ch/event/868940/contributions/3815641/attachments/2080353/3496097/RadoslavMarchevski_ICHEP_2020.pdf}{presentation} at the ICHEP2020 conference.}
\be\begin{split}
	\text{Br}(K^+ \to \pi^+ \nu\nu) & = (11.0^{+4.0}_{-3.5}) \times 10^{-11} ~,\\
	\text{Br}(K_L \to \pi^0 \nu \nu) &<3.57 \times 10^{-9} \, \,  \, (95\% \text{CL}) ~.\\
\end{split}\ee
Assuming that the contribution in $\nu_\tau$ is the dominant one we get the following approximate numerical expression:
\be\begin{split}
	10^{10} \text{Br}(K^+ \to \pi^+ \nu \nu) &\approx 0.61+0.25 \Bigg|1 + (1.56 - i 0.51) \frac{\lambda^{1L *}_{d\tau} \lambda^{1L}_{s\tau}}{|V_{td}||V_{ts}| m_1^2} + \\
	& + (1.39 - i 0.45) \frac{\lambda^{3L *}_{d\tau} \lambda^{3L}_{s\tau}}{|V_{td}||V_{ts}| m_3^2} - 0.016 \frac{| \lambda^{3L}_{b\tau}|^2}{m_3^2} + \ldots \Bigg|^2~, \\
	10^{10} \text{Br}(K_L \to \pi^0 \nu \nu) &\approx 0.23 + 0.11 \left( 1 - \Im \left[(16.4 - i 7.1) \frac{\lambda^{1L *}_{d\tau} \lambda^{1L}_{s\tau}}{|V_{td}||V_{ts}| m_1^2} \right] \right. \\
		& \left. - \Im \left[ (14.7 - i 6.3) \frac{\lambda^{3L *}_{d\tau} \lambda^{3L}_{s\tau}}{|V_{td}||V_{ts}| m_3^2} \right] + \ldots  \right)^2 ~.
\end{split}\ee

%%--------------------------------------------------
\subsection{$b\to d_i \tau \tau$ and $b\to d_i \tau \mu$ decays}
\label{app:BLFVnLFV}

We consider the leptonic branching ratios:
\be
	\Br(B_{d,s}\to \tau ^+ \tau ^-), \qquad 
	\Br(B_{d,s}\to \tau ^\pm \mu ^\mp)~,
\ee
and the semileptonic ones:
\be
	\Br(B^+\to K^+ \tau ^+ \tau ^-),\qquad 
	\Br(B^+\to K^+ \tau ^\pm \mu ^\mp)~.
\ee
At partonic level, these processes are induced by the same low-energy operators as $b \to s \ell \ell$, see Eqs. (\ref{eq:C910bsll},\ref{eq:Lbsll}), with obvious substitutions for the flavor indices. We limit ourselves to $C^{9,10}$ contributions; furthermore we restrict to a tree-level analysis for the $\tau\tau$ modes, whose current experimental bounds are too weak to be included in our global fit. In our model, the tree-level generated effective vertices lead to left-handed four-fermion interactions with $\mathcal{C}^9_{s b \tau \tau} = - \mathcal{C}^{10}_{s b \tau \tau}$, which are the ones that interfere with the SM processes.
Concerning the $\tau$ modes, we have:
\begin{align}
\text{Br}(B_{d_i}\to \tau^- \tau^+)&=\text{Br}(B_{d_i}\to \tau^- \tau^+)_{\text{SM}} \left|1 + \frac{\mathcal{C}^{10}_{d_i b \tau \tau}}{{\mathcal{C}^{10\,\text{SM}}_{d_i b \tau \tau}}}\right|^2~,\\
\text{Br}(B^+\to K^+ \tau ^+ \tau ^-) &=\text{Br}(B^+\to K^+ \tau ^+ \tau ^-)_{\text{SM}}\left|1 + \frac{\mathcal{C}^{(9-10)}_{d_i b \tau \tau}}{{\mathcal{C}^{(9-10)\,\text{SM}}_{d_i b \tau \tau}}}\right|^2~,
\end{align}
where $\mathcal{C} ^{9,10}$ are defined as in Eqs. (\ref{eq:C910bsll},\ref{eq:Lbsll}), $\mathcal{C} ^{(9-10)}\equiv\mathcal{C} ^{9}-\mathcal{C} ^{10}$ and we have
\be
\mathcal{C}^{9,SM}_{s b \tau \tau} = - \mathcal{C}^{10,SM}_{s b \tau \tau} = 4.3 ~,
\ee
and \cite{Bobeth:2013uxa,Du:2015tda}
\begin{align}
	\text{Br}(B_s \to \tau \tau)_{\text{SM}}& =  (7.73 \pm 0.49) \times 10^{-7}~,  \\
	\text{Br}(B^+\to K^+ \tau^- \tau^+)_{\text{SM}}& = (1.22\pm 0.10) \times 10^{-7}~. 
\end{align}
Numerically, we get:
\be
\dfrac{\text{Br}(B_{s}\to \tau^- \tau^+)}{\text{Br}(B_{s}\to \tau^- \tau^+)_{\text{SM}}}=\dfrac{\text{Br}(B^+\to K^+\tau^- \tau^+)}{\text{Br}(B^+\to K^+\tau^- \tau^+)_{\text{SM}}}=\left| 1- 137.5 \frac{\lambda^{3L }_{b \tau} \lambda^{3L *}_{s \tau}}{ m_3^2} \right|^2~.
\ee
The present 95\% CL limit on this branching ratio is $6.8 \times 10^{-3}$ \cite{Aaij:2017xqt}, so almost a factor $\sim 10^4$ larger than the SM prediction. \\

The lepton flavor violating modes are zero in the SM and we need the full branching ratio expressions. Neglecting the muon mass (amounting to an $\mathcal{O}(10\%)$ effect), the $B_{d_i}\to \tau^-  \mu^+$  branching ratios can be expressed as \cite{Becirevic:2016zri}
\be
	\text{Br}(B_{d_i}\to \tau^- \mu^+) = \dfrac{\tau_{B_{d_i}} f^2_{B_{d_i}} m_{B_{d_i}} \alpha ^2 G_f ^2 |V_{tb} V_{t d_{i}}^*|^2}{64\pi^3} m_\tau^2\left(1-\frac{m_\tau^2}{m_{B_{d_i}}^2}\right)^2 
		\left( \left| \mathcal{C}_{9}^{d_i b \tau \mu} \right|^2 + \left| \mathcal{C}_{10}^{d_i b \tau \mu} \right|^2 \right)~,
		\label{eq:Bstaumu}
\ee
where the input parameters are $f_{B_s}=0.224$GeV~\cite{Aoki:2016frl},  $m_{B_s}=5.37$GeV and $\tau_{B_s}=1.51\times 10^{-12} s$,
$f_{B_d}=0.190$GeV
$m_{B_d}=5.279$GeV and $\tau_{B_d}=1.5\times 10^{-12} s$~\cite{Tanabashi:2018oca}. 
The $  \text{Br}(B_{d_i}\to \mu^- \tau^+)$ case is analogous, via exchange of lepton indices.

In our analysis we keep $m_\mu$ effects as well as the complete one-loop matching contributions. Summing the $ \tau^- \mu^+$ and $\mu^- \tau^+$ modes, we get these approximate numerical expressions:
\be
\begin{split}
	\text{Br}(B_s \to \tau \mu)& \approx 1.25 \times 10^{-5} \left( \frac{|\lambda^{3L *}_{b \tau} \lambda^{3L}_{s \mu}|^2 }{|V_{ts}|^2 m_3^4} +   \frac{|\lambda^{3L *}_{b \mu} \lambda^{3L}_{s \tau}|^2}{|V_{ts}|^2 m_3^4} \right) + \ldots ~, \\
	\text{Br}(B_d \to \tau \mu)& \approx 8.67 \times 10^{-6} \left( \frac{|\lambda^{3L *}_{b \tau} \lambda^{3L}_{d \mu}|^2 }{|V_{td}|^2 m_3^4} +   \frac{|\lambda^{3L *}_{b \mu} \lambda^{3L}_{d \tau}|^2}{|V_{td}|^2 m_3^4} \right) + \ldots  ~. 
\end{split}
\ee

Finally, for the LFV three body decays $\text{Br}(B^+\to K^+ \tau ^\pm \mu ^\mp)$, we get the integrated expressions for the branching ratios for $B \to K^{(*)} \tau \mu$ from \cite{Becirevic:2016zri}.
The leading contribution is given by
\be
\begin{split}
	\text{Br}(B \to K^* \mu^- \tau^+)& \approx 2.4 \times 10^{-5} \frac{|\lambda^{3L *}_{b \tau} \lambda^{3L}_{s \mu}|^2 }{|V_{ts}|^2 m_3^4} + \ldots ~, \\
	\text{Br}(B \to K^* \tau^- \mu^+)& \approx 2.4 \times 10^{-5} \frac{|\lambda^{3L *}_{b \mu} \lambda^{3L}_{s \tau}|^2 }{|V_{ts}|^2 m_3^4} + \ldots ~, \\
	\text{Br}(B \to K \mu^- \tau^+)& \approx 1.4 \times 10^{-5} \frac{|\lambda^{3L *}_{b \tau} \lambda^{3L}_{s \mu}|^2 }{|V_{ts}|^2 m_3^4} + \ldots ~, \\
	\text{Br}(B \to K \tau^- \mu^+)& \approx 1.4 \times 10^{-5} \frac{|\lambda^{3L *}_{b \mu} \lambda^{3L}_{s \tau}|^2 }{|V_{ts}|^2 m_3^4} + \ldots ~. 
\end{split}
\ee
Present limits and future prospect for the observables discussed in this Section are reported in Table~\ref{tab:BelleIIprosp}.

%---------------------------------------------------------------------------------------
\subsection{$\Delta F=2$}
\label{app:DF2}

Here, we analyse the meson-mixing $\Delta F=2$ processes $B_{(s)}^{0}-\overline{B}_{(s)}^{0}$, $K^{0}-\overline{K}^{0}$ and $D^{0}-\overline{D}^{0}$.
They receive NP corrections, at one-loop level in the $S_{1,3}$ model, from the four-quark
SMEFT operators $\mathcal{O}_{qq}^{(1,3)}$, $\mathcal{O}_{qq_{R}}^{(1,8)}$
and $\mathcal{O}_{q_{R}q_{R}}$, where $q_{R}=d$ for $B_{(s)}^{0}-\overline{B}_{(s)}^{0}$
and $K^{0}-\overline{K}^{0}$ mixing, while $q_{R}=u$ for $D^{0}-\overline{D}^{0}$
mixing. In the $S_{1}+S_{3}$ model, the non-vanishing coefficients are:
\begin{align}
16\pi^{2} \, \left[C_{qq}^{(1)}\right]_{ijkl}  = & -\frac{9}{16}\frac{(\Lambda_{q}^{(3)})_{il}(\Lambda_{q}^{(3)})_{kj}}{M_{3}^{2}}-\frac{1}{16}\frac{(\Lambda_{q}^{(1)})_{il}(\Lambda_{q}^{(1)})_{kj}}{M_{1}^{2}}+ \nonumber\\
&-\frac{3}{16}\frac{\left[(\Lambda_{q}^{31})_{il}(\Lambda_{q}^{31\dagger})_{kj}+(\Lambda_{q}^{31})_{kj}(\Lambda_{q}^{31\dagger})_{il}\right]\ln(M_{3}^{2}/M_{1}^{2})}{M_{3}^{2}-M_{1}^{2}} +\cdots ~,\\
16\pi^{2} \,\left[C_{qq}^{(3)}\right]_{ijkl}  = & -\frac{1}{16}\frac{(\Lambda_{q}^{(3)})_{il}(\Lambda_{q}^{(3)})_{kj}}{M_{3}^{2}}-\frac{1}{16}\frac{(\Lambda_{q}^{(1)})_{il}(\Lambda_{q}^{(1)})_{kj}}{M_{1}^{2}}+ \nonumber\\
& +\frac{1}{16}\frac{\left[(\Lambda_{q}^{31})_{il}(\Lambda_{q}^{31\dagger})_{kj}+(\Lambda_{q}^{31})_{kj}(\Lambda_{q}^{31\dagger})_{il}\right]\ln(M_{3}^{2}/M_{1}^{2})}{M_{3}^{2}-M_{1}^{2}} +\cdots ~,\\
16\pi^{2} \,\left[C_{uu}\right]_{ijkl}  = & -\frac{1}{8}\frac{(\Lambda_{u})_{il}(\Lambda_{u})_{kj}}{M_{1}^{2}}+\cdots ~,\\
16\pi^{2} \,\left[C_{qu}^{(1)}\right]_{ijkl} = & -\frac{1}{12}\frac{(\Lambda_{q}^{(1)})_{ij}(\Lambda_{u})_{kl}}{M_{1}^{2}}+\cdots ~,\\
16\pi^{2} \,\left[C_{qu}^{(8)}\right]_{ijkl} = & -\frac{1}{2}\frac{(\Lambda_{q}^{(1)})_{ij}(\Lambda_{u})_{kl}}{M_{1}^{2}}+\cdots ~,
\end{align}
where the omitted terms that do not contribute to $\Delta F=2$ processes and the ones proportional to lepton Yukawa couplings. We also defined
\be\begin{array}{l l l}
	\Lambda_{q}^{(n)} \equiv \lambda^{nL*} \lambda^{nL \, T}, &\hspace{0.5cm}
	\Lambda_{q}^{(31)} \equiv \lambda^{3L*} \lambda^{1L \, T}, &\hspace{0.5cm}
	\Lambda_{u} \equiv \lambda^{1R*} \lambda^{1R \, T}, \\[1mm]
	\Lambda_{\ell}^{(n)}  \equiv\lambda^{nL\dagger}\lambda^{nL}, &\hspace{0.5cm}
	\Lambda_{\ell}^{(31)}  \equiv\lambda^{3L\dagger}\lambda^{1L}, &\hspace{0.5cm}
	\Lambda_{e} \equiv\lambda^{1R\dagger}\lambda^{1R}.
\end{array}\ee

In our fit we use the limits on $\Delta F = 2$ operators from the UTFit collaboration \cite{Bona:2007vi}.\footnote{We employ the results updated at the 2018 La Thuile conference \url{https://agenda.infn.it/event/14377/timetable/}.} The bounds on the $\Delta F = 2$ Wilson coefficients (or their real and imaginary parts, where appropriate) are given in terms of a scale 
$\Lambda_{\text{NMFV}}$, defined via $C = \frac{F}{\Lambda_{\text{NMFV}}^2}$, where
\be
	\sqrt{F}=\begin{cases}
		\left|V_{tb}V_{tq}^{*}\right| & \text{for \ensuremath{B_{q}^{0}}-\ensuremath{\overline{B}_{q}^{0}} mixing}\\
		\left|V_{ts}V_{td}^{*}\right| & \text{for \ensuremath{K^{0}}-\ensuremath{\overline{K}^{0}} and \ensuremath{D^{0}}-\ensuremath{\overline{D}^{0}} mixing}
	\end{cases}~.
\ee
The coefficients are evaluated at the same scale $\Lambda_{\text{NMFV}}$, which being close to the TeV in most cases is close enough to the LQ masses we are interested in that we can neglect the residual running to the LQ matching scale.
All constraints are collected in Table~\ref{tab:obs2}. In terms of SMEFT Wilson coefficients, and LQ couplings, the relevant non-zero combinations are 
\begin{align}
	C_{B_{i}}^{1} & =-\left[ C_{qq}^{(1)}+C_{qq}^{(3)} \right]_{b d_i b d_i} 
	=  \frac{1}{128 \pi^2}\left( 
		5 \frac{(\Lambda_{q}^{(3)})_{bi}^{2}}{M_{3}^{2}}
		+ \frac{(\Lambda_{q}^{(1)})_{bi}^{2}}{M_{1}^{2}} 
		+ 2 \frac{(\Lambda_{q}^{31})_{bi}(\Lambda_{q}^{31})_{ib}^{*}\ln(M_{3}^{2}/M_{1}^{2})}{M_{3}^{2}-M_{1}^{2}}\right)~,  \nonumber\\ 
	C_{K}^{1} & =-\left[C_{qq}^{(1)}+C_{qq}^{(3)}\right]_{sdsd} 
		=  \frac{1}{128 \pi^2}\left( 
		5 \frac{(\Lambda_{q}^{(3)})_{sd}^{2}}{M_{3}^{2}}
		+ \frac{(\Lambda_{q}^{(1)})_{sd}^{2}}{M_{1}^{2}} 
		+ 2 \frac{(\Lambda_{q}^{31})_{sd}(\Lambda_{q}^{31})_{ds}^{*}\ln(M_{3}^{2}/M_{1}^{2})}{M_{3}^{2}-M_{1}^{2}}\right)~,  \nonumber\\ 
	C_{D}^{1} & =-V_{ci}V_{uj}^{*}V_{ck}V_{ul}^{*}\left[C_{qq}^{(1)}+C_{qq}^{(3)}\right]_{ijkl} 
		=  \frac{V_{ci}V_{uj}^{*}V_{ck}V_{ul}^{*}}{128 \pi^2}\left( 
		5 \frac{(\Lambda_{q}^{(3)})_{il}(\Lambda_{q}^{(3)})_{kj}}{M_{3}^{2}} 
		+ \frac{(\Lambda_{q}^{(1)})_{il}(\Lambda_{q}^{(1)})_{kj}}{M_{1}^{2}} \right.  \nonumber\\
		& \left. \qquad \qquad \qquad \qquad \qquad \qquad \qquad
		+ 2 \frac{\left[(\Lambda_{q}^{31})_{il}(\Lambda_{q}^{31\dagger})_{kj}+(\Lambda_{q}^{31})_{kj}(\Lambda_{q}^{31\dagger})_{il}\right]\ln(M_{3}^{2}/M_{1}^{2})}{M_{3}^{2}-M_{1}^{2}} \right)~,  \nonumber\\ 
	\widetilde{C}_{D}^{1} & =-\left[C_{uu}\right]_{cucu} 
		= \frac{1}{128 \pi^2}\frac{(\Lambda_{u})_{il}(\Lambda_{u})_{kj}}{M_{1}^{2}} ,  \nonumber\\
	C_{D}^{4} & =V_{ci}V_{uj}^{*}\left[C_{qu}^{(8)}\right]_{ijcu} 
		= -\frac{V_{ci}V_{uj}^{*}}{32\pi^2}\frac{(\Lambda_{q}^{(1)})_{ij}(\Lambda_{u})_{cu}}{M_{1}^{2}},  \nonumber\\
	C_{D}^{5} & =2V_{ci}V_{uj}^{*}\left[C_{qu}^{(1)}-\frac{1}{6}C_{qu}^{(8)}\right]_{ijcu} 
		= 0.
		\label{eq:DeltaF2LQ}
\end{align}
We remark that the bounds in Ref.\cite{UTFIT:2016} assume a single non-vanishing
coefficient at a time, so that the latter bounds could in principle be evaded
by fine-tuning $C_{D}^{1}$ and $C_{D}^{4}$; we neglect this possibility, also because in the scenarios we focus on, the leading constraint is the one from $B_s$-mixing.

%------------------------------------------------------------------------------------ 
\subsection{LFU in $\tau$ decays}
\label{app:tauLFU}

Lepton flavor universality has been tested in $\tau$ decays at the permille level, in particular the strongest constraints are obtained from the ratios \cite{Pich:2013lsa}
\be\begin{split}
	|g_\mu / g_e|^2 &\equiv \frac{\Gamma(\tau \to \mu \nu \bar\nu)}{\Gamma(\tau \to e \nu \bar\nu)} \left( \frac{\Gamma_{\SM}(\tau \to \mu \nu \bar\nu)}{\Gamma_{\SM}(\tau \to e \nu \bar\nu)} \right)^{-1} = 1.0036 \pm 0.0028~, \\
	|g_\tau / g_\mu|^2 &\equiv \frac{\Gamma(\tau \to e \nu \bar\nu)}{\Gamma(\mu \to e \nu \bar\nu)} \left( \frac{\Gamma_{\SM}(\tau \to e \nu \bar\nu)}{\Gamma_{\SM}(\mu \to e \nu \bar\nu)}  \right)^{-1} = 1.0022 \pm 0.0030~,  \\
	|g_\tau / g_e|^2 &\equiv \frac{\Gamma(\tau \to \mu \nu \bar\nu)}{\Gamma(\mu \to e \nu \bar\nu)} \left( \frac{\Gamma_{\SM}(\tau \to \mu \nu \bar\nu)}{\Gamma_{\SM}(\mu \to e \nu \bar\nu)} \right)^{-1} = 1.0058 \pm 0.0030~.  \\
	\label{eq:tauLFU}
\end{split}\ee
The decay amplitudes can be described by four-lepton operators at the tau mass scale, which are one-loop suppressed in our model. Given this suppression and the experimental precision, we can confine ourselves to contributions interfering with the SM one, which comes from a $(V-A)^2$ chiral structure; we can thus write:
\be
	R_{\Gamma_{\beta\alpha}} \equiv \frac{\Gamma(\ell_\beta \to \ell_\alpha \nu \bar\nu)}{ \Gamma(\ell_\beta \to \ell_\alpha \nu \bar\nu)^{\SM}} 
	 \approx  1 + \frac{2 \Re[L^{V,LL}_{\nu e}]^{\rm NP}_{\alpha \beta \beta \alpha} (m_\tau) }{[L^{V,LL}_{\nu e}]^\SM_{\alpha \beta \beta \alpha}(m_\tau) } ~,
\ee
where $[L^{V,LL}_{\nu e}]^\SM_{\alpha \beta \beta \alpha}(m_W) = - \frac{2}{v^2}$. In terms of these ratios, the LFU quantities in Eq.~\eqref{eq:tauLFU} are given by:
\be
	|g_\mu / g_e|^2 = \frac{ R_{\Gamma_{\tau\mu}} }{ R_{\Gamma_{\tau e}} }~, \quad
	|g_\tau / g_\mu|^2 = \frac{ R_{\Gamma_{\tau e}} }{ R_{\Gamma_{\mu e}} }~, \quad
	|g_\tau / g_e|^2 = \frac{ R_{\Gamma_{\tau\mu}} }{ R_{\Gamma_{\mu e}} }~.
\ee
In the LEFT these operators do not run at the leading order in the EFT expansion.
The matching to the SMEFT at a scale $\mu_W$ is given by 
\begin{align}
	[L^{V,LL}_{\nu e}]^{\rm NP}_{\alpha \beta \beta \alpha}(\mu_W) &= - 2  [C_{Hl}^{(3)}]_{\alpha\alpha}(\mu_W)  - 2  [C_{Hl}^{(3)}]_{\beta\beta}(\mu_W)  + [C_{ll}]_{\alpha \beta \beta \alpha}(\mu_W)  + [C_{ll}]_{\beta \alpha \alpha \beta}(\mu_W) + \nonumber\\
	& - \frac{1}{16 \pi^2} \frac{2 m_t^2 N_c}{v^2} \left( [C_{lq}^{(3)}]^{(0)}_{\alpha \alpha t t} + [C_{lq}^{(3)}]^{(0)}_{\beta \beta t t} \right) \left( 1+ 2 \log \frac{\mu_W^2}{m_t^2} \right)~.
\end{align}
The leading terms in the one-loop matching from the LQ model to the SMEFT coefficients \cite{Gherardi:2020det} give:
\begin{align}
	[C_{Hl}^{(3)}]_{\alpha\alpha} &\approx \frac{N_c |y_t|^2 |V_{tb}|^2}{64 \pi^2} \left( \frac{|\lambda^{1L}_{b\alpha}|^2}{M_1^2} (1 + \log \frac{\mu_W^2}{M_1^2}) - \frac{|\lambda^{3L}_{b\alpha}|^2}{M_3^2}  (1 + \log \frac{\mu_W^2}{M_3^2})  \right) + \ldots ~, \nonumber \\
	[C_{ll}]_{\mu\tau\tau\mu} &\approx - \frac{N_c}{128 \pi^2} \left[ \frac{ |\lambda^{1L}_{b\mu}|^2 + |\lambda^{1L}_{b\tau}|^2 }{M_1^2}  + 5 \frac{ |\lambda^{3L}_{b\mu}|^2 + |\lambda^{3L}_{b\tau}|^2 }{M_3^2}  - \frac{\log M_3^2/M_1^2}{M_3^2 - M_1^2} \times \right. \\
	& \left.  ~ \times \left( |\lambda^{1L}_{b\tau}|^2 |\lambda^{3L}_{b\mu}|^2 + |\lambda^{3L}_{b\tau}|^2 |\lambda^{1L}_{b\mu}|^2 - 2 \lambda^{1L}_{b\tau}\lambda^{1L *}_{b\mu} \lambda^{3L}_{b\mu} \lambda^{3L *}_{b\tau} - 2 \lambda^{1L}_{b\mu}\lambda^{1L *}_{b\tau} \lambda^{3L *}_{b\mu} \lambda^{3L}_{b\tau} \right) \right] + \ldots .  \nonumber
\end{align}
From the complete expression of the matching, and neglecting at the end LQ couplings to electrons and to second generation left-handed quarks (which are expected to be suppressed due to quark flavor constraints), the leading numerical contributions are:
\begin{align}
	&	10^3  (R_{\Gamma_{\tau\mu}} - 1 ) \approx 
		- \frac{(\lambda^{1L}_{b\tau})^2 + (\lambda^{1L}_{b\mu})^2}{m_1^2} (1.62 + 0.74 \log m_1^2) 
		+ 0.29 \frac{(\lambda^{1L}_{b\tau})^2 (\lambda^{1L}_{b\mu})^2}{m_1^2} + \nonumber\\
	& \qquad \qquad 
		+ \frac{(\lambda^{3L}_{b\tau})^2 + (\lambda^{3L}_{b\mu})^2}{m_3^2} (1.46 + 0.74 \log m_3^2) 
		+ 1.43 \frac{(\lambda^{3L}_{b\tau})^2 (\lambda^{3L}_{b\mu})^2}{m_3^2} + \nonumber\\
	& \qquad \qquad
		+ 1.15  \frac{\lambda^{1L}_{b\mu} \lambda^{1L}_{b\tau} \lambda^{3L}_{b\mu} \lambda^{3L}_{b\tau} }{m_3^2 - m_1^2} \log \frac{m_3^2}{m_1^2} +\nonumber \\
	& \qquad \qquad - 0.29 ((\lambda^{1L}_{b\tau})^2 (\lambda^{3L}_{b\mu})^2 + (\lambda^{3L}_{b\tau})^2 (\lambda^{1L}_{b\mu})^2 ) \frac{\log m_3^2 / m_1^2}{m_3^2 - m_1^2} + \ldots ~, \\
	&	10^3  (R_{\Gamma_{\tau e}} - 1 ) \approx 
		 - \frac{(\lambda^{1L}_{b\tau})^2}{m_1^2} (1.61 + 0.67 \log m_1^2)  
		+ \frac{(\lambda^{3L}_{b\tau})^2}{m_1^2} (1.50 + 0.67 \log m_1^2) + \ldots ~, \\
	&	10^3  (R_{\Gamma_{\mu e}} - 1 ) \approx 
		 - \frac{(\lambda^{1L}_{b\mu})^2}{m_1^2} (1.60 + 0.61 \log m_1^2)  
		+ \frac{(\lambda^{3L}_{b\tau})^2}{m_1^2} (1.53 + 0.61 \log m_1^2) + \ldots ~.
\end{align}

%------------------------------------------------------------------------------------ 
\subsection{LFV decay $\tau \to \mu \phi $} 
\label{app:taumuphi}

The relevant effective operators for the $\tau \to \mu \bar s s $ transition in the LEFT are:
\be\begin{split}
	\LL_{\rm LEFT} \supset & 
	[L^{V,LL}_{ed}]_{\tau\mu s s} (\bar \tau_L \gamma_\alpha \mu_L) (\bar s_L \gamma_\alpha s_L) +
	[L^{V,LR}_{de}]_{s s \tau\mu } (\bar \tau_R \gamma_\alpha \mu_R) (\bar s_L \gamma_\alpha s_L) + h.c. \\
	& [L^{V,LR}_{ed}]_{\tau\mu s s}  (\bar \tau_L \gamma_\alpha \mu_L) (\bar s_R \gamma_\alpha s_R) +
	[L^{V,RR}_{ed}]_{\tau\mu s s} (\bar \tau_R \gamma_\alpha \mu_R) (\bar s_R \gamma_\alpha s_R) + h.c. \\
	& [L^{S,RL}_{ed}]_{\tau\mu s s}(\bar \tau_R \mu_L)  (\bar s_R s_L)  + h.c.
	\label{eq:Ltaumuphi}
\end{split}\ee

At the tree-level, in our model the only non-vanishing contribution is
\be
	[L^{V,LL}_{ed}]_{\tau\mu s s}^{\rm tree} = \frac{\lambda^{3L *}_{s\tau} \lambda^{3L}_{s\mu}}{M_3^2}~.
\ee
Since the coupling to the $s$ quark might be flavor-suppressed, loop contributions from LQ couplings to third generation quarks can potentially give a sizeable contribution. We evaluate the Wilson coefficients at the $\tau$ lepton mass scale, including all contributions up to one-loop arising from the matching to the UV model, the RGE, and the matching between SMEFT and LEFT.

Simplified analytical formulae, which can help to understand the physics underlying this observable, can be derived by putting to zero electroweak corrections (except for the QED RG contribution) and keeping only couplings to third generation (as well as $\lambda^{1L}_{s\mu}$, which can be large in some scenarios). In this case we get:
\begin{align}
	[L^{V,LL}_{ed}]_{\tau\mu s s}^{\rm loop}(m_\tau) &= 
		\frac{N_c y_t^2 |V_{tb}|^2}{96\pi^2} \left[ \frac{ \lambda^{1L *}_{b\tau} \lambda^{1L}_{b\mu} }{M_1^2} \left( \log \frac{M_1^2}{m_t^2} - 1 \right) 
			+ \frac{ \lambda^{3L *}_{b\tau} \lambda^{3L}_{b\mu} }{M_3^2} \left( \log \frac{M_3^2}{m_t^2} - 1 \right) \right] + \nonumber \\
		& - \frac{ \lambda^{1L *}_{b\tau} \lambda^{1L}_{b\mu}  |\lambda^{1L}_{s\mu}|^2 }{64 \pi^2 M_1^2 } 
		+ \frac{N_c \alpha Q_d^2}{6 \pi} \frac{ \lambda^{3L *}_{b\tau} \lambda^{3L}_{b\mu}}{M_3^2} \log \frac{m_b^2}{M_3^2} + \ldots , \\
	[L^{V,LR}_{ed}]_{\tau\mu s s }^{\rm loop}(m_\tau) &= 
		- \frac{N_c y_t^2 |V_{tb}|^2}{48\pi^2} \left[ \frac{ \lambda^{1L *}_{b\tau} \lambda^{1L}_{b\mu} }{M_1^2} \left( \log \frac{M_1^2}{m_t^2} - 1 \right) 
			+ \frac{ \lambda^{3L *}_{b\tau} \lambda^{3L}_{b\mu} }{M_3^2} \left( \log \frac{M_3^2}{m_t^2} - 1 \right) \right] + \nonumber \\
		& + \frac{N_c \alpha Q_d^2}{6 \pi} \frac{ \lambda^{3L *}_{b\tau} \lambda^{3L}_{b\mu}}{M_3^2} \log \frac{m_b^2}{M_3^2} + \ldots , \\
	[L^{V,LR}_{de}]_{s s \tau\mu}^{\rm loop}(m_\tau) &= 
		\frac{N_c y_t^2 |V_{tb}|^2}{96\pi^2} \frac{ \lambda^{1R *}_{t\tau} \lambda^{1R}_{t\mu} }{M_1^2} 
		+ \frac{N_c \alpha Q_d Q_u}{12 \pi} \frac{ \lambda^{1R *}_{c\tau} \lambda^{1R}_{c\mu}}{M_1^2} \log \frac{m_c^2}{M_1^2} + \nonumber \\
		& - \frac{ ( \lambda^{1R *}_{t\tau} \lambda^{1R}_{t\mu} +  \lambda^{1R *}_{c\tau} \lambda^{1R}_{c\mu} ) |\lambda^{1L}_{s\mu}|^2 }{64 \pi^2 M_1^2 }  + \ldots , \\
	[L^{V,RR}_{ed}]_{\tau\mu s s }^{\rm loop}(m_\tau) &=
		\frac{N_c y_t^2 |V_{tb}|^2}{48\pi^2}  \frac{ \lambda^{1R *}_{t\tau} \lambda^{1R}_{t\mu} }{M_1^2} \left( \log \frac{M_1^2}{m_t^2} - 1 \right) 
			+ \frac{N_c \alpha Q_d Q_u}{12 \pi} \frac{ \lambda^{1R *}_{c\tau} \lambda^{1R}_{c\mu}}{M_1^2} \log \frac{m_c^2}{M_1^2} + \ldots ~,
\end{align}
while contributions to $[L^{S,RL}_{ed}]_{\tau\mu s s}$ are proportional to small Yukawa couplings and thus negligible.
It should be noted that the electroweak contributions, not included above, can modify the final expression by a non negligible amount. From the complete expressions, we get the following numerical dependence:
\be\begin{split}
	[L^{V,LL}_{ed}]_{\tau\mu s s}(m_\tau) [\TeV^{-2}] &\approx 
		0.0017 \frac{\lambda^{3L *}_{s\tau} \lambda^{3L}_{s\mu}}{|V_{ts}|^2 M_3^2}
		+ 0.019 \frac{\lambda^{1L *}_{b\tau} \lambda^{1L}_{b\mu}}{M_1^2} 
		+ 0.018 \frac{\lambda^{3L *}_{b\tau} \lambda^{3L}_{b\mu} }{M_3^2}
		+ \ldots , \\
	[L^{V,LR}_{ed}]_{\tau\mu s s }(m_\tau)  [\TeV^{-2}] &\approx 
		 - 0.0027 \frac{\lambda^{1L *}_{b\tau} \lambda^{1L}_{b\mu}}{M_1^2}
		 - 0.0042 \frac{\lambda^{3L *}_{b\tau} \lambda^{3L}_{b\mu} }{M_3^2}
		 + \ldots , \\
	[L^{V,LR}_{de}]_{s s \tau\mu}(m_\tau)  [\TeV^{-2}] &\approx 
		+ 0.0016 \frac{\lambda^{1R *}_{c\tau} \lambda^{1R}_{c\mu} }{M_1^2}
		+ 0.0067 \frac{\lambda^{1R *}_{t\tau} \lambda^{1R}_{t\mu}  }{M_1^2} (1 - \log m_1^2)
		+ \ldots , \\
	[L^{V,RR}_{ed}]_{\tau\mu s s }(m_\tau)  [\TeV^{-2}] &\approx 
		+ 0.0016 \frac{\lambda^{1R *}_{c\tau} \lambda^{1R}_{c\mu} }{M_1^2}
		+ 0.0036 \frac{\lambda^{1R *}_{t\tau} \lambda^{1R}_{t\mu}  }{M_1^2} (1+ 0.4 \log m_1^2)
		+ \ldots  ,
\end{split}\ee
where terms with coefficients smaller than $10^{-4}$ are not shown.

Among the several observables testing the $\tau \to \mu s s$ contact interaction, the branching ratio of $\tau \to \mu \phi$ gives the most stringent constraints \cite{Mandal:2019gff}. It is given by (see also \cite{Bhattacharya:2016mcc,Mandal:2019gff})
\be
	\text{\Br}(\tau \to \mu \phi) = \frac{f_\phi^2 m_\tau^3}{128 \pi \Gamma_\tau} C_{\tau \to \mu \phi} \left( 1 - \frac{m_\phi^2}{m_\tau^2} \right)^2 \left( 1 + 2 \frac{m_\phi^2}{m_\tau^2} \right) \approx 2.32 \times 10^{-4} \; C_{\tau \to \mu \phi} \TeV^4~,
\ee
where $m_\phi = 1019 \MeV$, $f_\phi \approx 225 \MeV$~\cite{Bhattacharya:2016mcc,Tanabashi:2018oca}, and, at tree-level,
\be
	C_{\tau \to \mu \phi} \equiv \left( |[L^{V,LL}_{ed}]_{\tau\mu s s} + [L^{V,LR}_{ed}]_{\tau\mu s s }|^2 + |[L^{V,RR}_{ed}]_{\tau\mu s s} + [L^{V,LR}_{de}]_{s s \tau\mu}|^2 \right)~.
\ee
In principle we should also include one-loop contributions in the LEFT, from operators generated by integrating out the leptoquarks at tree-level, to the expression above in order to consistently obtain finite corrections. This might require calculation of non-local hadronic form factors, which is beyond the purpose of this work. However, one can realize that these are necessarily finite QED corrections, thus very suppressed with respect to the logarithmic correction from the QED RGE, which has a factor $\sim \log M_{\rm LQ}^2 / m_\tau^2 \sim 13$ for a 1 TeV leptoquark. For this reason we neglect them.
Taking all couplings real we can obtain the approximate expression:
\be\begin{split}
	10^9 \Br(\tau \to \mu \phi) &\approx 
		(62 + 46 \log m_1^2)  \frac{(\lambda^{1L}_{b\tau} \lambda^{1L}_{b\mu})^2}{m_1^4} 
		+ (42 + 37 \log m_1^2)  \frac{(\lambda^{3L}_{b\tau} \lambda^{3L}_{b\mu})^2}{m_3^4} +\\
	& 	+ (103 + 38 \log m_1^2 + 45 \log m_3^2 + 16 \log m_1^2 \log m_3^2)  \frac{\lambda^{1L}_{b\tau} \lambda^{1L}_{b\mu}\lambda^{3L}_{b\tau} \lambda^{3L}_{b\mu}}{m_1^2 m_3^2} + \\
	& 	+ (24 - 27 \log m_1^2) \frac{(\lambda^{1R}_{t\tau} \lambda^{1R}_{t\mu})^2}{m_1^4} 
		+ (2.3 - 0.2 \log m_1^2)  \frac{(\lambda^{1R}_{c\tau} \lambda^{1R}_{c\mu})^2}{m_1^4} + \\
	&	+ (15 - 7.7 \log m_1^2)  \frac{\lambda^{1R}_{t\tau} \lambda^{1R}_{t\mu} \lambda^{1R}_{c\tau} \lambda^{1R}_{c\mu}}{m_1^4} + \\	
	&	+ (13 + 4.8 \log m_1^2)  \frac{\lambda^{3L}_{s\tau} \lambda^{3L}_{s\mu}\lambda^{1L}_{b\tau} \lambda^{1L}_{b\mu}}{|V_{ts}|^2 m_1^2m_3^2}
		+ (11 + 4.7 \log m_3^2) \frac{\lambda^{3L}_{s\tau} \lambda^{3L}_{s\mu}\lambda^{3L}_{b\tau} \lambda^{3L}_{b\mu}}{|V_{ts}|^2 m_3^4} + \\
	&	+ 0.69  \frac{(\lambda^{3L}_{s\tau} \lambda^{3L}_{s\mu})^2 }{|V_{ts}|^4 m_3^4} + \ldots , 
	\label{eq:taumuphi_num}
\end{split}\ee
where we normalize the LH couplings to the strange quark with $|V_{ts}| \approx 0.041$ and the last line is dominated by the tree-level contribution.
The experimental limit on this branching ratio at 95\% C.L. is~\cite{Tanabashi:2018oca}
\be
	\text{\Br}(\tau \to \mu \phi) <1.00 \times 10^{-7} ~.
\ee

%---------------------------------------------------
\subsection{LFV decay $\tau \to 3 \mu$}
\label{app:tau3mu}

New physics contributions to the three body LFV decay $\tau \to 3\mu$ can be described by the effective Lagrangian at the tau energy scale:
\be
	\LL_{\rm eff} = \sum_i C_i Q_i~,
\ee
where
\begin{align}
	Q_{V,LL} &=(\bar {\mu_L} \gamma_\mu \tau_L )\,( \bar \mu_L \gamma^\mu \mu_L) , & 
	Q_{V,RR}&=(\bar {\mu_R} \gamma_{\mu} \tau_R)( \, \bar \mu_R \gamma^{\mu}\mu_R) , \nonumber\\
	Q_{V,LR} &=(\bar {\mu_L }\gamma_\mu \tau_L)( \, \bar \mu_R \gamma^\mu \mu_R) , & 
	Q_{V,LR} &=(\bar {\mu_L}\gamma_\mu  \mu_L) \,( \bar \mu_R\gamma^\mu \tau_R), \\
	Q_{S,LL} &=(\bar {\mu_R}  \tau_L )\,( \bar \mu_R \mu_L) , & 
	Q_{S,RR}&=(\bar {\mu_L} \tau_R)( \, \bar \mu_L \mu_R) , \nonumber\\
	Q_{D,L} &= e m_\tau (\bar{\mu}_R \sigma^{\mu\nu} \tau_L) F_{\mu\nu},  & 
	Q_{D,R} &= e m_\tau (\bar{\mu}_L \sigma^{\mu\nu} \tau_R) F_{\mu\nu}.  \nonumber
\end{align}
The branching ratio is given by \cite{Kuno:1999jp,Crivellin:2017rmk}
\be\begin{split}
	\text{Br}(\tau \to 3\mu) &= \dfrac{m_\tau^5}{3 (16\pi)^3 \Gamma_\tau}  \Big( 16 |C_{V,LL}|^2 + 16 |C_{V,RR}|^2 + 8 |C_{V,LR}|^2 + 8 |C_{V,RL}|^2 + \\
		& \qquad \qquad \qquad  + |C_{S,LL}|^2  + |C_{S,RR}|^2   \Big) + \\
		& + \dfrac{\alpha_e^2 m_\tau^5}{12 \pi \Gamma_\tau} \left( |C_{D,L}|^2  + |C_{D,R}|^2 \right) \left( 4 \log\frac{m_\tau^2}{m_\mu^2} - 11 \right) + X_\gamma~,
\end{split}\ee
where
\be
	X_\gamma = +\dfrac{\alpha_e m_\tau^5}{3 (4 \pi)^2 \Gamma_\tau} 
		\left( \Re [ C_{D,L} (C_{V,RL} + 2 C_{V,RR})^*]  
			+ \Re [ C_{D,R} (C_{V,LR} + 2 C_{V,LL})^*]
		\right)
\ee
In terms of the LEFT operator basis in App.~\ref{sec:LEFTbasis}, the EFT coefficients above are given by:
\be\begin{split}
	C_{V,LL} &= [ L^{V,LL}_{ee} ]_{\mu\tau\mu\mu} + [ L^{V,LL}_{ee} ]_{\mu\mu\mu\tau} = 2 [ L^{V,LL}_{ee} ]_{\mu\tau\mu\mu} ~, \\
	C_{V,RR} &= [ L^{V,RR}_{ee} ]_{\mu\tau\mu\mu} + [ L^{V,RR}_{ee} ]_{\mu\mu\mu\tau} = 2 [ L^{V,RR}_{ee} ]_{\mu\tau\mu\mu} ~, \\
	C_{V,LR} &= [ L^{V,LR}_{ee} ]_{\mu\tau\mu\mu}  ~, \\
	C_{V,RL} &= [ L^{V,LR}_{ee} ]_{\mu\mu\mu\tau}  ~, \\
	C_{S,RR} &= [ L^{S,RR}_{ee} ]_{\mu\tau\mu\mu} + [ L^{S,RR}_{ee} ]_{\mu\mu\mu\tau} = 2 [ L^{S,RR}_{ee} ]_{\mu\tau\mu\mu}  ~, \\
	C_{S,LL} &= [ L^{S,RR}_{ee} ]_{\tau\mu\mu\mu}^* + [ L^{S,RR}_{ee} ]_{\mu\mu\tau\mu}^* = 2 [ L^{S,RR}_{ee} ]_{\tau\mu\mu\mu}^*  ~, \\
	C_{D,R} &=  [ L_{e \gamma}]_{\mu\tau} / (e m_\tau ) ~, \\
	C_{D,L} &=  [ L_{e \gamma}]_{\tau\mu}^* / (e m_\tau ) ~,
\end{split}\ee
where in the second equalities we took into account the symmetry of flavor indices in the LEFT coefficients.
Note also that the overall sign in $X_\gamma$ depends on the convention used for covariant derivatives, our is the same as \cite{Kuno:1999jp}.

In our LQ model, all these operators arise at loop level, except the scalar ones $Q_{S,LL (RR)}$ which are not generated at dimension-six in the SMEFT and therefore we will not consider in the following.
There are three different types of contributions: the SMEFT-LEFT matching at tree-level, involving SMEFT operators that in the model are generated at one-loop level; the one-loop SMEFT-LEFT matching induced by tree-level four-fermion semileptonic operators; one-loop contributions in the LEFT, generated by penguin diagrams from the tree-level generated operators $O_{ed}^{V,LL}$ and $O_{eu}$. The latter are purely QED loops, so the only phenomenologically relevant contribution is the log-enhanced one, corresponding to the QED-induced RG evolution. Taking into account these different contributions and retaining the terms that should be dominant in most of the relevant region of parameter space, which is to say the $\sim \lambda^4$, $\sim y_t^2 \lambda^2$ ones and the LEFT RG running involving third quark generation, we can express the LEFT Wilson coefficients, evaluated at $\tau$ mass scale, as a function of the UV model parameters:
\begin{align}
16\pi^2 [L_{ee}^{V,LL}]_{\mu\tau\mu\mu}(m_\tau)&=
	\frac{N_{c}}{8} \Big[  - \frac{5 ( \laLdag{3} \laL{3})_{\mu\tau} (\laLdag{3} \laL{3})_{\mu\mu}}{M_3^2}  - \frac{(\laLdag{1} \laL{1})_{\mu\tau} (\laLdag{1} \laL{1})_{\mu\mu}}{M_1^2}  + \nonumber \\
	& + \frac{\log M_3^2 / M_1^2}{M_3^2 - M_1^2} \Big( 
		-(\laLdag{1} \laL{3})_{\mu\tau} (\laLdag{3} \laL{1})_{\mu\mu} 
		- (\laLdag{3} \laL{1})_{\mu\tau} (\laLdag{1} \laL{3})_{\mu\mu} \Big) \Big] \nonumber \\
	&+\left( \frac{1}{2} - c_W^2\right) N_{c}y_t^2\left[\frac{\laLst{1}_{b\mu}  \laL{1}_{b\tau}}{2M_{1}^{2}}+\frac{\laLst{3}_{b\mu}  \laL{3}_{b\tau}}{2M_{3}^{2}}\right] |V_{tb}|^2 \left(1+\log ( m_t^2/M^2) \right)+ \nonumber \\
	&+\frac{1}{2} e^2 N_c \frac{2}{9}  \frac{\laLst{3}_{b\mu}  \laL{3}_{b\tau}}{M_{3}^{2}} |V_{tb}|^2\log(m_b^2/M^2) + \ldots~, \label{eq:LVLL}
\end{align}
\begin{align}
16\pi^2 [L_{ee}^{V,LR}]_{\mu\tau\mu\mu}(m_\tau)&=- \frac{N_{c}}{4 M_1^2}  (\laLdag{1} \laL{1})_{\mu\tau} (\laRdag \laR)_{\mu\mu} +\nonumber\\
	&+2\left( 1 - c_W^2\right)  N_{c}y_t^2\left[\frac{\laLst{1}_{b\mu}  \laL{1}_{b\tau}}{2M_{1}^{2}}+\frac{\laLst{3}_{b\mu}  \laL{3}_{b\tau}}{2M_{3}^{2}}\right]|V_{tb}|^2\left(1+\log ( m_t^2/M^2) \right)+ \nonumber\\
 	&+ e^2 N_c \frac{2}{9}   \frac{\laLst{3}_{b\mu}  \laL{3}_{b\tau}}{M_{3}^{2}} |V_{tb}|^2\log(m_b^2/M^2) +\ldots ~,
\end{align}
\begin{align}
16\pi^2 [L_{ee}^{V,LR}]_{\mu\mu\mu\tau}(m_\tau)&=- \frac{N_{c}}{4 M_1^2}  (\laLdag{1} \laL{1})_{\mu\mu} (\laRdag \laR)_{\mu\tau}+\nonumber\\
	& -\left( 1 - 2 c_W^2\right)   N_{c} y_t^2 \frac{\laRst_{t\mu}  \laR_{t\tau}}{2M_{1}^{2}}\left(1+\log ( m_t^2/M^2) \right) +\ldots ~,
\end{align}
\begin{align}
16\pi^2[L_{ee}^{V,RR}]_{\mu\tau\mu\mu}(m_\tau)&=- \frac{N_{c}}{8 M_1^2} (\laRdag \laR)_{\mu\tau} (\laRdag \laR)_{\mu\mu} +\nonumber\\
	&-\left( 1 - c_W^2\right) N_{c} y_t^2\frac{\laRst_{t\mu}  \laR_{t\tau}}{2M_{1}^{2}} \left(1+\log ( m_t^2/M^2) \right)+\ldots ~, 
\end{align}
\begin{align}
16 \pi^2 \, [L_{e\gamma}]_{\mu\tau}(m_\tau)  = &- N_{c}e m_{t} \frac{V_{tk} \lambda_{k\mu}^{1L*}\lambda_{t\tau}^{1R}}{3M_{1}^{2}}\left(\log ( m_t^2/M_1^2)+\frac{7}{4}\right)+ \nonumber  \\
	& - N_{c}e m_{c} \frac{V_{ck} \lambda_{k\mu}^{1L*}\lambda_{c\tau}^{1R}}{3M_{1}^{2}} \left( \log ( m_c^2/M_1^2) + \frac{7}{4} \right)  + \label{eq:LegammaR} \\
	& + \frac{e N_c }{8} \left( \frac{m_\tau (\lambda^{1L \dagger} \lambda^{1L})_{\mu \tau}}{3 M_1^2} - \frac{m_\tau (\lambda^{3L \dagger} \lambda^{3L})_{\mu \tau}}{M_3^2} - \frac{m_\mu (\lambda^{1R \dagger} \lambda^{1R})_{\mu \tau}}{M_1^2} \right) + \ldots ~, \nonumber 
\end{align}
\begin{align}
16 \pi^2 \, [L_{e\gamma}^*]_{\tau\mu}(m_\tau) =& - N_{c}e m_{t} \frac{V_{tk}^* \lambda_{k\tau}^{1L}\lambda_{t\mu}^{1R*}}{3M_{1}^{2}}\left(\log ( m_t^2/M_1^2)+\frac{7}{4}\right)+ \nonumber  \\
	&- N_{c}e m_{c} \frac{V_{ck}^* \lambda_{k\tau}^{1L}\lambda_{c\mu}^{1R*}}{3M_{1}^{2}} \left( \log ( m_c^2/M_1^2) + \frac{7}{4} \right) + \label{eq:LegammaL} \\
	&  + \frac{e N_c }{8} \left( \frac{m_\mu (\lambda^{1L \dagger} \lambda^{1L})_{\mu \tau}}{3 M_1^2} -  \frac{m_\mu (\lambda^{3L \dagger} \lambda^{3L})_{\mu \tau}}{M_3^2} - \frac{m_\tau (\lambda^{1R \dagger} \lambda^{1R})_{\mu \tau}}{M_1^2} \right) + \ldots ~. \nonumber 
\end{align}
In most cases, unless $\lambda_{c\mu}^{1R}$ is large, the contributions proportional to the squared top Yukawa are the largest ones.

Assuming real couplings we get the following numerical expression in terms of LQ couplings:
\be\begin{split}
	10^4 \text{Br}(\tau \to 3\mu) & \approx \frac{(\lambda^{1L}_{b\mu})^2 (\lambda^{1R}_{t\tau})^2 + (\lambda^{1L}_{b\tau})^2 (\lambda^{1R}_{t\mu})^2 }{m_1^4} \left(1.76 +1,87 \log m_1^2 + 0.50 (\log m_1^2)^2 \right) + \\
	& 	- (0.083+0.046 \log m_1^2) \frac{\lambda^{1L}_{b\tau} \lambda^{1R}_{t\mu}  \lambda^{1R}_{c\tau} \lambda^{1R}_{c\mu} }{m_1^4} + \ldots~.
\end{split}\ee

If the $S_1$ couplings to RH fermions vanish, the contributions to the process are much smaller, the leading terms being:
\begin{align}
	10^{10}  \, \text{Br}(\tau \to 3\mu) |_{\lambda^{1R}_{i\alpha}=0} & \approx 
		\frac{(\lambda^{3L}_{b\mu})^2 (\lambda^{3L}_{b\tau})^2}{m_3^4} (12 + 4.4 \log m_1^2) + \frac{(\lambda^{1L}_{b\mu})^2 (\lambda^{1L}_{b\tau})^2}{m_1^4} (7.2 + 6.6 \log m_1^2) + \nonumber \\
		& + \frac{\lambda^{3L}_{b\mu} \lambda^{3L}_{b\tau} \lambda^{1L}_{b\mu} \lambda^{1L}_{b\tau}}{m_1^2 m_3^2}(12.7 + 5.4 \log m_1^2 + 3.9 \log m_3^2) + \ldots~.
\end{align}
Presently, the experimental bound on this LFV decay, at 95\% C.L. is~\cite{Tanabashi:2018oca}
\be
	\text{Br}(\tau \to 3\mu) < 2.5\times 10^{-8}~.
\ee

%------------------------------------------------------------------------------------ 
\subsection{LFV decays $\ell^\prime \to \ell \gamma$, magnetic and electric dipole moments}
\label{app:CLFV}

We analyze in this Section another set of LFV interactions, namely $\tau \to \mu \gamma$, $\tau \to e \gamma$ and $\mu \to e \gamma$.
Following \cite{Crivellin:2013hpa}, we can parametrize the $\ell_{\alpha}(p)\to\ell_{\beta}(p-q)\gamma(q)$
vertex in terms of Lorentz invariant form factors:
\be
\label{eq:Vll}
\mathcal{M}_{\beta\alpha}^{\rho}=i\left\{ \gamma^{\rho}\left(V_{\beta\alpha}^{L}P_{L}+V_{\beta\alpha}^{R}P_{R}\right)+q^{\rho}\left(S_{\beta\alpha}^{L}P_{L}+S_{\beta\alpha}^{R}P_{R}\right)+i\sigma^{\rho\nu}q_{\nu}\left(T_{\beta\alpha}^{L}P_{L}+T_{\beta\alpha}^{R}P_{R}\right)\right\} ,
\ee
which involves functions of $p^{2},q^{2}$ and $pq$.
On-shell one has $p^{2}=m_{\alpha}^{2}$, $q^{2}=0$, $pq=(m_{\alpha}^{2}-m_{\beta}^{2})/2$,
so that the form factors are just constants.
In terms of the form factors the observables we are interested in can be written as
\be\begin{split}
	\text{Br}(\ell_{\alpha} & \to\ell_{\beta}\gamma)=\frac{m_{\ell_{\alpha}}^{3}}{16\pi\Gamma_{\ell_{\alpha}}}\left(\left|T_{\beta\alpha}^{L}\right|^{2}+\left|T_{\beta\alpha}^{R}\right|^{2}\right),\\
	d_{\ell_{\alpha}} & =-\text{Im}(T_{\alpha\alpha}^{R}),\\
	\Delta a_{\ell_{\alpha}} & =\frac{2m_{\ell_{\alpha}}}{e}\text{Re}(T_{\alpha\alpha}^{R})
 \label{eq:dipoles},
\end{split}\ee
where $d$ and $a$ stand for electric and (anomalous) magnetic dipole
moment, respectively. In terms of LEFT coefficients we have
\be
	T^{R}_{\mu\tau}= 2 \, [L_{e\gamma}]_{\mu\tau}(m_\tau) \qquad \text{and} \qquad T^{L}_{\mu\tau}=2[L_{e\gamma}^*]_{\tau\mu}(m_\tau)~.
	\label{eq:TRL}
\ee
In terms of parameters of the UV model, the leading terms are given by
\begin{align}
16 \pi^2 \, T^{R}_{\beta\alpha} = &- N_{c}e m_{t} \frac{2 V_{tk} \lambda_{k\beta}^{1L*}\lambda_{t\alpha}^{1R}}{3M_{1}^{2}}\left(\log ( m_t^2/M_1^2)+\frac{7}{4}\right)+ \nonumber  \\
	& - N_{c}e m_{c} \frac{2 V_{ck} \lambda_{k\beta}^{1L*}\lambda_{c\alpha}^{1R}}{3M_{1}^{2}} \left( \log ( m_c^2/M_1^2) + \frac{7}{4} \right) + \label{eq:TRbetaalfa} \\
	& + \frac{e N_c }{4} \left( \frac{m_{\ell_\alpha} (\lambda^{1L \dagger} \lambda^{1L})_{\beta \alpha}}{3 M_1^2} - \frac{m_{\ell_\alpha} (\lambda^{3L \dagger} \lambda^{3L})_{\beta \alpha}}{M_3^2} - \frac{m_{\ell_\beta} (\lambda^{1R \dagger} \lambda^{1R})_{\beta \alpha}}{M_1^2} \right) + \ldots ~, \nonumber  
\end{align}
\begin{align}
16 \pi^2 \, T^{L}_{\beta\alpha}  =& - N_{c}e m_{t} \frac{2 V_{tk}^* \lambda_{k\alpha}^{1L}\lambda_{t\beta}^{1R*}}{3M_{1}^{2}}\left(\log ( m_t^2/M_1^2)+\frac{7}{4}\right)+ \nonumber  \\
	&- N_{c}e m_{c} \frac{2 V_{ck}^* \lambda_{k\alpha}^{1L}\lambda_{c\beta}^{1R*}}{3M_{1}^{2}} \left( \log ( m_c^2/M_1^2) + \frac{7}{4} \right) + \label{eq:TLbetaalfa} \\
	&  + \frac{e N_c }{4} \left( \frac{m_{\ell_\beta} (\lambda^{1L \dagger} \lambda^{1L})_{\beta \alpha}}{3 M_1^2} -  \frac{m_{\ell_\beta} (\lambda^{3L \dagger} \lambda^{3L})_{\beta \alpha}}{M_3^2} - \frac{m_{\ell_\alpha} (\lambda^{1R \dagger} \lambda^{1R})_{\beta \alpha}}{M_1^2} \right) + \ldots ~. \nonumber 
\end{align}

Particularly relevant for our model is the $\tau\to\mu\gamma$ constraint. Assuming TeV-scale masses (to fix the logarithmic dependence on $m_1$) we get this approximate numerical expression:
\begin{align}
	\text{Br}&(\tau\to\mu\gamma) \approx \frac{4.9 \times 10^{-8}}{m_1^4} \Big( 
	\left| 1220 (V^* \lambda^{1L})_{t\tau} \lambda^{1R}_{t\mu} + 58 (V^* \lambda^{1L})_{c\tau} \lambda^{1R}_{c\mu} - 2.6 (\lambda^{1R \dagger} \lambda^{1R})_{\mu\tau} \right|^2 + \nonumber \\
	& ~ + \left| 1220 (V \lambda^{1L *})_{t\mu} \lambda^{1R}_{t\tau} + 58 (V \lambda^{1L *})_{c\mu} \lambda^{1R}_{c\tau} + 0.88 (\lambda^{1L \dagger} \lambda^{1L})_{\mu\tau}  - 2.6 \frac{m_1^2}{m_3^2}  (\lambda^{3L \dagger} \lambda^{3L})_{\mu\tau}   \right|^2 \Big) . \label{eq:taumuga_num} 
\end{align}
\begin{align}
	\text{Br}&(\mu\to e\gamma) \approx \frac{7.8 \times 10^{-5}}{m_1^4} \Big( 
	\left| 1220 (V^* \lambda^{1L})_{t\mu} \lambda^{1R}_{t e} + 58 (V^* \lambda^{1L})_{c\mu} \lambda^{1R}_{c e} - 0.16 (\lambda^{1R \dagger} \lambda^{1R})_{e \mu} \right|^2 + \nonumber \\
	& ~ + \left| 1220 (V \lambda^{1L *})_{te } \lambda^{1R}_{t \mu} + 58 (V \lambda^{1L *})_{ce} \lambda^{1R}_{c\mu} + 0.05 (\lambda^{1L \dagger} \lambda^{1L})_{e\mu}  - 0.15 \frac{m_1^2}{m_3^2}  (\lambda^{3L \dagger} \lambda^{3L})_{e\mu}   \right|^2 \Big) . \label{eq:muega_num} 
\end{align}
The experimental 95\% CL bounds are \cite{Tanabashi:2018oca}
\be\begin{split}
	\text{Br}(\mu\to e\gamma) & <5.00\times10^{-13} \, ,\\
	\text{Br}(\tau\to\mu\gamma) & <5.24\times10^{-8} \, ,\\
	\text{Br}(\tau\to e\gamma) & <3.93\times10^{-8} \, .
\end{split}\ee

The imaginary and real part of the diagonal terms $T_{\alpha\alpha}^{R}$ define respectively the Electric Dipole Moment (EDM) $d_\alpha$ and the anomalous magnetic moment $a_\alpha$ for the three charged leptons, as seen in Eq.\eqref{eq:dipoles}.
For the particularly relevant cases of the muon and electron magnetic moments, good numerical expressions, for TeV-scale masses and real couplings, are
\be\begin{split}
	\Delta a_\mu \approx& \frac{10^{-9}}{m_1^2} \Re \left[ 819 (V \lambda^{1L *})_{t\mu} \lambda^{1R}_{t\mu} + 39 (V \lambda^{1L *})_{c\mu} \lambda^{1R}_{c\mu}  + \ldots \right] ~, \\
	\Delta a_e \approx& \frac{10^{-9}}{m_1^2} \Re \left[ 4.0 (V \lambda^{1L *})_{te} \lambda^{1R}_{te} + 0.19 (V \lambda^{1L *})_{ce} \lambda^{1R}_{ce}  + \ldots \right] ~.
	\label{eq:amu_num}
\end{split}\ee
The combined SM prediction from the \emph{Muon $g-2$ Theory Initiative} \cite{Aoyama:2020ynm} is
\be
	a_\mu^\SM = 116 \, 591 \, 810 (43) \times 10^{-11},
\ee
while the Brookhaven E821 measurement \cite{Bennett:2006fi} finds
\be
	a_\mu^{\text{exp}} = 116 \, 592 \, 089 (63) \times 10^{-11},
\ee
corresponding to a $3.7 \sigma$ discrepancy: $\Delta a_\mu = (2.79 \pm 0.76) \times 10^{-9}$. Updated measurements will be available from the E989 experiment at Fermilab and E34 at J-PARC, which aim to reduce the experimental uncertainty by a factor of four. 

Other potentially interesting observables are quark EDM, which receive a strong bound via the constraint on the neutron EDM. Since these are not directly relevant for the parameter space relevant for our fits, we do not consider them in this work. A comprehensive analysis of EDMs in scalar leptoquark models is given in Ref.~\cite{Dekens:2018bci}; for a recent analysis in the context of a (vector) $U_1$-leptoquark model see Ref.~\cite{Altmannshofer:2020ywf}.

%--------------------------------------------------------------------
\subsection{$Z$ boson couplings}
\label{app:Zcouplings}

The couplings of the $Z$ boson have been measured very precisely at LEP 1. At one-loop, the LQ model generates contributions to these very well measured quantities, which pose strong constraints on the model.
The RGE-induced contributions in models aimed at addressing the $B$ anomalies have first been studied in Ref.~\cite{Feruglio:2016gvd,Feruglio:2017rjo,Cornella:2018tfd}. Here we include the effect of finite corrections from the one-loop matching.

The pseudo-observables corresponding to the pole couplings of the Z boson to fermions correspond to these effective Lagrangian couplings:
\be
	\mathcal{L}_{eff} = - \frac{g}{c_\theta} ( [g^{Z, \SM}_{\psi}]_{ij} +  [\delta g^Z_{\psi}]_{ij} ) Z_\mu (\bar \psi_i \gamma^\mu \psi_j)~,
	\label{eq:LZpole}
\ee
where $[g^{Z, \SM}_{\psi}]_{ij} = \delta_{ij} (T_{3L}^\psi - Q_\psi s_\theta^2)$. The measurements of these pseudo-observables and the predictions for the SM contributions can be found in \cite{ALEPH:2005ab}. Also often used is the effective number of neutrino species at the $Z$ peak, which depends on the couplings to neutrinos as
\be
	N_\nu = \sum_{\alpha\beta} \left| \delta_{\alpha\beta} + \frac{[\delta g^Z_{\nu}]_{\alpha\beta}}{ g^{Z, \SM}_{\nu}} \right|^2~,
\ee
where $g^{Z, \SM}_{\nu} \approx 0.502$. The latest update on the extraction of $N_\nu$ from LEP data is given in \cite{Janot:2019oyi}. We collect in Table~\ref{tab:ZcouplLEP} the limits used in our fit.

Working at one-loop accuracy, there are two possible contributions to these pseudo-observables in our setup: tree-level contributions from the SMEFT operators\footnote{There might be some indirect contributions via modifications to $G_F$, but are negligible in our model.}, one-loop matrix elements from the tree-level generated semileptonic operators. The result is:
\be\begin{split}
	[\delta g^Z_{e_L}]_{\alpha\beta}  \approx&   - v^2 \left[ \frac{1}{2} [C^{(1-3)}_{H\ell}]^{(1)}_{\alpha\beta}(\mu) + \frac{N_c}{16\pi^2} [C_{\ell q}^{(1 - 3)}]^{(0)}_{\alpha\beta i i} I_L^{{MS}}(u_i,m_Z^2, \mu) \right] ~,\\
	[\delta g^Z_{e_R}]_{\alpha\beta}  \approx& - v^2 \left[ \frac{1}{2} [C_{He}]^{(1)}_{\alpha\beta}(\mu) + \frac{N_c}{16\pi^2} [C_{eu}]^{(0)}_{\alpha\beta i i} I_R^{\overline{MS}}(u_i,m_Z^2, \mu) \right] ~, \\
	[\delta g^Z_{\nu}]_{\alpha\beta}  \approx& - v^2 \left[ \frac{1}{2} [C^{(1+3)}_{H\ell}]^{(1)}_{\alpha\beta}(\mu) + \frac{N_c}{16\pi^2} [C_{\ell q}^{(1 + 3)}]^{(0)}_{\alpha\beta i i} I_L^{\overline{MS}}(u_i,m_Z^2, \mu) \right] ~, \\
	[\delta g^Z_{d_L}]_{i j}  \approx&   - \frac{v^2}{2}  [C^{(1+3)}_{Hq}]^{(1)}_{i j}(\mu) ~,\\
	[\delta g^Z_{u_L}]_{i j}  \approx&   - \frac{v^2}{2}  V_{i k} V^*_{j l} [C^{(1-3)}_{Hq}]^{(1)}_{k l}(\mu) ~,\\
	[\delta g^Z_{u_R}]_{i j}  \approx&   - \frac{v^2}{2}  [C_{Hu}]^{(1)}_{i j}(\mu) ~,\\
	[\delta g^Z_{d_R}]_{i j}  \approx&   - \frac{v^2}{2}  [C_{Hd}]^{(1)}_{i j}(\mu) ~,
\end{split}\ee
where $\mu$ is the scale at which the operators are evaluated, $C^{(1\pm 3)}_X = C^{(1)}_X \pm C^{(3)}_X$, and we included only one-loop matrix elements from up-type quark loops, since the top quark gives the dominant effect. The expressions for the one-loop matching of the SMEFT operators to the LQ model can be found in \cite{Gherardi:2020det}. The loop functions are given by
\be\begin{split}
	 v^2 I_L^{\overline{MS}}(u_i, q^2, \mu) &= \frac{1}{9} \Big[ 5 q^2 (1 - 3 s_\theta^2) - 6 m_{u_i}^2 (1+6 s_\theta^2) + \\
		& \qquad - 3 \left(m_{u_i}^2 - q^2 + 3 (2 m_{u_i}^2 + q^2) s_\theta^2 \right) \text{DiscB}(q^2, m_{u_i}, m_{u_i}) + \\
		& \qquad + 3 (-3 m_{u_i}^2 + q^2 - 3 q^2 s_\theta^2) \log \frac{\mu^2}{m_{u_i}^2} \Big]~, \\
	 v^2 I_R^{\overline{MS}}(u_i, q^2, \mu) &= -\frac{5}{3} q^2 s_\theta^2 + m_{u_i}^2 (2 - 4 s_\theta^2) + \\
		& \qquad + \left(m_{u_i}^2 - (2 m_{u_i}^2 + q^2) s_\theta^2 \right) \text{DiscB}(q^2, m_{u_i}, m_{u_i}) + \\
		& \qquad + ( m_{u_i}^2 - q^2 s_\theta^2) \log \frac{\mu^2}{m_{u_i}^2} \Big]~, \\
	\text{DiscB}(q^2, m, m) &= \frac{\sqrt{q^2(q^2 - 4 m^2)}}{q^2} \log \frac{2 m^2 -q^2 + \sqrt{q^2(q^2 - 4 m^2)}}{2 m^2}~.
\end{split}\ee
Note that $I_L^{\overline{MS}}(c ,m_Z^2, \mu = 1.5 \TeV) = 0.102 + i 0.044$, $I_R^{\overline{MS}}(c ,m_Z^2, \mu = 1.5 \TeV) = - 0.23 - i 0.10$ while $I_L^{\overline{MS}}(t ,m_Z^2, \mu = 1.5 \TeV) = - 1.91$ and $I_R^{\overline{MS}}(t ,m_Z^2, \mu = 1.5 \TeV) = 1.83$. Furthermore, in the limit $q^2 \to 0$,
\be
	I_{R,L}^{\overline{MS}}( t, 0, \mu) \approx \pm \frac{m_t^2}{v^2} \log \frac{\mu^2}{m_Z^2} = \pm 1.95  \log \left( \frac{\mu^2}{1.5 \TeV^2} \right)~.
\ee

When all contributions are included, numerical expressions are:
\be\begin{split}
	10^3 \delta g^Z_{e_{\alpha L}}  \approx&  1.58 \frac{(\lambda^{3L}_{b\alpha})^2}{m_3^2} (1 + 0.31 \log m_3^2) - 0.16 \frac{(\lambda^{1L}_{b\alpha})^2}{m_1^2} - 0.16 \frac{\lambda^{3L}_{b\alpha} \lambda^{3L}_{s\alpha}}{m_3^2} + \ldots , \\
	10^3 \delta g^Z_{\nu_{\alpha}}  \approx&  0.86 \frac{(\lambda^{1L}_{b\alpha})^2}{m_1^2} (1 + 0.27 \log m_1^2) + 0.53 \frac{(\lambda^{3L}_{b\alpha})^2}{m_3^2} (1 + 0.52 \log m_3^2) + \ldots , \\
	10^3 \delta g^Z_{e_{\alpha R}}  \approx&  -0.67 \frac{(\lambda^{1R}_{t\alpha})^2}{m_1^2} (1 + 0.37 \log m_1^2) + 0.059 \frac{(\lambda^{1R}_{c\alpha})^2}{m_1^2} + 0.030 \frac{(\lambda^{1R}_{u\alpha})^2}{m_1^2}  + \ldots , \\
	10^3 \delta g^Z_{b_{L}}  \approx&  - 0.044 \frac{(\lambda^{1L}_{b \tau})^2 + (\lambda^{1L}_{b \mu})^2}{m_1^2} + \ldots , \\
	10^3 \delta g^Z_{c_{R}}  \approx&  - 0.014 \frac{(\lambda^{1R}_{c \tau})^2 + (\lambda^{1R}_{c \mu})^2}{m_1^2} + \ldots , 
\end{split}\ee
where the dots represent smaller, thus negligible, contributions. The largest radiative contributions are those due to the top quark, which get a $y_t^2$ factor, while all others arise only due to light Yukawas or gauge couplings and are thus much smaller.
The present measurements are reported in Table~\ref{tab:ZcouplLEP}.

\subsection{Oblique corrections and Higgs couplings}
\label{app:EWPO}

Let us now consider a set of constraints for the potential couplings between leptoquark and the SM Higgs.
Since these are universal corrections we focus on constraints independent on the LQ couplings to fermions.
Considering that all effects related to the couplings in the potential arise at one-loop, and that we also have a scale suppression from the TeV-scale LQ masses, it is clear we need to consider only high precision constraints for universal theories: EW precision observables, and Higgs processes which arise at one-loop in the SM.

Using the expressions of the oblique corrections in terms of Warsaw basis operators from \cite{Wells:2015uba} we get
\be\begin{split}
	\hat{S} &= \frac{\alpha}{4 s_W^2} S = v^2 g^2 \left( \frac{C_{HWB}}{g g^\prime} + \frac{1}{4} C_{HJW} + \frac{1}{4} C_{HJB} - \frac{1}{2} C_{2JW} - \frac{1}{2} C_{2JB} \right) = \\
		& = - \frac{g^2 N_c v^2 Y_{S_3}}{48 \pi^2} \frac{\lambda_{\epsilon H3}}{M_3^2} \approx - 5.4 \times 10^{-5} \lambda_{\epsilon H3} / m^2~,\\
	\hat{T} &= \alpha T = -\frac{v^2}{2} \left( C_{HD} - g^{\prime 2} (C_{HJB} - C_{2JB}) \right) = \\
		& = \frac{N_c v^2 \lambda_{\epsilon H3}^2}{48 \pi^2 M_3^2}  + \frac{N_c v^2 }{16 \pi^2} |\lambda_{H13}|^2 \frac{M_1^4 - M_3^4 - 2 M_1^2 M_3^2 \log M_1^2/M_3^2}{(M_1^2 - M_3^2)^3} =\\
		& \approx 3.8 \times 10^{-4} \lambda^2_{\epsilon H3} / m^2 + 3.8 \times 10^{-4} |\lambda_{H13}|^2 / m^2 ~, \\
	Y &= -\frac{v^2 g^2}{2} C_{2JB} = \frac{N_c v^2 g^2 g^{\prime 2} }{1920 \pi^2} \left( \frac{Y_{S_1}^2}{M_1^2} + 3 \frac{Y_{S_3}^2}{M_3^2} \right) \approx 2.3 \times 10^{-7} / m^2~,\\
	W &= -\frac{v^2 g^2}{2} C_{2JW} = \frac{N_c v^2 g^4 }{960 \pi^2 M_3^2} \approx 3.5 \times 10^{-6} / m^2~,\\
	Z &= -\frac{v^2 g^2}{2} C_{2JG} = \frac{v^2 g^2 g_s^2 }{3840 \pi^2} \left( \frac{1}{M_1^2} + \frac{3}{M_3^2} \right)  \approx 4.0 \times 10^{-6} / m^2~,\\
\end{split}\ee
where in the numerical expressions we set $M_1 = M_3 = m \TeV$ and ($\OO_{HJW}$, $\OO_{HJB}$, $\OO_{2JW}$, $\OO_{2JB}$, $\OO_{2JG}$) are universal combination of Warsaw basis operators as defined in \cite{Wells:2015uba}.
Given the smallness of the contributions to $Y,W,Z$, we neglect these and use the $S,T$ fit from \cite{Haller:2018nnx} with $U = 0$: $S = 0.04 \pm 0.08$, $T = 0.08 \pm 0.07$, with correlation $\rho_{ST} = 0.92$.

All deviations in Higgs couplings arise at loop level in this model and are thus very suppressed. The only channels potentially sensitive to such deviations are those for which also the SM contribution arises at one-loop: Higgs couplings to photons and gluons. For the couplings to photons and gluons we use the latest combination of Higgs measurements done by ATLAS with 80 fb$^{-1}$ of luminosity \cite{Aad:2019mbh} while for the $Z \gamma$ channel we use the ATLAS \cite{Aad:2020plj} and CMS \cite{Sirunyan:2018tbk} constraints:
\be\begin{split}
	&\kappa_\gamma = 1.00 \pm 0.06~, \quad
	\kappa_g = 1.030 \pm 0.065~, \quad \rho_{\gamma - g} = -0.44~, \\
	&\sigma/\sigma_{\SM}(Z\gamma) = \kappa_g^2 \kappa_{Z\gamma}^2 = 2.0^{+1.0}_{-0.9} |_{\rm ATLAS} ~ \left(< 3.9  ~@95\% {\rm CL} \right)_{\rm CMS}~,
\end{split}\ee
where $\kappa_i$ are coupling-modifiers in the so-called \emph{kappa}-framework, defined for each channel by $\kappa_i^2 \equiv \Gamma_i / \Gamma_i^{\SM}$ or $\kappa_i^2 \equiv \sigma_i / \sigma_i^{\SM}$.
We take the HL-LHC prospects from \cite{Cepeda:2019klc} (Table 38), which give the following uncertainties for the $\kappa$ parameters we are interested in:
\be
	\sigma_{\kappa_\gamma} = 2.0 \%, \quad
	\sigma_{\kappa_Z\gamma} = 12.4 \%, \quad
	\sigma_{\kappa_g} = 2.5 \%.
\ee
Defining the phenomenological effective Lagrangian
\be
	\LL = - \frac{c_{Z\gamma}}{v} h Z_{\mu\nu} A^{\mu\nu} - \frac{c_{\gamma\gamma}}{2 v} h A_{\mu\nu} A^{\mu\nu} - \frac{c_{gg}}{2 v} h G^{A}_{\mu\nu} G^{A \mu\nu}~,
\ee
the $\kappa$ parameters are given by $\kappa_i = 1 + c^{\rm NP}_i / c_i^{\SM}$, with $c_{Z\gamma}^{\SM} = 6.9 \times 10^{-3}$, $c_{\gamma\gamma}^{\SM} = 3.8 \times 10^{-3}$, $c_{gg}^{\SM} = 8.1 \times 10^{-3}$.
The matching to the SMEFT, and to our LQ model, is given by
\be\begin{split}
	c^{\rm NP}_{Z\gamma} &= v^2 \left[ 2 c_W s_W (C_{HB} - C_{HW}) + (c_W^2 - s_W^2) C_{HWB} \right]~, \\
	c^{\rm NP}_{\gamma\gamma} &= - 2 v^2 \left[ L_W^2 C_{HB} + s_W ( s_W C_{HW} - c_W C_{HWB}) \right]~, \\
	c^{\rm NP}_{gg} &= - 2 v^2 C_{HG} ~,
\end{split}\ee
and
\be\begin{split}
	C_{HB} &= \frac{g^{\prime2}N_{c}}{96 \pi^2}\left(3\frac{\lambda_{H3}Y_{S_{3}}^{2}}{M_{3}^{2}}+\frac{\lambda_{H1}Y_{S_{1}}^{2}}{M_{1}^{2}}\right) ~, \\
	C_{HW} &= \frac{g^{2}N_{c}\lambda_{H3}}{48 \pi^2 M_{3}^{2}},\\
	C_{HWB} &= -N_{c}\frac{gg^\prime Y_{S_3}\lambda_{\epsilon H3}}{48 \pi^2  M_{3}^{2}}, \\
	C_{HG} &= \frac{g_{s}^{2}}{192 \pi^2 }\left(\frac{3\lambda_{H3}}{M_{3}^{2}}+\frac{\lambda_{H1}}{M_{1}^{2}}\right).
\end{split}\ee
Numerically, fixing both LQ masses to 1 TeV, we get:
\be\begin{split}
	\kappa_{Z\gamma} - 1 &= -(1.89 \lambda_{H3} + 0.23 \lambda_{\epsilon H3}  - 0.033 \lambda_{H1} ) \times 10^{-2} ~, \\
	\kappa_{\gamma} - 1 &= -(2.32 \lambda_{H3} + 0.66 \lambda_{\epsilon H3}  - 0.11 \lambda_{H1} ) \times 10^{-2} ~, \\
	\kappa_{g} - 1 &= - (3.51 \lambda_{H3} + 1.17 \lambda_{H1} ) \times 10^{-2} ~.
\end{split}\ee

\section{SMEFT and LEFT operator bases}
\label{sec:op_bases}
For ease of reference, we report in this Section the SMEFT \cite{Grzadkowski:2010es} and LEFT \cite{Jenkins:2017jig} operator bases used throughout our analyses. We denote SMEFT and LEFT
operators by calligraphic $\mathcal{O}$ and $O$ respectively. Flavor
indices are suppressed.

\subsection{SMEFT basis}
\label{sec:SMEFTbasis}

\begin{center}
\begin{table}[H]
\begin{centering}
\begin{tabular}{|c|c|c|c|c|c|}
\hline 
\multicolumn{2}{|c|}{$X^{3}$} & \multicolumn{2}{c|}{$X^{2}H^{2}$} & \multicolumn{2}{c|}{$H^{4}D^{2}$}\tabularnewline
\hline 
$\mathcal{O}_{3G}$ & $f^{ABC}G_{\mu}^{A\nu}G_{\nu}^{B\rho}G_{\rho}^{C\mu}$ & $\mathcal{O}_{HG}$ & $G_{\mu\nu}^{A}G^{A\mu\nu}(H^{\dagger}H)$ & $\mathcal{O}_{H\square}$ & $(H^{\dagger}H)\square(H^{\dagger}H)$\tabularnewline
$\mathcal{O}_{3\widetilde{G}}$ & $f^{ABC}\widetilde{G}_{\mu}^{A\nu}G_{\nu}^{B\rho}G_{\rho}^{C\mu}$ & $\mathcal{O}_{H\widetilde{G}}$ & $\widetilde{G}_{\mu\nu}^{A}G^{A\mu\nu}(H^{\dagger}H)$ & $\mathcal{O}_{HD}$ & $(H^{\dagger}D^{\mu}H)^{\dagger}(H^{\dagger}D_{\mu}H)$\tabularnewline
\cline{5-6} \cline{6-6} 
$\mathcal{O}_{3W}$ & $\epsilon^{IJK}W_{\mu}^{I\nu}W_{\nu}^{J\rho}W_{\rho}^{K\mu}$ & $\mathcal{O}_{HW}$ & $W_{\mu\nu}^{I}W^{I\mu\nu}(H^{\dagger}H)$ & \multicolumn{2}{c|}{$H^{6}$}\tabularnewline
\cline{5-6} \cline{6-6} 
$\mathcal{O}_{3\widetilde{W}}$ & $\epsilon^{IJK}\widetilde{W}_{\mu}^{I\nu}W_{\nu}^{J\rho}W_{\rho}^{K\mu}$ & $\mathcal{O}_{H\widetilde{W}}$ & $\widetilde{W}_{\mu\nu}^{I}W^{I\mu\nu}(H^{\dagger}H)$ & $\mathcal{O}_{H}$ & $(H^{\dagger}H)^{3}$\tabularnewline
 &  & $\mathcal{O}_{HB}$ & $B_{\mu\nu}B^{\mu\nu}(H^{\dagger}H)$ &  & \tabularnewline
 &  & $\mathcal{O}_{H\widetilde{B}}$ & $\widetilde{B}_{\mu\nu}B^{\mu\nu}(H^{\dagger}H)$ &  & \tabularnewline
 &  & $\mathcal{O}_{HWB}$ & $W_{\mu\nu}^{I}B^{\mu\nu}(H^{\dagger}\tau^{I}H)$ &  & \tabularnewline
 &  & $\mathcal{O}_{H\widetilde{W}B}$ & $\widetilde{W}_{\mu\nu}^{I}B^{\mu\nu}(H^{\dagger}\tau^{I}H)$ &  & \tabularnewline
\hline 
\end{tabular}
\par\end{centering}
\caption{SMEFT bosonic operators.}
\end{table}
\par\end{center}

\begin{center}
\begin{table}[H]
\begin{centering}
\begin{tabular}{|c|c|c|c|c|c|}
\hline 
\multicolumn{2}{|c|}{$\psi^{2}XH+\text{h.c.}$} & \multicolumn{2}{c|}{$\psi^{2}H^{3}+\text{h.c.}$} & \multicolumn{2}{c|}{$\psi^{2}DH^{2}$}\tabularnewline
\hline 
$\mathcal{O}_{uG}$ & $(\overline{q}T^{A}\sigma^{\mu\nu}u)\widetilde{H}G_{\mu\nu}^{A}$ & $\mathcal{O}_{uH}$ & $(H^{\dagger}H)\overline{q}\widetilde{H}u$ & $\mathcal{O}_{Hq}^{(1)}$ & $(\overline{q}\gamma^{\mu}q)(H^{\dagger}i\overleftrightarrow{D}_{\mu}H)$\tabularnewline
$\mathcal{O}_{uW}$ & $(\overline{q}\sigma^{\mu\nu}u)\tau^{I}\widetilde{H}W_{\mu\nu}^{I}$ & $\mathcal{O}_{dH}$ & $(H^{\dagger}H)\overline{q}Hd$ & $\mathcal{O}_{Hq}^{(3)}$ & $(\overline{q}\tau^{I}\gamma^{\mu}q)(H^{\dagger}i\overleftrightarrow{D}_{\mu}^{I}H)$\tabularnewline
$\mathcal{O}_{uB}$ & $(\overline{q}\sigma^{\mu\nu}u)\widetilde{H}B_{\mu\nu}$ & $\mathcal{O}_{eH}$ & $(H^{\dagger}H)\overline{\ell}He$ & $\mathcal{O}_{Hu}$ & $(\overline{u}\gamma^{\mu}u)(H^{\dagger}i\overleftrightarrow{D}_{\mu}H)$\tabularnewline
$\mathcal{O}_{dG}$ & $(\overline{q}T^{A}\sigma^{\mu\nu}d)HG_{\mu\nu}^{A}$ &  &  & $\mathcal{O}_{Hd}$ & $(\overline{d}\gamma^{\mu}d)(H^{\dagger}i\overleftrightarrow{D}_{\mu}H)$\tabularnewline
$\mathcal{O}_{dW}$ & $(\overline{q}\sigma^{\mu\nu}d)\tau^{I}HW_{\mu\nu}^{I}$ &  &  & $\mathcal{O}_{Hud}$ & $(\overline{u}\gamma^{\mu}d)(\widetilde{H}^{\dagger}iD_{\mu}H)$\tabularnewline
$\mathcal{O}_{dB}$ & $(\overline{q}\sigma^{\mu\nu}d)HB_{\mu\nu}$ &  &  & $\mathcal{O}_{H\ell}^{(1)}$ & $(\overline{\ell}\gamma^{\mu}\ell)(H^{\dagger}i\overleftrightarrow{D}_{\mu}H)$\tabularnewline
$\mathcal{O}_{eW}$ & $(\overline{\ell}\sigma^{\mu\nu}e)\tau^{I}HW_{\mu\nu}^{I}$ &  &  & $\mathcal{O}_{H\ell}^{(3)}$ & $(\overline{\ell}\tau^{I}\gamma^{\mu}\ell)(H^{\dagger}i\overleftrightarrow{D}_{\mu}^{I}H)$\tabularnewline
$\mathcal{O}_{eB}$ & $(\overline{\ell}\sigma^{\mu\nu}e)HB_{\mu\nu}$ &  &  & $\mathcal{O}_{He}$ & $(\overline{e}\gamma^{\mu}e)(H^{\dagger}i\overleftrightarrow{D}_{\mu}H)$\tabularnewline
\hline 
\end{tabular}
\par\end{centering}
\caption{SMEFT two-fermion operators.}
\end{table}
\par\end{center}

\begin{center}
\begin{table}[H]
\begin{centering}
\begin{tabular}{|c|c|c|c|c|c|}
\hline 
\multicolumn{2}{|c|}{Four quark} & \multicolumn{2}{c|}{Four lepton} & \multicolumn{2}{c|}{Semileptonic}\tabularnewline
\hline 
$\mathcal{O}_{qq}^{(1)}$ & $(\overline{q}\gamma^{\mu}q)(\overline{q}\gamma_{\mu}q)$ & $\mathcal{O}_{\ell\ell}$ & $(\overline{\ell}\gamma^{\mu}\ell)(\overline{\ell}\gamma_{\mu}\ell)$ & $\mathcal{O}_{\ell q}^{(1)}$ & $(\overline{\ell}\gamma^{\mu}\ell)(\overline{q}\gamma_{\mu}q)$\tabularnewline
$\mathcal{O}_{qq}^{(3)}$ & $(\overline{q}\gamma^{\mu}\sigma^{I}q)(\overline{q}\gamma_{\mu}\sigma^{I}q)$ & $\mathcal{O}_{ee}$ & $(\overline{e}\gamma^{\mu}e)(\overline{e}\gamma_{\mu}e)$ & $\mathcal{O}_{\ell q}^{(3)}$ & $(\overline{\ell}\gamma^{\mu}\sigma^{I}\ell)(\overline{q}\gamma_{\mu}\sigma^{I}q)$\tabularnewline
$\mathcal{O}_{uu}$ & $(\overline{u}\gamma^{\mu}u)(\overline{u}\gamma_{\mu}u)$ & $\mathcal{O}_{\ell e}$ & $(\overline{\ell}\gamma^{\mu}\ell)(\overline{e}\gamma_{\mu}e)$ & $\mathcal{O}_{eu}$ & $(\overline{e}\gamma^{\mu}e)(\overline{u}\gamma_{\mu}u)$\tabularnewline
$\mathcal{O}_{dd}$ & $(\overline{d}\gamma^{\mu}d)(\overline{d}\gamma_{\mu}d)$ &  &  & $\mathcal{O}_{ed}$ & $(\overline{e}\gamma^{\mu}e)(\overline{d}\gamma_{\mu}d)$\tabularnewline
$\mathcal{O}_{ud}^{(1)}$ & $(\overline{u}\gamma^{\mu}u)(\overline{d}\gamma_{\mu}d)$ &  &  & $\mathcal{O}_{qe}$ & $(\overline{q}\gamma^{\mu}q)(\overline{e}\gamma_{\mu}e)$\tabularnewline
$\mathcal{O}_{ud}^{(8)}$ & $(\overline{u}\gamma^{\mu}T^{A}u)(\overline{d}\gamma_{\mu}T^{A}d)$ &  &  & $\mathcal{O}_{\ell u}$ & $(\overline{\ell}\gamma^{\mu}\ell)(\overline{u}\gamma_{\mu}u)$\tabularnewline
$\mathcal{O}_{qu}^{(1)}$ & $(\overline{q}\gamma^{\mu}q)(\overline{u}\gamma_{\mu}u)$ &  &  & $\mathcal{O}_{\ell d}$ & $(\overline{\ell}\gamma^{\mu}\ell)(\overline{d}\gamma_{\mu}d)$\tabularnewline
$\mathcal{O}_{qu}^{(8)}$ & $(\overline{q}\gamma^{\mu}T^{A}q)(\overline{u}\gamma_{\mu}T^{A}u)$ &  &  & $\mathcal{O}_{\ell edq}$ & $(\overline{\ell}e)(\overline{d}q)$\tabularnewline
$\mathcal{O}_{qd}^{(1)}$ & $(\overline{q}\gamma^{\mu}q)(\overline{d}\gamma_{\mu}d)$ &  &  & $\mathcal{O}_{\ell equ}^{(1)}$ & $(\overline{\ell}^{r}e)\epsilon_{rs}(\overline{q}^{s}u)$\tabularnewline
$\mathcal{O}_{qd}^{(8)}$ & $(\overline{q}\gamma^{\mu}T^{A}q)(\overline{d}\gamma_{\mu}T^{A}d)$ &  &  & $\mathcal{O}_{\ell equ}^{(3)}$ & $(\overline{\ell}^{r}\sigma^{\mu\nu}e)\epsilon_{rs}(\overline{q}^{s}\sigma_{\mu\nu}u)$\tabularnewline
$\mathcal{O}_{quqd}^{(1)}$ & $(\overline{q}^{r}u)\epsilon_{rs}(\overline{q}^{s}d)$ &  &  &  & \tabularnewline
$\mathcal{O}_{quqd}^{(8)}$ & $(\overline{q}^{r}T^{A}u)\epsilon_{rs}(\overline{q}^{s}T^{A}d)$ &  &  &  & \tabularnewline
\hline 
\end{tabular}
\par\end{centering}
\caption{SMEFT four-fermion baryon number conserving operators.}
\end{table}
\par\end{center}

\subsection{LEFT basis}
\label{sec:LEFTbasis}

\begin{center}
\begin{table}[H]
\begin{centering}
\begin{tabular}{|c|c|}
\hline 
\multicolumn{2}{|c|}{$(\overline{L}R)X$}\tabularnewline
\hline 
$O_{e\gamma}$ & $\overline{e}_{L}\sigma^{\mu\nu}e_{R}F_{\mu\nu}$\tabularnewline
$O_{u\gamma}$ & $\overline{u}_{L}\sigma^{\mu\nu}u_{R}F_{\mu\nu}$\tabularnewline
$O_{d\gamma}$ & $\overline{d}_{L}\sigma^{\mu\nu}d_{R}F_{\mu\nu}$\tabularnewline
$O_{uG}$ & $\overline{u}_{L}T^{A}\sigma^{\mu\nu}u_{R}G_{\mu\nu}^{A}$\tabularnewline
$O_{d\gamma}$ & $\overline{d}_{L}\sigma^{\mu\nu}d_{R}F_{\mu\nu}$\tabularnewline
\hline 
\end{tabular}
\par\end{centering}
\caption{LEFT dipole (dimension five) operators.}
\end{table}
\par\end{center}

\begin{center}
\begin{table}[H]
\begin{centering}
\begin{tabular}{|c|c|}
\hline 
\multicolumn{2}{|c|}{$X^{3}$}\tabularnewline
\hline 
$O_{3G}$ & $f^{ABC}G_{\mu}^{A\nu}G_{\nu}^{B\rho}G_{\rho}^{C\mu}$\tabularnewline
$O_{3\widetilde{G}}$ & $f^{ABC}\widetilde{G}_{\mu}^{A\nu}G_{\nu}^{B\rho}G_{\rho}^{C\mu}$\tabularnewline
\hline 
\end{tabular}
\par\end{centering}
\caption{LEFT bosonic operators.}
\end{table}
\par\end{center}

\begin{center}
% {\small{}}
\begin{table}[H]
%\begin{centering}
%{\small{}}%
\hskip-1.0cm
\begin{tabular}{|c|c|c|c|c|c|}
\hline 
\multicolumn{2}{|c|}{{\small{}$(\overline{L}L)(\overline{L}L)$}} & \multicolumn{2}{c|}{{\small{}$(\overline{L}L)(\overline{R}R)$}} & \multicolumn{2}{c|}{{\small{}($\overline{L}R)(\overline{L}R)$}}\tabularnewline
\hline 
{\small{}$O_{\nu\nu}^{V,LL}$} & {\small{}$(\overline{\nu}_{L}\gamma^{\mu}\nu_{L})(\overline{\nu}_{L}\gamma_{\mu}\nu_{L})$} & {\small{}$O_{\nu e}^{V,LR}$} & {\small{}$(\overline{\nu}_{L}\gamma^{\mu}\nu_{L})(\overline{e}_{R}\gamma_{\mu}e_{R})$} & {\small{}$O_{ee}^{S,RR}$} & {\small{}$(\overline{e}_{L}e_{R})(\overline{e}_{L}e_{R})$}\tabularnewline
{\small{}$O_{ee}^{V,LL}$} & {\small{}$(\overline{e}_{L}\gamma^{\mu}e_{L})(\overline{e}_{L}\gamma_{\mu}e_{L})$} & {\small{}$O_{ee}^{V,LR}$} & {\small{}$(\overline{e}_{L}\gamma^{\mu}e_{L})(\overline{e}_{R}\gamma_{\mu}e_{R})$} & {\small{}$O_{eu}^{S,RR}$} & {\small{}$(\overline{e}_{L}e_{R})(\overline{u}_{L}u_{R})$}\tabularnewline
{\small{}$O_{\nu e}^{V,LL}$} & {\small{}$(\overline{\nu}_{L}\gamma^{\mu}\nu_{L})(\overline{e}_{L}\gamma_{\mu}e_{L})$} & {\small{}$O_{\nu u}^{V,LR}$} & {\small{}$(\overline{\nu}_{L}\gamma^{\mu}\nu_{L})(\overline{u}_{R}\gamma_{\mu}u_{R})$} & {\small{}$O_{eu}^{T,RR}$} & {\small{}$(\overline{e}_{L}\sigma^{\mu\nu}e_{R})(\overline{u}_{L}\sigma_{\mu\nu}u_{R})$}\tabularnewline
{\small{}$O_{\nu u}^{V,LL}$} & {\small{}$(\overline{\nu}_{L}\gamma^{\mu}\nu_{L})(\overline{u}_{L}\gamma_{\mu}u_{L})$} & {\small{}$O_{\nu d}^{V,LR}$} & {\small{}$(\overline{\nu}_{L}\gamma^{\mu}\nu_{L})(\overline{d}_{R}\gamma_{\mu}d_{R})$} & {\small{}$O_{ed}^{S,RR}$} & {\small{}$(\overline{e}_{L}e_{R})(\overline{d}_{L}d_{R})$}\tabularnewline
{\small{}$O_{\nu d}^{V,LL}$} & {\small{}$(\overline{\nu}_{L}\gamma^{\mu}\nu_{L})(\overline{d}_{L}\gamma_{\mu}d_{L})$} & {\small{}$O_{eu}^{V,LR}$} & {\small{}$(\overline{e}_{L}\gamma^{\mu}e_{L})(\overline{u}_{R}\gamma_{\mu}u_{R})$} & {\small{}$O_{ed}^{T,RR}$} & {\small{}$(\overline{e}_{L}\sigma^{\mu\nu}e_{R})(\overline{d}_{L}\sigma_{\mu\nu}d_{R})$}\tabularnewline
{\small{}$O_{eu}^{V,LL}$} & {\small{}$(\overline{e}_{L}\gamma^{\mu}e_{L})(\overline{u}_{L}\gamma_{\mu}u_{L})$} & {\small{}$O_{ed}^{V,LR}$} & {\small{}$(\overline{e}_{L}\gamma^{\mu}e_{L})(\overline{d}_{R}\gamma_{\mu}d_{R})$} & {\small{}$O_{\nu edu}^{S,RR}$} & {\small{}$(\overline{\nu}_{L}e_{R})(\overline{d}_{L}u_{R})$}\tabularnewline
{\small{}$O_{ed}^{V,LL}$} & {\small{}$(\overline{e}_{L}\gamma^{\mu}e_{L})(\overline{d}_{L}\gamma_{\mu}d_{L})$} & {\small{}$O_{ue}^{V,LR}$} & {\small{}$(\overline{u}_{L}\gamma_{\mu}u_{L})(\overline{e}_{R}\gamma^{\mu}e_{R})$} & {\small{}$O_{\nu edu}^{T,RR}$} & {\small{}$(\overline{\nu}_{L}\sigma^{\mu\nu}e_{R})(\overline{d}_{L}\sigma_{\mu\nu}u_{R})$}\tabularnewline
{\small{}$O_{\nu edu}^{V,LL}$} & {\small{}$(\overline{\nu}_{L}\gamma^{\mu}e_{L})(\overline{d}_{L}\gamma_{\mu}u_{L})$} & {\small{}$O_{de}^{V,LR}$} & {\small{}$(\overline{d}_{L}\gamma_{\mu}d_{L})(\overline{e}_{R}\gamma^{\mu}e_{R})$} & {\small{}$O_{uu}^{S1,RR}$} & {\small{}$(\overline{u}_{L}u_{R})(\overline{u}_{L}u_{R})$}\tabularnewline
{\small{}$O_{uu}^{V,LL}$} & {\small{}$(\overline{u}_{L}\gamma^{\mu}u_{L})(\overline{u}_{L}\gamma_{\mu}u_{L})$} & {\small{}$O_{\nu edu}^{V,LR}$} & {\small{}$(\overline{\nu}_{L}\gamma^{\mu}e_{L})(\overline{d}_{R}\gamma_{\mu}u_{R})$} & {\small{}$O_{uu}^{S8,RR}$} & {\small{}$(\overline{u}_{L}T^{A}u_{R})(\overline{u}_{L}T^{A}u_{R})$}\tabularnewline
{\small{}$O_{dd}^{V,LL}$} & {\small{}$(\overline{d}_{L}\gamma^{\mu}d_{L})(\overline{d}_{L}\gamma_{\mu}d_{L})$} & {\small{}$O_{uu}^{V1,LR}$} & {\small{}$(\overline{u}_{L}\gamma^{\mu}u_{L})(\overline{u}_{R}\gamma_{\mu}u_{R})$} & {\small{}$O_{ud}^{S1,RR}$} & {\small{}$(\overline{u}_{L}u_{R})(\overline{d}_{L}d_{R})$}\tabularnewline
{\small{}$O_{ud}^{V1,LL}$} & {\small{}$(\overline{u}_{L}\gamma^{\mu}u_{L})(\overline{d}_{L}\gamma_{\mu}d_{L})$} & {\small{}$O_{uu}^{V8,LR}$} & {\small{}$(\overline{u}_{L}T^{A}\gamma^{\mu}u_{L})(\overline{u}_{R}T^{A}\gamma_{\mu}u_{R})$} & {\small{}$O_{ud}^{S8,RR}$} & {\small{}$(\overline{u}_{L}T^{A}u_{R})(\overline{d}_{L}T^{A}d_{R})$}\tabularnewline
{\small{}$O_{ud}^{V8,LL}$} & {\small{}$(\overline{u}_{L}T^{A}\gamma^{\mu}u_{L})(\overline{d}_{L}T^{A}\gamma_{\mu}d_{L})$} & {\small{}$O_{ud}^{V1,LR}$} & {\small{}$(\overline{u}_{L}\gamma^{\mu}u_{L})(\overline{d}_{R}\gamma_{\mu}d_{R})$} & {\small{}$O_{dd}^{S1,RR}$} & {\small{}$(\overline{d}_{L}d_{R})(\overline{d}_{L}d_{R})$}\tabularnewline
\cline{1-2} \cline{2-2} 
\multicolumn{2}{|c|}{{\small{}$(\overline{R}R)(\overline{R}R)$}} & {\small{}$O_{ud}^{V8,LR}$} & {\small{}$(\overline{u}_{L}T^{A}\gamma^{\mu}u_{L})(\overline{d}_{R}T^{A}\gamma_{\mu}d_{R})$} & {\small{}$O_{dd}^{S8,RR}$} & {\small{}$(\overline{d}_{L}T^{A}d_{R})(\overline{d}_{L}T^{A}d_{R})$}\tabularnewline
\cline{1-2} \cline{2-2} 
{\small{}$O_{ee}^{V,RR}$} & {\small{}$(\overline{e}_{R}\gamma^{\mu}e_{R})(\overline{e}_{R}\gamma_{\mu}e_{R})$} & {\small{}$O_{du}^{V1,LR}$} & {\small{}$(\overline{d}_{L}\gamma^{\mu}d_{L})(\overline{u}_{R}\gamma_{\mu}u_{R})$} & {\small{}$O_{uddu}^{S1,RR}$} & {\small{}$(\overline{u}_{L}d_{R})(\overline{d}_{L}u_{R})$}\tabularnewline
{\small{}$O_{eu}^{V,RR}$} & {\small{}$(\overline{e}_{R}\gamma^{\mu}e_{R})(\overline{u}_{R}\gamma_{\mu}u_{R})$} & {\small{}$O_{du}^{V8,LR}$} & {\small{}$(\overline{d}_{L}T^{A}\gamma^{\mu}d_{L})(\overline{u}_{R}T^{A}\gamma_{\mu}u_{R})$} & {\small{}$O_{uddu}^{S8,RR}$} & {\small{}$(\overline{u}_{L}T^{A}d_{R})(\overline{d}_{L}T^{A}u_{R})$}\tabularnewline
\cline{5-6} \cline{6-6} 
{\small{}$O_{ed}^{V,RR}$} & {\small{}$(\overline{e}_{R}\gamma^{\mu}e_{R})(\overline{d}_{R}\gamma_{\mu}d_{R})$} & {\small{}$O_{dd}^{V1,LR}$} & {\small{}$(\overline{d}_{L}\gamma^{\mu}d_{L})(\overline{d}_{R}\gamma_{\mu}d_{R})$} & \multicolumn{2}{c|}{{\small{}$(\overline{L}R)(\overline{R}L)$}}\tabularnewline
\cline{5-6} \cline{6-6} 
{\small{}$O_{uu}^{V,RR}$} & {\small{}$(\overline{u}_{R}\gamma^{\mu}u_{R})(\overline{u}_{R}\gamma_{\mu}u_{R})$} & {\small{}$O_{dd}^{V8,LR}$} & {\small{}$(\overline{d}_{L}T^{A}\gamma^{\mu}d_{L})(\overline{d}_{R}T^{A}\gamma_{\mu}d_{R})$} & {\small{}$O_{eu}^{S,RL}$} & {\small{}$(\overline{e}_{L}e_{R})(\overline{u}_{R}u_{L})$}\tabularnewline
{\small{}$O_{dd}^{V,RR}$} & {\small{}$(\overline{d}_{R}\gamma^{\mu}d_{R})(\overline{d}_{R}\gamma_{\mu}d_{R})$} & {\small{}$O_{uddu}^{V1,LR}$} & {\small{}$(\overline{u}_{L}\gamma^{\mu}d_{L})(\overline{d}_{R}\gamma_{\mu}u_{R})$} & {\small{}$O_{ed}^{S,RL}$} & {\small{}$(\overline{e}_{L}e_{R})(\overline{d}_{R}d_{L})$}\tabularnewline
{\small{}$O_{ud}^{V1,RR}$} & {\small{}$(\overline{u}_{R}\gamma^{\mu}u_{R})(\overline{d}_{R}\gamma_{\mu}d_{R})$} & {\small{}$O_{uddu}^{V8,LR}$} & {\small{}$(\overline{u}_{L}T^{A}\gamma^{\mu}d_{L})(\overline{d}_{R}T^{A}\gamma_{\mu}u_{R})$} & {\small{}$O_{\nu edu}^{S,RL}$} & {\small{}$(\overline{\nu}_{L}e_{R})(\overline{d}_{R}u_{L})$}\tabularnewline
{\small{}$O_{ud}^{V8,RR}$} & {\small{}$(\overline{u}_{R}T^{A}\gamma^{\mu}u_{R})(\overline{d}_{R}T^{A}\gamma_{\mu}d_{R})$} &  &  &  & \tabularnewline
\hline 
\end{tabular}
%{\small\par}
%\par
%\end{centering}
{\small{}\caption{LEFT four-fermion lepton and baryon number conserving operators.}
}
%{\small\par}
\end{table}
%{\small\par}
%\par
\end{center}

\newpage

%%%%%%%%%%%%%%%%%%%%%%%%%%%%%%%%
%%%%%%%%%%%%%%%%%%%%%%%%%%%%%%%%
{\small
\bibliography{Biblio}{}
\bibliographystyle{JHEP}}

\end{document}